\documentclass[useAMS,usenatbib,usegraphicx,fleqn]{mn2e}

\usepackage{color}
\usepackage{psfig}
\usepackage{amssymb}
\usepackage{savesym}
\usepackage{amsmath}
\usepackage{hyperref}
\usepackage{multirow}
\hypersetup{final=false}

\newcommand{\ssc}{_c} 
 
\newcommand{\ssi}{_i} 
\newcommand{\ssj}{_j} 
\newcommand{\sss}{_s} 
\newcommand{\sst}{_t}

\newcommand{\ssij}{_{ij}}  
\newcommand{\ssis}{_{is}}

\newcommand{\sssi}{_{si}}  
  
\newcommand{\ssst}{_{st}}

\newcommand{\meanhij}{\overline{h}\ssij}

\newcommand{\meanhst}{\overline{h}\ssst}

\newcommand{\refone}{}

\def\be{\begin{equation}}
\def\ee{\end{equation}}

\newcommand{\arepo}{{\small{AREPO}} }
\newcommand{\gandalf}{{\small{GANDALF}}}
\newcommand{\seren}{{\small{SEREN}} }
\newcommand{\gizmo}{{\small{GIZMO}} }

\newcommand{\deriv}[2]{\ensuremath{\frac{d#1}{d#2}}}
\newcommand{\pderiv}[2]{\ensuremath{\frac{\partial #1}{\partial #2}}}

\title[GANDALF]{GANDALF - Graphical Astrophysics code for N-body Dynamics And Lagrangian Fluids}

\author[Hubber, Rosotti \& Booth]{D. A. Hubber$^{1,2}$\thanks{E:mail:dhubber@usm.lmu.de}, G. P. Rosotti$^{3}$, R. A. Booth$^{3}$ \\
$^{1}$Universitats-Sternwarte M\"unchen, Scheinerstra\ss{}e 1, D-81679 M\"unchen, Germany \\
$^{2}$Excellence Cluster Universe, Boltzmannstr. 2, D-85748 Garching, Germany\\
$^{3}$Institute of Astronomy, Madingley Rd, Cambridge, CB3 0HA, UK}

\begin{document}
           
\date{28/06/2017}

\pagerange{\pageref{firstpage}--\pageref{lastpage}} \pubyear{2017}

\label{firstpage}

\maketitle

\begin{abstract}
{\small GANDALF} is a new hydrodynamics and N-body dynamics code designed for investigating planet formation, star formation and star cluster problems.  {\small GANDALF} is written in C++, parallelised with both OpenMP and MPI and contains a python library for analysis and visualisation.  The code has been written with a fully object-oriented approach to easily allow user-defined implementations of physics modules or other algorithms.  The code currently contains implementations of Smoothed Particle Hydrodynamics, Meshless Finite-Volume and collisional N-body schemes, but can easily be adapted to include additional particle schemes.  We present in this paper the details of its implementation, results from the test suite, serial and parallel performance results and discuss the planned future development.   The code is freely available as an open source project on the code-hosting website github at \href{https://github.com/gandalfcode/gandalf}{https://github.com/gandalfcode/gandalf} and is available under the GPLv2 license.
\end{abstract}


\section{Introduction} \label{S:INTRO}

Numerical simulations are becoming increasingly more important in modern astrophysics research. They allow us to study systems where analytical solutions do not exist and explore the complex (non-linear) interplay due to the multiple physical processes that are normally present in astrophysical problems.  In recent years more attention has been given to exploring which algorithms give the most accurate and reliable results and comparing different algorithms to one another, as well as the development of brand new or hybrid algorithms.  While many specialist codes exist with single hard-wired implementations of particular physical processes (e.g. Hydrodynamics), the current desire for flexibility in algorithm choice is not always fulfilled with a single code and may often require using multiple codes for a single project.

In this paper we present {\small GANDALF} ({G}raphical {A}strophysics code for {N}-body {D}ynamics {A}nd {L}agrangian {F}luids), a new multi-purpose hydrodynamics, N-body and analysis code.   {\small GANDALF} has been designed with Star and Planet Formation problems in mind, but with the flexibility to be extended with different physics algorithms to simulate other kinds of astrophysical problems.

{\small GANDALF} was developed with a heavy object-oriented design philosophy in order to improve code maintainability and simplify the process of implementing new features in the future.  C++ was chosen as the main development language as a low-level, high-performance computing (HPC) object-oriented language that is easy to bind with other (often C-based) external libraries and can easily be parallelised with both OpenMP and MPI (either individually or combined with a hybrid OpenMP-MPI approach).  {\small GANDALF} also contains an optional Python library, which can be used for analysis and visualisation of whole simulations or single snapshots.  It is also possible to generate initial conditions and set-up and run the simulation from a Python script making it easier for users not accustomed with C++.

{\small GANDALF} contains implementations of two particle-based hydrodynamics schemes, Smoothed Particle Hydrodynamics \citep[SPH; e.g.][]{Monaghan1992} and the Meshless Finite-Volume scheme \citep[MFV;][]{LV2008, GN2011, GIZMO}. Many algorithms (e.g. gravity, the tree used for neighbour finding) are shared between the two implementations, minimising the amount of code duplication.  {\small GANDALF} also includes algorithms for collisional N-body dynamics.

This paper is structured as follows.  In Section \ref{S:HYDRO}, we discuss the Hydrodynamical algorithms that we have implemented into {\small GANDALF}, including any differences from traditional implementations.  In Section \ref{S:NBODY}, we discuss our implementations of the collisional N-body and sink particle algorithms.  In Section \ref{S:MISC}, we discuss other miscellaneous algorithms such as implementing boundary conditions and trees.  In Section \ref{S:IMPLEMENTATION}, we discuss the class structure of the code, how to add new classes on top of the existing framework, the python library and how it can be easily used to perform analysis and run the code.   In Section \ref{S:TESTS}, we present results from our test suite comparing all methods against each other and against other published codes.  We also show the serial and parallel performance of the code.  In Section \ref{S:PERFORMANCE}, we discuss the performance and parallel scaling of the code, both with OpenMP and hybrid OpenMP/MPI. In Section \ref{S:FUTUREWORK}, we briefly discuss ongoing work with the code and planned features for the future.

\section{Hydrodynamical methods in GANDALF} \label{S:HYDRO}

{\refone {\small GANDALF} solves the traditional Euler Equations of Hydrodynamics with additional physics terms such as gravitational accelerations.  In Lagrangian form, these are 
\begin{eqnarray}
\frac{d\rho}{dt} &=& - \rho \,\nabla \cdot {\bf v} \\
\frac{d{\bf v}}{dt} &=& -\frac{\nabla P}{\rho} - \nabla \Phi \\
\frac{du}{dt} &=& -\frac{P}{\rho} \nabla \cdot {\bf v} \\
\nabla^2 \Phi &=& 4\,\pi\,G\,\rho\,, \label{Eqn:Poisson}
\end{eqnarray}
where $\rho$ is the fluid density, {\bf v} is the fluid velocity, $u$ is the specific internal, $P$ the thermal pressure and $\Phi$ is the gravitational potential and $G$ is Newton's constant. 
}

{\small GANDALF} contains implementations of two particle-based hydrodynamical schemes that use the {\it smoothing kernel} as a fundamental quantity in {\refone solving the numerical form of these equations}.  The fluid properties of all particles are smoothed over a length scale $h$, called the {\it smoothing length}, with a weighting function $W({\bf r},h)$ called the {\it kernel function}.  Each particle occupies/influences a spherical volume called the {\it smoothing kernel} of total radius ${\cal R}\,h$.  The fluid particles interact with neighbouring particles, i.e. particles whose smoothing kernels overlap, where the interaction is weighted somewhat by the kernel function.  The exact details of how the smoothing kernel influences the hydrodynamical equations are explained in each scheme's implementation.

{\small GANDALF} contains two principal kernel functions which have a finite extent of ${\cal R}\,h$; (i) the M4 cubic spline kernel \citep{ML1985} with ${\cal R} = 2$ and (ii) the quintic spline kernel \citep{MorrisPhD} with ${\cal R} = 3$.  The complete mathematical description of all these kernels, plus related derivative and integrated quantities, are given in Appendix A of \citet{Hubber2011}.

\subsection{Smoothed Particle Hydrodynamics} \label{S:SPH}

SPH \citep{Lucy1977,GM1977} is a popular Lagrangian hydrodynamics scheme that has been implemented in many astrophysical hydrodynamics codes, such as {\small GADGET2} \citep{Gadget2}, {\small VINE} \citep{VINE2009}, {\small SEREN} \citep{Hubber2011} and {\small PHANTOM} \citep{Phantom2017}.  The main advantages of SPH are (i) it is simple conceptually and to code, and (ii) its Lagrangian nature which provides various advantages over Eulerian methods, {\refone such as having an in-built adaptivity to the wide range of densities found in gravitational collapse problems, (iii) it can be derived from the Euler-Lagrange equations so is naturally conservative, (iv) it can be integrated with symplectic equations such as the Leapfrog resulting in good orbital conservation properties (e.g. angular-momentum conservation) and (v) it can be easily coupled to the N-body equations of motion when including point gravitational sources (e.g. stars, planets)}.  SPH has been derived in many mathematical forms, each with different assumptions, different integration variables or different methods of computing hydrodynamical quantities.  \gandalf{} currently uses the standard conservative conservative `grad-h' SPH following \citet{SH2002} and \citet{Price2012}, with the pressure-entropy scheme of \citet{SM2013} planned for the future.

\subsubsection{Conservative 'grad-h' SPH} \label{SS:GRADHSPH}
Conservative `grad-h' SPH \citep{SH2002} is one of the standard derivations of the SPH equations that is used in astrophysical codes, such as {\small GADGET2}.  The fluid equations are derived from Lagrangian mechanics and hence guarantee conservation of mass, momentum, angular momentum and energy to at least integration error.  However, it should be noted that the use of the tree {\refone in calculating gravitational accelerations} and block time-stepping algorithms introduces additional sources of error meaning `perfect' conservation is not achieved in practice.

The algorithm described here is similar to that implemented in \seren \citep{Hubber2011}.  We first compute the density, $\rho$, and smoothing length, $h$ of each SPH particle.  The smoothed density for particle $i$ is given by 
\begin{equation} \label{EQN:SPHRHO}
\rho\ssi = \sum \limits_{j=1}^{N}  m\ssj W({\bf r}\ssij,h\ssi)\,,
\end{equation}
where ${\bf r}\ssij = {\bf r}\ssi - {\bf r}\ssj$, $W({\bf r}\ssij, h\ssi)$ is the smoothing kernel and $m\ssj$ is the mass of particle $j$.  The density and smoothing length are related by the simple relation
\begin{equation} \label{EQN:HRHO}
h\ssi = \eta_{_{\rm SPH}} \left( \frac{m\ssi }{\rho\ssi } \right)^{\frac{1}{D}}\,,
\end{equation}
where $D$ is the dimensionality of the simulation and $\eta_{_{\rm SPH}}$ is a dimensionless parameter that relates the smoothing length to the local inter-particle spacing (default value $\eta_{_{\rm SPH}} = 1.2$).  Since $h$ and $\rho$ depend on each other, we must iterate their values until Equations \ref{EQN:SPHRHO} and \ref{EQN:HRHO} converge to some tolerance, {\refone usually to within about $\sim 1\,\%$.}

The SPH momentum equation is given by 
\begin{equation} \label{EQN:GRADHMOMEQN}
\frac{d{\bf v}\ssi }{dt} = -
\sum \limits_{j=1}^{N}  m\ssj  \left\{ 
\frac{P\ssi}{\Omega\ssi \rho\ssi^2} \nabla\ssi W({\bf r}\ssij ,h\ssi) + \frac{P\ssj}{\Omega\ssj \rho\ssj^2} \nabla\ssi W({\bf r}\ssij ,h\ssj) \right\}
\,,
\end{equation}
where $P\ssi = (\gamma - 1)\,\rho\ssi\,u\ssi$ is the thermal pressure, $u\ssi$ is the specific internal energy, {\refone $\gamma$ is the ratio of specific heats for an ideal gas}, $\nabla\ssi W$ is the kernel gradient and 
\begin{equation} \label{EQN:OMEGA}
\Omega\ssi = 1 - \frac{\partial h\ssi }{\partial \rho\ssi } 
\sum \limits_{j=1}^{N}  m\ssj  \frac{\partial W}{\partial h}
({\bf r}\ssij , h\ssi )
\end{equation}
is a dimensionless correction term that accounts for the spatial variability of $h$ amongst its neighbouring particles.

{If the temperature is not prescribed (e.g. by an isothermal {\refone equation of state; hereafter EOS}), we integrate an energy equation of the form}
\begin{equation} \label{EQN:GRADHENEQN}
\frac{du\ssi }{dt} = 
\frac{P\ssi }{\Omega\ssi  \rho\ssi ^2} \sum \limits_{j=1}^{N}
m\ssj  {\bf v}\ssij \cdot \nabla W\ssij ({\bf r}\ssij , h\ssi )\,,
\end{equation}
where ${\bf v}\ssij = {\bf v}\ssi - {\bf v}\ssj$.

The SPH equations presented so far describe a fluid without dissipation, where the fluid quantities are always continuous. However, many astrophysical problems contain shocks, which lead to dissipation and need to be handled properly.  We use the \citet{Monaghan1997} formulation of artificial viscosity for shock-capturing,
\begin{eqnarray} 
\frac{d{\bf v}\ssi}{dt} &=& 
\sum\limits_{j=1}^{N}\,\frac{m\ssj}{\overline{\rho}\ssij}\,\left\{\alpha_{_{\rm AV}}\,v_{_{\rm SIG}}\mu\ssij\right\}\,\overline{\nabla\ssi W}\ssij\,,\label{EQN:MON97ARTVISC} \\
\frac{du\ssi}{dt} &=& -\,\sum\limits_{j=1}^{N}\,\frac{m\ssj}{\overline{\rho}\ssij}\,\frac{\alpha_{_{\rm AV}}\,v_{_{\rm SIG}}\mu\ssij^2}{2} \,\hat{\bf r}\ssij\cdot\overline{\nabla\ssi W}\ssij\, \nonumber \\
&& +\,\sum\limits_{j=1}^{N}\,\frac{m\ssj}{\overline{\rho}\ssij}\, \alpha_{_{\rm AC}}\,v_{_{\rm SIG}}'(u\ssi - u\ssj) \,\hat{\bf r}\ssij\cdot\overline{\nabla\ssi W}\ssij\,,
\label{EQN:MON97ENERGYDISS}
\end{eqnarray}
where $\alpha_{_{\rm AV}}$ and $\alpha_{_{\rm AC}}$ are constants of order unity that control the dissipation strength, $v_{_{\rm SIG}}$ and $v_{_{\rm SIG}}'$ are the signal speeds for artificial viscosity and conductivity respectively, $\hat{\bf r}\ssij = {\bf r}\ssij/|{\bf r}\ssij|$ and $\overline{\nabla\ssi W}\ssij = \frac{1}{2}\,\left(\nabla\ssi W({\bf r}\ssij ,h\ssi) + \nabla\ssi W({\bf r}\ssij ,h\ssj) \right)$ and $\mu\ssij = {\rm MIN}(0, {\bf v}\ssij \cdot {\bf r}\ssij)$.  For artificial viscosity, we use $v_{_{\rm SIG}} = c\ssi + c\ssj - \beta_{_{\rm AV}}\,{\bf v}\ssij \cdot \hat{\bf r}\ssij$, {\refone where $c\ssi$ and $c\ssj$ are the sound speeds of particles $i$ and $j$ respectively} and $\beta_{_{\rm AV}} = 2\,\alpha_{_{\rm AV}}$.  The signal speed for artificial conductivity is problem and physics dependent. By default, we chose the \citet{Wadsley2008} prescription, where $v_{_{\rm SIG}}' = |{\bf v}\ssij\cdot\hat{\bf r}\ssij|$ {\refone although the \citet{Price2008} conductivity, $v_{_{\rm SIG}}' = \sqrt{|P\ssi - P\ssj|/\overline{\rho}\ssij}$, is also available in the code.}

{\refone Since excessive dissipation is undesirable in hydrodynamical codes, we have implemented two artificial viscosity switches, \citet{1997JCoPh.136...41M} and \citet{2010MNRAS.408..669C}, in order to reduce the artificial viscosity as much as possible in regions away from shocks.}

\subsubsection{Self-gravity}

Computing self-gravity in SPH can be done consistently by considering the continuous density field given by Equation \ref{EQN:SPHRHO} in the Poisson Equation (Equation \ref{Eqn:Poisson}), instead of solving the N-body problem with each particle representing a discrete point mass  \citep{PM2007}.  Deriving the Equations of Motion via Lagrangian mechanics leads to a conservative set of Equations with self-gravity. The SPH gravitational acceleration is given by  
\begin{flalign}
{\bf g}\ssi =& -\,G\sum\limits_{j=1}^{N}{m\ssj\,
\frac{\phi'({\bf r}\ssij , h\ssi ) + \phi'({\bf r}\ssij , h\ssj )}{2} \,\hat{\bf r}\ssij}\, \nonumber \\ 
 &- \,\frac{G}{2}\sum\limits_{j=1}^{N}m\ssj\left\{\frac{\zeta_i}{\Omega_i}\nabla W\ssi({\bf r}\ssij,h\ssi)+\frac{\zeta_j}{\Omega_j}\nabla W\ssi({\bf r}\ssij,h\ssj)\right\}\,,
\label{EQN:SPHSELFGRAV}
\end{flalign}
where 
\begin{equation} \label{EQN:GRAVFORCEKERNEL}
\phi'({\bf r},h) = \frac{4\,\pi}{r^2}\int\limits_0^r{W({\bf r}',h)\,r'^2\,dr'}\,,
\end{equation}
\begin{equation} \label{EQN:GRADHZETA}
\zeta_i = \frac{\partial h\ssi}{\partial \rho\ssi} 
\sum \limits_{j=1}^{N} m_j \frac{\partial \phi}{\partial h}({\bf r}\ssij, h\ssi)\,,
\end{equation}
and $\Omega\ssi$ is given by Eqn. \ref{EQN:OMEGA}.  $\phi'({\bf r},h)$ is often called the gravitational force or gravitational acceleration kernel and in effect calculates the gravitational force between SPH particles accounting for the smoothed density distribution.  {\refone Similarly $\phi({\bf r},h)$ is the gravitational potential kernel which gives the smoothed gravitational potential.}  The $\zeta$ term is an additional term to $\Omega$ in accounting for the spatial variation of $h$ for self-gravity.

\subsubsection{Time integration} \label{SS:TIMESTEPPING}

The SPH particles can be integrated with two related integration schemes, the Leapfrog kick-drift-kick (KDK) and the Leapfrog drift-kick-drift (DKD) schemes.  Leapfrog schemes are symplectic schemes that exhibit accurate but stable integration of gravitational orbits.  The KDK and DKD schemes are mathematically equivalent in the case of global, constant time-steps with similar integration errors.  However, in the case of non-constant, individual time-steps (see Section \ref{SS:BLOCKTIMESTEPS}), they can behave differently with different rates of error growth.

The position and velocity of a particle integrated with the KDK scheme is described by :
\begin{eqnarray}
{\bf r}\ssi^{n+1} &=& {\bf r}\ssi^{n} + {\bf v}\ssi^{n}\,\Delta t + \tfrac{1}{2}\,{\bf a}\ssi^{n}\,\Delta t^2\,, \label{EQN:LFKDK1} \\
{\bf v}\ssi^{n+1} &=& {\bf v}\ssi^{n} + \tfrac{1}{2}\,\left( {\bf a}\ssi^{n} + {\bf a}\ssi^{n+1} \right)\,\Delta t\,. \label{EQN:LFKDK2}
\end{eqnarray}
where $\Delta t$ is the time-step.  Although the acceleration appears twice in Equation \ref{EQN:LFKDK2}, we only compute it once per step since the second acceleration term, ${\bf a}\ssi^{n+1}$, then becomes the first acceleration term for the next step.

In the DKD scheme, the updates to the positions and velocities are shifted by half a step : 
\begin{eqnarray}
{\bf r}\ssi^{n+1/2} &=& {\bf r}\ssi^{n} + \tfrac{1}{2}{\bf v}\ssi^{n}\,\Delta t\,, \label{EQN:LFDKD1} \\
{\bf v}\ssi^{n+1/2} &=& {\bf v}\ssi^{n} + \tfrac{1}{2}\,{\bf a}\ssi^{n-1/2}\,\Delta t\,, \label{EQN:LFDKD2} \\
{\bf v}\ssi^{n+1} &=& {\bf v}\ssi^{n} + {\bf a}\ssi^{n+1/2}\,\Delta t\,, \label{EQN:LFDKD3} \\
{\bf r}\ssi^{n+1} &=& {\bf r}\ssi^{n} + \tfrac{1}{2}\,\left( {\bf v}\ssi^{n} + {\bf v}\ssi^{n+1} \right)\,\Delta t\,. \label{EQN:LFDKD4}
\end{eqnarray}
The acceleration is computed only once, at the midpoint of the step.  This requires in practice the DKD scheme to be computed as a two-step scheme, where particles are `drifted' to the mid-point, the acceleration is computed and then the second half of the step is computed with the updated acceleration.

\subsubsection{Time-stepping} \label{SS:BLOCKTIMESTEPS}

All SPH schemes use a {\refone Courant-Friedrichs-Lewy (CFL)}-like condition to compute the time-steps, $\Delta\,t\ssi$, of the form : 
\begin{equation} \label {EQN:SPH_DT}
\Delta\,t\ssi = C_{_{\rm CFL}}\,\frac{h\ssi}{|v_{\rm sig,i}|}\,,
\end{equation}
where {\refone $C_{_{\rm CFL}}$ is a dimensionless timestep multiplier (typically $\sim 0.2$) analogous to the Courant number in grid codes and the signal speed is }
\begin{equation}
v_{\rm sig,i} = {\rm MAX}\ssj \left[ { c\ssi + c\ssj - \beta_{_{\rm AV}}{\rm MIN} \left(0, {\bf v}\ssij \cdot \hat{\bf r}\ssij \right)  }  \right]\,.
\end{equation}
The signal velocity, $v_{\rm sig,i}$, is the speed of propagation of information either through sound waves or translational velocity.  In effect, Eqn. \ref{EQN:SPH_DT} prevents information from crossing the smoothing kernel in a single time-step.  The $\beta_{_{\rm AV}}$ term exists to ensure strong shocks are captured adequately.  If additional physics (e.g. self-gravity) are employed, then we use a second criterion called the {\it acceleration condition}, i.e. 
\begin{equation} \label {EQN:SPH_DTACCEL}
\Delta\,t\ssi =  C_{_{\rm GRAV}} \frac{h\ssi}{\sqrt{|{\bf a}\ssi|}}\,
\end{equation}
{\refone where $C_{_{\rm GRAV}}$ is the dimensionless gravitational acceleration timestep multiplier (typically $\sim 0.5$).}

{\small GANDALF} uses a hierarchical block time-stepping scheme, similar to many other SPH and N-body codes {\refone like {\small GADGET} \citep{Gadget2} and {\small NBODY6} \citep{Aarseth2003}}.  The basic principle is that all time-steps are integer power-of-two multiples of some base time-step.  In {\small GANDALF}, we fix the maximum time-step, $\Delta\,t_{_{\rm MAX}}$, based on the time-steps available whenever the block time-steps are recomputed.  By default, particles on the maximum time-step occupy level $l = 0$.  Particles on higher levels $l$ therefore occupy shorter time-steps, i.e.
\begin{eqnarray}
\Delta\,t_l &=& \frac{\Delta\,t_{_{\rm MAX}}} {2^{l}}\;\;\; {\rm where\;} l = 0, 1, 2, ..., l_{_{\rm MAX}}\,.
\end{eqnarray}
{\refone The time-step level for a given particle} is allowed to increase to an arbitrarily high number based on the given time-step criterion when required.  However, {\refone the time-step level can only be reduced (a) by one level at a time, and (b) when the new time-step level} is correctly synchronised within the time-step hierarchy.  When we have completed exactly one full time-step (on the lowest level) then all particles are synchronised and we can recompute the full time-step hierarchy again.

\subsubsection{Time-step limiter} \label{SS:TIMESTEPLIMITER}

Block time-steps can introduce numerical artifacts in the results of a simulation if particles on very different time-steps are allowed to interact with each other. {\refone As an extreme example, particles in a cold, low density region may have too long time steps to react to the passage of a shock front. 
In \gandalf{} we solve this problem similarly to \citet{SM2009}, using a dual approach including both a predictive and reactive component. In the predictive component, for each particle we keep track of the minimum time-step of its neighbours during the hydrodynamic force calculation. When assigning new time-steps to the particle, we ensure that the particle does not have a time-step more than a fixed factor longer than the minimum of its neighbours.}

{\refone Additionally, we apply a reactive limiter for two reasons: 1) in the predictive component we employ the \textit{old} time-step of the neighbours. This does not guarantee that the \textit{current} time-step obeys the level constraint, once the new time-step has been computed; 2) the time-step of the neighbours may reduce rapidly, e.g. due to an approaching shock. The reactive limiter works by checking whether the minimum time-step of its neighbours has reduced below the acceptable level. This is achieved by using a scatter gather operation, i.e. active particles inform their inactive neighbours of their time-step during the hydrodynamic force calculation.  If the neighbour time-step criterion is found to be violated, the inactive particle's time-step is reduced and it becomes active as soon as its new time-step is synchronized with the time-step hierarchy.}

{\refone We note that the predictive tree-based limiter based on \citet{AREPO} included in the meshless scheme \autoref{SSS:MFV_TIMING} is not currently included in SPH. This is for pragmatic reasons: the primary advantage of the tree-based limiter is in maintaining exact conservation, which is already not maintained in SPH when block timesteps are used. Given that it is more expensive than the \citet{SM2009} type limiter (which already performs well) and can introduce unnecessarily small time-steps when gravity is included, we see no clear reason to use it in SPH. However, there is no fundamental reason it could not be easily added.}

\subsection{Meshless Finite-Volume scheme} \label{SS:MFV}

The Meshless Finite-Volume (MFV) scheme is a Hydrodynamical scheme developed originally by \citet{LV2008} and further developed for Astrophysical applications by \citet{GN2011} and \citet{GIZMO}.  The MFV scheme combines elements of both SPH and traditional Finite-Volume schemes \citep[see][]{ToroBook} where freely-moving particles interact and exchange mass, momentum and energy using a 2nd-order Godunov approach but weighted with a smoothing kernel.  We provide here a summary derivation presenting the main assumptions and equations as implemented in {\small GANDALF}.

\subsubsection{Volume discretisation} \label{SSS:MFVVOLUME}

Similar to SPH, the MFV scheme uses the smoothing kernel to compute various smoothed quantities.  We first compute the smoothing length of all the particles using the number density, $n$, instead of the mass density, $\rho$, i.e. 
\begin{equation} \label{EQN:MFV_NDENS}
n\ssi = \sum \limits_{j=1}^{N} W({\bf r}\ssij,h\ssi)\,,
\end{equation}
where the $n\ssi$ and $h\ssi$ are related by 
\begin{equation} \label{EQN:MFV_HN}
h\ssi = \eta_{_{\rm MFV}} {n\ssi}^{-\frac{1}{D}}\,,
\end{equation}
and $\eta_{_{\rm MFV}}$ is a dimensionless parameter analogous to $\eta_{_{\rm SPH}}$ controlling the number of neighbours. For comparison with our results in Section \ref{S:TESTS}, \cite{GIZMO} presents results consistent with $\eta_{_{\rm MFV}} = 1$ in 1 and 3D but with a larger value $\eta_{_{\rm MFV}} \approx 1.13$ in 2D.

In order to discretise the fluid onto a set of $N$ particles, we must chose a method of partitioning the surrounding fluid volume between the different particles.  \cite{AREPO} uses a Voronoi tessellation, which assigns a volume element to its nearest particle.  \citet{LV2008} instead use the SPH kernel to calculate the fraction of a volume element $d^{\mu}\,{\bf r}$ that is assigned to particle $i$, $\psi\ssi({\bf r}) = W({\bf r} - {\bf r}\ssi, h({\bf r}))\,n({\bf r})^{-1}$.  In effect, the particles `share' the surrounding volume in a similar way to SPH, resulting in an ensemble of overlapping `fuzzy' volume elements \citep[see Figure 1 of][for a useful visual aid]{GIZMO}.  The partition function should be normalised such that $\sum_{1} {\psi\ssi({\bf r})} = 1$ everywhere.  The numerical volume of a particle becomes the integral of all the partial volume elements, i.e. $V\ssi = \int { \psi\ssi({\bf r})\,d^{\mu}{\bf r}}$. Since this integral cannot be computed analytically for arbitrary particle distributions, we follow \cite{GIZMO} in using the second order accurate approximation, 
$V\ssi \sim n\ssi^{-1} = \left( \eta_{_{\rm MFV}} / h\ssi \right)^{D}$.

\subsubsection{Gradient operators} \label{SSS:MFVGRADIENTS}

Instead of using a SPH-type gradient operator, \citet{LV2008} use a least-squares matrix operator which is accurate to second-order and is relatively inexpensive to calculate.  In this form, the gradient of a general function $f\ssi$ for particle $i$ is given by : 
\begin{equation}
\left( \nabla^{\alpha}\,f \right)\ssi = \sum \limits_{j}^{} { \left( f\ssj - f\ssi \right)\,\tilde{\psi}^{\alpha}_{j}({\bf r}_i) }\,, \label{EQN:MFVGRADIENT}
\end{equation}
where $j$ is the summation over all (overlapping) neighbouring particles, 
\begin{equation}
\tilde{\psi}^{\alpha}_{j}({\bf r}_i) = \sum_{\beta = 1}^{\beta = \mu} { {\bf B}^{\alpha\,\beta}_{i}\, \left( {\bf r}_j - {\bf r}_i \right)^{\beta}\, \psi_j ({\bf r}_i) }\,, \label{EQN:MFVPSI}
\end{equation}
where ${\bf B} \equiv {\bf E}^{-1}$ and 
\begin{equation}
{\bf E}^{\alpha\,\beta}\ssi = \sum \limits_{j} {
\left( {\bf r}_j - {\bf r}_i \right)^{\alpha}\,\, \left( {\bf r}_j - {\bf r}_i \right)^{\beta}\,\psi\ssj({\bf r}_i) }\,. \label{EQN:MFVEMATRIX}
\end{equation}
In rare cases with pathological particle distributions, the gradient matrix can become close to singular resulting in poor gradient estimation. We follow \citet{GIZMO} in using the condition number of the matrix ${\bf E}$ to detect the occurrence of bad gradients. When the condition number exceeds 100, we switch to a direct SPH estimate of the gradient. We use a constant exact linear gradient estimate (equation 72, \citealt{Price2012}), which is equivalent to making the substitution
\begin{equation}
\tilde{\psi}^\alpha\ssj(r) \rightarrow V_i \nabla^\alpha\ssi W\ssij.
\end{equation}
This substitution is made in both the gradient computation and the face area (${\bf A}\ssij$ below).

\subsubsection{The Euler equations in conserved form} \label{SSS:MFVEULER}

In traditional Finite-Volume schemes, each fluid cell is a discrete volume where mass, momentum and energy is exchanged at well-defined boundaries between adjacent cells.  Traditional grid codes often use the vector ${\bf U} = \left( \rho\ssi, \rho\ssi{\bf v}\ssi, \rho\ssi e\ssi \right)$, which are the conserved quantities (mass, momentum and energy) per unit volume. Since the particle volume can change in MFV, the vector ${\bf Q} \equiv V\,{\bf U} = \left( m\ssi, m\ssi {\bf v}\ssi, E\ssi \right)$ is more appropriate.  We also use the vector, ${\bf W} = \left( \rho\ssi, {\bf v}\ssi, P\ssi \right)$, which is the vector of primitive quantities given to the Riemann solver. 

The general conservation laws for Hydrodynamics in a moving frame ${\bf v}_{\rm frame}$ are
\begin{equation}
\frac{\partial {\bf U}} {\partial t} + \nabla \cdot {\bf F}({\bf U}) = {\bf S}\,, \label{EQN:FV}
\end{equation}
where ${\bf U}$ is the vector of conserved variables, ${\bf F} = \left( \rho\,{\bf v}, \rho\,{\bf v} \otimes {\bf v} + P\,{\cal I}, (\rho\,e + P) \right)$ is the flux matrix, ${\cal I}$ is the identity matrix, and ${\bf S}$ is the source vector.  Following \citet{LV2008} who discretise these Equations using Galerkin-methods with the least-squares gradient operators \citep[see][for a complete derivation]{LV2008, GN2011, GIZMO}, we obtain the discrete Euler Equations, 
\begin{equation}
\frac{d{\bf Q}\ssi}{dt} + \sum \limits_{j}{} {\left[ V_i{\bf F}^{\alpha}_{i}\tilde\psi^{\alpha}_{j}({\bf r}_i) - V_j{\bf F}^{\alpha}_{j}\tilde\psi^{\alpha}_{i}({\bf r}_j) \right] } = {\bf S}\ssi\,V_i\,. \label{EQN:FVEULER}
\end{equation}

By replacing the two individual fluxes, ${\bf F\ssi}$ and ${\bf F\ssj}$, with a single flux across the interface between the two particles, ${\bf }F\ssij$, we obtain an exactly conservative scheme,  
\begin{equation}
\frac{d{\bf Q}\ssi}{dt} + \sum \limits_{j}{} {{\bf F}_{ij} \cdot {\bf A}_{ij} } = {\bf S}\ssi\,V_i \,, \label{EQN:MFV_FV}
\end{equation}
where the quantity ${\bf A}^{\alpha}\ssij \equiv V\ssi\tilde{\psi}^{\alpha}\ssj({\bf r}\ssi) - V\ssj\tilde{\psi}^{\alpha}\ssi({\bf r}\ssj)$ is the effective area of the face between the particles. 

The flux, ${\bf F}\ssij$, can be found by solving one dimensional Riemann problems between pairs of particles, where we assume that the interface is aligned with the face vector, ${\bf A}\ssij$. We have implemented two Riemann solvers in {\small GANDALF}; (i) the Exact Riemann solver for adiabatic gases \citep[e.g.][]{ToroBook}, and (ii) the HLLC approximate solver \citep{HLLC, ToroBook}, using the wave-speed estimate of \citet{Batten1996}. For isothermal equations of state, the HLLC solver has been modified to ensure that the density is constant across the contact discontinuity as well as the pressure, while still resolving shear waves \citep[e.g.][]{Mignone2007}.

\subsubsection{Face reconstruction}

Equation \ref{EQN:FVEULER} alone can be used to construct a first order Godunov method without specifying any further information about the location of the face (although its velocity is still needed in a Lagrangian scheme, see below); however, such a scheme is quite diffusive. Second order accuracy in space can be achieved following the standard MUSCL approach \citep{LV2008,GN2011,Hopkins2013}, in which the primitive variables evaluated at the cell faces are passed to the Riemann solver (instead of using the particle values). We do this using a slope-limited linear reconstruction to avoid oscillations near discontinuities,
\begin{equation}
{\bf W}\ssi({\bf r}_{_{\rm face}}) = {\bf W}\ssi({\bf r}\ssi) + \left( {\bf r}_{_{\rm face}} - {\bf r}\ssi \right) \cdot (\chi\nabla  {\bf W})\,, \label{EQN:RECONSTRUCTION}
\end{equation}
where $\chi \nabla  {\bf W}$ is the slope-limited gradient and $\nabla  {\bf W}$ is computed using Equation \ref{EQN:MFVGRADIENT}. The limiters are applied to each primitive variable independently. Both first order and second order (linear) reconstructions are available, including a wide range of slope limiters such as those suggested by \citet{AREPO}, \citet{GN2011}, \citet{Hess2010} and \citet{GIZMO}. The TVD limiter of \citet{Hess2010} is the most diffusive, while the non-TVD limiters of \citet{AREPO} and \citet{GN2011} are the least diffusive. The limiter suggested by \citet{GIZMO} falls in between.

In the second-order scheme, it is necessary to specify the location of the face. Following \citet{LV2008} and \cite{GN2011} we take
\begin{equation}
{\bf r}_{_{\rm face}} = \frac{1}{2} \left( {\bf r}\ssi + {\bf r}\ssj \right).
\end{equation}
We are free to choose how the particle positions, ${\bf r}\ssi$, are updated. By default we choose to move the particles at the local fluid velocity, ${\bf v}\ssi$. Finally, the Riemann problem must be solved in a frame that is consistent with the motion of the effective faces\footnote{This is done as described in Appendix A of \citet{GIZMO}.}, which moves along with the particles. An obvious choice for this is
\begin{equation}
{\bf v}_{_{\rm face}} = \deriv{{\bf r}_{_{\rm face}}}{t} = \frac{1}{2} \left( {\bf \dot{r}}\ssi + {\bf \dot{r}}\ssj \right),
\end{equation}
where ${\bf \dot{r}}\ssi$ are the velocities with which the particles are moved. This results in the Meshless Finite Volume (MFV) scheme as described by \cite{GIZMO}. Since this choice of face velocity may differ from the fluid velocity that comes from solving the Riemann problem, this results in a small amount of mass transferred between neighbouring particles. To construct a fully Lagrangian scheme, \cite{GIZMO} suggests using the speed of the contact discontinuity in place of ${\bf v}_{_{\rm face}}$.  This approach is similar to that employed by \citet{GodunovSPH2002} and ensures that no mass is advected between neighbouring particles.  Following \citet{GIZMO}, we refer to this modified scheme as the Meshless Finite-Mass (MFM) scheme, which is used by default in \gandalf{}.

\subsubsection{Time integration: Second-order MUSCL-Hancock}
To achieve second order accurate integration in time we employ an unsplit second order MUSCL-Hancock scheme \citep{VanLeer1979,ToroBook}. The conserved quantities are updated according to ${\bf Q}^{n+1}\ssi{} = {\bf Q}^n\ssi{} + \sum \limits_{j}{} d{\bf Q}\ssij$, where
\begin{equation} 
d{\bf Q}\ssij = - \Delta t \, {{\bf F}\ssij^{n + 1/2} \cdot {\bf A}_{ij}}\,, \label{EQN:MFVdQ}
\end{equation}
and ${\bf F}\ssij^{n + 1/2}$ is the time-centred estimate of the flux. This is calculated by predicting the primitive quantities passed to the Riemann solver to the mid-point of the time-step along with reconstructing them to the faces. This is done via the Taylor-series expansion,
\begin{equation}
{\bf W}^{n+1/2}\ssi = {\bf W}^n\ssi + \left( {\bf r}_{_{\rm face}} - {\bf r}\ssi \right) \cdot \nabla  {\bf W} + \frac{\Delta t}{2} \pderiv{{\bf W}\ssi}{t},
\end{equation}
and the primitive form of the Euler equations,
\begin{equation}
\pderiv{{\bf W}}{t} + {\bf A}({\bf W}) \cdot \nabla {\bf W} = 0.
\label{eq:primitve}
\end{equation}
Eq. \ref{eq:primitve} is used with the slope-limited gradients to replace the time derivative, giving
\begin{equation}
{\bf W}^{n+1/2}\ssi = {\bf W}^n\ssi +
\left[ ({\bf r}_{_{\rm face}} - {\bf r}\ssi) - \frac{\Delta t}{2} {\bf A}({\bf W}^n\ssi) \right] \cdot (\chi\nabla  {\bf W}).  \label{EQN:EdgeStates}
\end{equation}
See e.g. Appendix A of \citet{GIZMO} for the form of ${\bf A}({\bf W})$.

In the Lagrangian mode, the particle positions are then updated via
\begin{equation}
{\bf r}^{n+1}\ssi = {\bf r}^n\ssi + \frac{\Delta t}{2} \left( {\bf v}^n \ssi + {\bf v}^{*}\ssi \right), \label{EQN:MFVDRIFT}
\end{equation}
where ${\bf v}^*\ssi = (m^n\ssi {\bf v}^n\ssi + \Delta {\bf p}\ssi + m^n\ssi {\bf g}^n\ssi \Delta t) / m^{n+1}\ssi$ and $\Delta {\bf p}\ssi$ is the change in momentum due the fluxes and ${\bf g}^n$ is the gravitational acceleration (see below).

\subsubsection{Self-gravity}
We adopt the approach of \citet{GIZMO} in treating self-gravity, which is itself an adaption of those used by \citet{AREPO} and \citet{PM2007} applied to the MFV schemes.  We have only implemented self-gravity for the MFM scheme and present this implementation here.  Similar to SPH, the gravitational softening can be calculated self-consistently following \citet{PM2007} but using the MFV definition for the density. The gravitational force, $m\ssi{\bf g}\ssi$, on a particle is then
\begin{flalign}
m\ssi{\bf g}\ssi =& -\,G\sum\limits_{j=1}^{N}{m\ssi\,m\ssj\,
\frac{\phi'({\bf r}\ssij , h\ssi ) + \phi'({\bf r}\ssij , h\ssj )}{2}
\,\hat{\bf r}\ssij}\, \nonumber \\ 
 &- \,\frac{G}{2}\sum\limits_{j=1}^{N}\left\{\frac{\zeta'_i}{\Omega'_i}\nabla W\ssi({\bf r}\ssij,h\ssi)+\frac{\zeta'_j}{\Omega'_j}\nabla W\ssi({\bf r}\ssij,h\ssj)\right\}\,,
\label{EQN:MFVSELFGRAV}
\end{flalign}
where the definitions of $\Omega'\ssi$ and $\zeta'\ssi$ for the MFV schemes are  
\begin{equation} \label{EQN:MFV_OMEGA}
\Omega'\ssi = 1 - \frac{\partial h\ssi }{\partial n\ssi } 
\sum \limits_{j=1}^{N} \frac{\partial W}{\partial h}({\bf r}\ssij , h\ssi )\,,
\end{equation}
\begin{equation} \label{EQN:MFVZETA}
\zeta'\ssi = m\ssi \frac{\partial h\ssi}{\partial n\ssi} 
\sum \limits_{j=1}^{N} m_j \frac{\partial \phi}{\partial h}({\bf r}\ssij, h\ssi)\,.
\end{equation}
We apply the gravitational force in a similar way to \citet{GIZMO}, updating the new momentum, ${\bf p}\ssi$, and energy, $E\ssi$, according to
\begin{equation} \label{EQN:MFV_MOM_GRAV}
{\bf p}\ssi^{n+1} = {\bf p}\ssi^{n} + \Delta\,{\bf p}\ssi + \frac{\Delta\,t}{2} \left( m\ssi^{n}{\bf g}^n\ssi + m\ssi^{n+1} {\bf g}^{n+1}\ssi \right)\,,
\end{equation}
\begin{flalign}
E\ssi^{n+1} =\,& E\ssi^{n} + \Delta\,E\ssi \nonumber \\ 
& + \frac{\Delta\,t}{2} \left( m\ssi^{n}{\bf v}^n\ssi \cdot {\bf g}^n\ssi + m^{n+1}\ssi {\bf v}^{n+1}\cdot {\bf g}^{n+1}\ssi \right) 
\label{EQN:MFV_ENERGY_GRAV}
\end{flalign}
For the MFM scheme, since there is no mass-flux (i.e. $m^{n} \equiv m^{n+1}$), the gravitational update (along with the update of particle positions, Equation \ref{EQN:MFVDRIFT}) reduces exactly to a leapfrog scheme when the pressure forces are negligible.  In the original MFV derivation \citep{GIZMO}, there are extra terms relating to the mass flux between neighbouring particles, $dm\ssij/dt$, but these also reduce to zero for the MFM scheme.

\subsubsection{Physical viscosity}

Since it is possible to achieve numerical viscosities that are smaller than the physical viscosity in real systems such as accretion discs, we have implemented a physical viscosity in the MFV schemes. The source term in Equation \ref{EQN:FV} due to viscosity can be written as,
\begin{flalign}
{\bf S}  =& \nabla \cdot (0, {\bf \Pi}, {\bf \Pi} \cdot {\bf v}), \label{EQN:ViscSource} \\
{\bf \Pi} =& \eta \left\{ \left[ \nabla {\bf v} + (\nabla {\bf v})^{\rm T} \right] - \frac{2}{3} {\cal I} (\nabla \cdot {\bf v}) \right\} + \zeta  {\cal I} (\nabla \cdot {\bf v}),
\end{flalign}
where $\eta$ and $\zeta$ are the shear and bulk viscosity coefficients. Since Equation \ref{EQN:ViscSource} takes the form of the divergence of a flux (with ${\bf F}_{\rm visc} =  - (0, {\bf \Pi}, {\bf \Pi} \cdot {\bf v})$), we follow \citet{Munoz2013} in discretising this term using a finite volume approach, which simply amounts to including the diffusive flux in Equation \ref{EQN:MFV_FV}. To compute the viscous flux, one needs to specify a `viscous Riemann solver' along with the edge states to pass to the Riemann solver. \citet{Munoz2013} suggest using a  slope-limited reconstruction of both the primitive variables and the velocity gradients, which are also needed to compute the viscous flux. However, \citet{Hopkins2017} found that reconstructing the velocity gradients makes only a very small difference to the solution (typically less than 1 per cent).
{\refone Thus, we take a pragmatic approach in using the primitive variables reconstructed at the edges and the particle-centred velocity gradients, which are already available (Equations \ref{EQN:MFVGRADIENT} and \ref{EQN:EdgeStates}).} For the Riemann solver we simply compute the arithmetic average of the face states and use those to compute the flux. 

\subsubsection{Time-stepping} \label{SSS:MFV_TIMING}

The MFV scheme uses a similar CFL time-stepping condition as used in SPH (ignoring any artificial viscosity terms), i.e. 
\begin{equation} \label{EQN:DT_MFV}
\Delta\,t\ssi = C_{\rm CFL}\,\frac{h\ssi}{{\rm MAX}\ssj |v_{\rm sig,ij}|}\,
\end{equation}
where
\begin{equation}
v_{\rm sig,j} = c\ssi + c\ssj - {\rm MIN} \left(0, {\bf v}\ssij \cdot \hat{\bf r}\ssij \right) \,.
\end{equation}
Similarly, when viscosity is included the time-step is limited according to 
\begin{equation}
\Delta\,t\ssi = C_{_{\rm VISC}}\,\frac{h^2\ssi}{2 \nu\ssi}\,
\end{equation}
where {\refone $C_{_{\rm VISC}}$ is the dimensionless viscosity timestep factor and } $\nu\ssi = (\eta\ssi + \zeta\ssi)/\rho\ssi$ is the total kinematic viscosity for the particles. Finally, when gravity is included the time-step is limited according to the acceleration condition,
\begin{equation}
\Delta\,t\ssi = C_{\rm GRAV}\,\frac{h\ssi}{\sqrt{{\bf g}\ssi}}\,.
\end{equation}

Similarly to SPH, block time-stepping can also be used with MFV. In order to ensure exact conservation, the changes to conserved quantities, $d{\bf Q}\ssij$, are computed on the smallest time-step of the particle pair and built up over the full time-step, following \citet{AREPO}. Since particles may be interacting with neighbours both on larger and smaller time-steps, the contribution to the fluxes from some particles will be computed once while others may contribute multiple sub-steps. This means that the conserved quantities only take meaningful values at the beginning and end of the time-steps. Since the primitive quantities may be needed at any point during the particle's time-step to compute the fluxes with a neighbour on a shorter time-step, we also record
\begin{equation}
\deriv{{\bf Q}_i}{t} = - \sum_j {\bf F}\ssij \cdot {\bf A}\ssij, \label{Eqn:dQdt}
\end{equation}
at the start of the time-step and use it to predict the primitive quantities throughout the time-step. Once the particle reaches the end of its time-step these are then replaced by the conserved fluxes built up throughout the time-step.

The block time-stepping scheme can suffer from the same problems with the Meshless scheme as in SPH when particles are allowed to interact with neighbours on much longer time-steps. We provide two time-step limiters to solve this problem. Firstly, we have implemented a simple limiter similar to the one used by SPH. When a particle detects that a neighbour is on a time-step lower than the accepted ratio, the particle is `woken up'. At this time the fluxes built up during the block time-stepping scheme are likely to be too large as some neighbours may be on the same time-step level as the particle, or longer. For this reason we use $\Delta t\deriv{{\bf Q}_i}{t}$ to estimate the new conserved quantities when the particle is woken up. We note that while this breaks the exact conservation, we find that it works well in practice.

Secondly, for cases when exact conservation is required we have also included the more expensive predictive time-step limiter of \citet{AREPO}, in which the CFL condition is evaluated for distant particles using a tree walk. By limiting the time-step based upon $|{\bf r}\ssij|/|{\bf v}_{_{\rm sig,ij}}|$ this ensures that particles `wake up' from long time-steps before shocks reach them. In simulations dominated by gravity, pathological configurations can occur where the predictive limiter forces the particles to have much lower time-steps than necessary. 
In this case the simple limiter will likely work well since the energy conservation errors are likely dominated by the gravitational forces.

\section{N-body methods in GANDALF} \label{S:NBODY}

N-body dynamics has been implemented into {\small GANDALF} as an independent class separate from the Hydrodynamical algorithms.  {\small GANDALF} can therefore be run for pure N-body problems, albeit not as efficiently compared as dedicated and optimised N-body codes such as NBODY6 \citep{Aarseth2003} or STARLAB/kira \citep{kira2001}.  In most simulations, the N-body module will be used in tandem with the Hydrodynamics to represent stars in the guise of sink particles (see Section \ref{S:SINKS}).  Nevertheless, there are situations where one is interested in the outcome of a simulation if there was no gas present, or as a pure N-body simulation after the gas has been removed.

In order to make the N-body algorithms compatible with the Hydrodynamical algorithms and to prevent unphysical 2- or 3-body ejections and/or large energy errors, we give each N-body particle a (constant) smoothing length.  The acceleration of an N-body particle due to all other N-body particles is simply : 
\begin{equation}
{\bf a}\sss = -G\,\sum_{t=1}^{N} { m\sst\,\phi'({\bf r}\ssst, \overline{h}\sss)\,\hat{\bf r}\ssst }\,,
\end{equation}
where $\overline{h}\ssst \equiv \frac{1}{2}\left(h\sss + h\sst \right)$.

\subsection{Integration schemes} \label{SS:NBODYINTEGRATION}

{\small GANDALF} can use several integration schemes for simulating N-body dynamics {\refone independent of the choice of Hydrodynamics scheme}.  For simple problems or when using accreting sink particles (see Section \ref{S:SINKS}), we can use the Leapfrog KDK and DKD schemes outlined in Section \ref{SS:TIMESTEPPING} using the same sets of Equations (\ref{EQN:LFKDK1} to \ref{EQN:LFDKD4}) together with the acceleration time-step condition (Eqn. \ref{EQN:SPH_DTACCEL}).  For pure N-body simulations, or hybrid simulations that require higher accuracy, we can use other higher-order schemes.

\subsubsection{4th-order Hermite scheme}
In the  4th-order Hermite scheme \citep{Hermite1992}, we explicitly calculate the 1st time derivative of the acceleration (often called the {\it jerk}), $\dot{\bf a}$, in order to achieve higher integration accuracy.  At the beginning of the step, we calculate both the acceleration and the jerk, where the jerk is given by : 
\begin{flalign}
\dot{\bf a}^n\sss =& -\,G\,\sum \limits_{t=1}^{N}   
{\frac{m\sst\,\phi'({\bf r}\ssst , \meanhst )}{|{\bf r}\ssst |}{\bf v}\ssst} \; \nonumber \\
& + \;3\,G\, \sum \limits_{t=1}^{N}  
{\frac{m\sst\,({\bf r}\ssst  \cdot {\bf v}\ssst)\,\phi'({\bf r}\ssst , \meanhst )}{|{\bf r}\ssst |^3} {\bf r}\ssst } \; \nonumber \\  
& - \;4\,\pi\,G\, \sum \limits_{t=1}^{N} {\frac{m\sst\,\,({\bf r}\ssst  \cdot {\bf v}\ssst)\,  W({\bf r}\ssst , \meanhst )}{|{\bf r}\ssst |^2}{\bf r}\ssst}\,.
\label{EQN:JERK}  
\end{flalign}
Once calculated for all stars, we predict the star positions and velocities to the end of the step with a Taylor expansion, 
\begin{eqnarray}
{\bf r}\sss^{n+1} &=& {\bf r}\sss^{n} + {\bf v}\sss^{n}\,\Delta t + \frac{1}{2}{\bf a}\sss^{n}\,\Delta t^2 + \frac{1}{6}\dot{\bf a}\sss^{n}\,\Delta t^3 \,, \\  
{\bf v}\sss^{n+1} &=& {\bf v}\sss^{n} + {\bf a}\sss^{n}\,\Delta t + \frac{1}{2}\dot{\bf a}\sss^{n}\,\Delta t^2\,.  
\end{eqnarray}
We then calculate the acceleration jerk again using Equation \ref{EQN:JERK} using the predicted positions and velocities at the end of the step, i.e. ${\bf a}\sss^{n+1}$ and  $\dot{\bf a}\sss^{n+1}$ .  This allows us to construct the higher-order time derivatives for the step, 
\begin{eqnarray}  
\ddot{\bf a}\sss^{n} &=& \frac{2 \left( -3({\bf a}\sss^{n} - {\bf a}\sss^{n+1}) - (2\dot{\bf a}\sss^{n} + \dot{\bf a}\sss^{n+1})\Delta t \right)}{\Delta t^2}\,,  \label{EQN:A2} \\  
\dddot{\bf a}\sss^{n} &=& \frac{6 \left( 2({\bf a}\sss^{n} - {\bf a}\sss^{n+1}) + (\dot{\bf a}\sss^{n} + \dot{\bf a}\sss^{n+1})\Delta t \right)}{\Delta t^3}\,. \label{EQN:A3}  
\end{eqnarray}
where $\ddot{\bf a}^{n}$ and $\dddot{\bf a}^{n}$ are the 2nd and 3rd time derivatives of the acceleration respectively.  Finally we apply these higher-order derivatives as a correction step to calculate the position and velocity to high-order, 
\begin{eqnarray}  
{\bf r}\sss^{n+1} &=& {\bf r}\sss^{n+1} + \frac{1}{24}\ddot{\bf a}\sss^{n}\,\Delta t^4 + \frac{1}{120}\,\dddot{\bf a}\sss^{n}\,\Delta t^5 \,, \\  
{\bf v}\sss^{n+1} &=& {\bf v}\sss^{n+1} + \frac{1}{6}\ddot{\bf a}\sss^{n}\,\Delta t^3 + \frac{1}{24}\,\dddot{\bf a}\sss^{n}\,\Delta t^4 \,.  
\end{eqnarray} 
To compute the time-step for each star, we use the Aarseth criterion as used in the NBODY codes \citep[e.g.][]{Aarseth2003}, 
\begin{eqnarray}  
\Delta t\sss &=& \gamma\sss \, \sqrt{\frac{|{\bf a}\sss| |\ddot{\bf a}\sss| +   
|\dot{\bf a}\sss|^2}{|\dot{\bf a}\sss| |\dddot{\bf a}\sss| +   
|\ddot{\bf a}\sss|^2}}\,. \label{EQN:DTAARSETH}  
\end{eqnarray}

\subsubsection{4th-order time-symmetric integration scheme}
For simulations which require higher stability or more accuracy, particularly with long-term orbital integration (e.g. binary or multiple systems), we can use the \citet{Hut1995} time-symmetric 4th-order Hermite scheme.  In this variant, we compute the acceleration and jerk at the beginning of the time-step similar to the standard Hermite scheme.  We then predict the position and velocities at the end of the time-step.  The corrected position and jerk are recomputed using 
\begin{flalign}
{\bf r}\sss^{n+1} &= {\bf r}\sss^{n} + \tfrac{1}{2}\left({\bf v}\sss^{n+1} + {\bf v}\sss^{n} \right)\Delta t - \tfrac{1}{12}\left( {\bf a}\sss^{n+1} - {\bf a}\sss^{n} \right)\Delta t^2\,, \label{EQN:H4TS_RCORR} \\
{\bf v}\sss^{n+1} &= {\bf v}\sss^{n} + \tfrac{1}{2}\left({\bf a}\sss^{n+1} + {\bf a}\sss^{n} \right)\Delta t - \tfrac{1}{12}\left( \dot{\bf a}\sss^{n+1} - \dot{\bf a}\sss^{n} \right) \Delta t^2\,. \label{EQN:H4TS_VCORR}
\end{flalign}
A more accurate solution is obtained by iterating the evaluate-correction step until the particle's position and velocity are converged.  Such schemes are often called ${\rm P}\left( {\rm EC} \right)^{n}$ where $n$ is the number of correction iterations.  In practice, even using $n = 2$ gives improved results.  We note that despite its name, a truly time-symmetric integration is only possible for constant time-steps whereas most N-body codes use adaptive time-steps.

\subsection{Hybrid SPH and N-body dynamics} \label{S:HYBRID}

{\small GANDALF} contains an implementation of the \citet{Hybrid2013} hybrid SPH/N-body algorithm.  This is designed to simulate small to intermediate size clusters which also have a live gaseous background.  One noticeable difference between this and the original \citet{Hybrid2013} implementation is the mode of symmetrising the particle-particle interactions.  In \citet{Hybrid2013}, the gravitational interactions between all particle pairs (gas-gas, gas-star and star-star) were smoothed using the average smoothing length, i.e. $W({\bf r}, \tfrac{1}{2}(h\ssi + h\ssj))$.  In {\small GANDALF}, this has been modified so gas-gas interactions use the standard \citet{PM2007} form in grad-h SPH with the average of the kernels (Equation \ref{EQN:SPHSELFGRAV}), whereas only the gas-star and star-star interactions use the average smoothing length approach.  Smoothing the gas-star interactions with the average smoothing length is designed to prevent the situation where the smoothing lengths of gas and star particles are hugely different leading to the unphysical 2-body scattering which softening is designed to prevent.  The full equation of motion for gas particles becomes
\begin{flalign} \label{EQN:SPHEOM}  
{\bf a}\ssi =&   
- \sum \limits_{j=1}^{N_g}   
m\ssj \,\left[ \frac{P\ssi}{\rho^2\ssi\Omega\ssi}\frac{\partial W\ssij}{\partial {\bf r}\ssi}(h\ssi)  + \frac{P\ssj}{\rho^2\ssj\Omega\ssj}\frac{\partial W\ssij}{\partial {\bf r}\ssi}(h\ssj) \right]  \nonumber \\
& - G \sum \limits_{j=1}^{N_g} m\ssj\, \frac{\phi'({\bf r}\ssij , h\ssi ) + \phi'({\bf r}\ssij , h\ssj )}{2}\,\hat{\bf r}\ssij   \nonumber \\
& - \frac{G}{2} \sum \limits_{j=1}^{N_g}   
m\ssj \,\left[ \frac{\zeta\ssi' + \bar{\chi}\ssi}{\Omega\ssi}\frac{\partial W\ssij}{\partial {\bf r}\ssi}(h\ssi)  + \frac{\zeta\ssj' + \bar{\chi}\ssj}{\Omega\ssj}\frac{\partial W\ssij}{\partial {\bf r}\ssi}(h\ssj) \right] \nonumber \\
& - G \sum \limits_{s=1}^{N_s} \,m\sss\,\phi'\ssis(\overline{h}\ssis)\,\hat{\bf r}\ssis \,,
\end{flalign}
where
\begin{equation}  
\bar{\chi}\ssi = \frac{\partial h\ssi}{\partial \rho\ssi}   
\sum \limits_{j=1}^{N} m\ssi \frac{\partial \phi\ssij}{\partial \meanhij}(\overline{h}\ssij)\,.
\label{EQN:GRADHCHI}  
\end{equation} 
{\refone These equations are then numerically integrated using the 2nd-order Leapfrog KDK scheme (Section \ref{SS:TIMESTEPPING}.}
The total equation of motion for stars becomes
\begin{equation} \label{EQN:STARGRAV}  
{\bf a}\sss = -G \sum \limits_{t=1}^{N_s} {  
m\sst\,\phi'\ssst(\overline{h}\ssst)\,\hat{\bf r}\ssst} \,   
- G \sum \limits_{i=1}^{N_g}\,{m\ssi\,\phi'\sssi(\overline{h}\sssi)\,\hat{\bf r}\sssi } \,.  
\end{equation}
This modification removes the need for an additional loop over SPH neighbours to calculate the values for $\zeta\ssi$ using averaged smoothing lengths.  

{\refone We note that this conservative scheme is not formally implemented to work with the MFV/MFM schemes although the basic 4th-order Hermite scheme can still be utilised together in tandem with the MFV/MFM Hydrodynamics integration scheme.}

\subsection{Sink particles} \label{S:SINKS}

Sink particles \citep{Bate1995} are used in self-gravitating hydrodynamics codes to relieve the problem of high density condensations (e.g. protostars) leading to very short time-steps and prohibitively long CPU run times.  In their most basic form, sink particles replace the forming protostar (or other accreting object) with a single particle with an accretion radius $R\sss$ that accretes any gas particles that enter the accretion radius by adding their mass and momentum to the sink.  \citet{NewSinks2013} introduced an improved sink particle algorithm in SPH which computed the accretion rate based on an internal sub-grid model leading to better convergence of results.  {\small GANDALF} implements both the simpler `vacuum-cleaner' sink particles and the improved sinks of \citep{NewSinks2013}, both for SPH and for the MFV/MFM schemes.

\subsubsection{Sink formation criteria}

A new sink particle is created from an existing gas particle that satisfies a number of criteria.  These criteria are designed to ensure that sinks are only formed in genuinely self-gravitating entities, such as in collapsing prestellar cores and protostars.  When a sink particle is formed, it is given an accretion radius that is some multiple of the original particle's smoothing length, 
\begin{equation}
R\sss = X_{_{\rm SINK}}\,h\ssi
\end{equation}
where $X_{_{\rm SINK}}$ is a user-defined factor of order unity and $h\ssi$ is the smoothing length of the original gas particle.  For consistency, $X_{_{\rm SINK}}$ is normally chosen so that the sink accretion volume is the same as the smoothing kernel volume (e.g. for the M4-kernel, $X_{_{\rm SINK}} = 2$).

The formation criteria are : 
\begin{enumerate}
\item The density of a gas particle should exceed the user-defined sink creation density, $\rho_{_{\rm SINK}}$, i.e. 
\begin{equation}
\rho\ssi > \rho_{_{\rm SINK}}\,.
\end{equation}
\item A new sink particle formed from a hydrodynamical particle does not overlap any existing sinks upon creation, i.e. 
\begin{equation}
|{\bf r}\ssi - {\bf r}\sss| > X_{_{\rm SINK}}\,h\ssi + R\sss\,.
\end{equation}
\item The gravitational potential of a hydrodynamical particle is the minimum (as in most negative) of all of its hydrodynamical neighbours, i.e. 
\begin{equation}
\phi\ssi < {\rm MIN}\,\left\{ \phi\ssj \right\}\,.
\end{equation}
\item 
The density is sufficiently large so local condensations do not lie within the Hill sphere (or equivalently the Roche limit) of all existing sinks, i.e. 
\begin{equation}
\rho\ssi > \frac{3\,X_{_{\rm HILL}}\,\Delta{\bf r}\ssis \cdot \Delta{\bf a}\ssis}{4\pi G |\Delta{\bf r}\ssis|^2}\,.
\end{equation}
\item A condensation can undergo freefall collapse before approaching any existing sinks, i.e. 
\begin{equation}
t_{_{\rm FF}} < \frac{|\Delta{\bf r}\ssis|^2}{\Delta{\bf v}\ssis \cdot \Delta{\bf r}\ssis}\,.
\end{equation}
\end{enumerate}

\subsubsection{Sink accretion}

In the simplest case, accretion of gas particles onto sink particles can be achieved simply by adding the mass, momentum and energy of every gas particle entering the sink radius. Additional criteria may be employed, such as checking if the gas particles are gravitationally bound to the sink particle.  \citet{NewSinks2013} introduced a simple two-mode sub-grid model of accretion which we have implemented into \gandalf{}.  The first mode treats the case of purely spherical collapse, i.e. inward radial velocities.  The (smoothed average) radial infall timescale in terms of the particle properties is 
\begin{equation}\label{EQN:TRAD}
\left<t_{_{\rm RAD}}\right>_s = \frac{\sum_j\!\left\{m_j\right\}\,{\cal W}}{4\pi\sum\limits_{j}\!\left\{|\Delta{\bf r}_{js}|\Delta{\bf r}_{js}\!\cdot\!\Delta{\bf v}_{js}m_jW(|\Delta{\bf r}_{js}|,H_s)\right\}}\,,
\end{equation}
where
\begin{equation}
{\cal W}=\sum_j\!\left\{m_jW(|\Delta{\bf r}_{js}|,H_s)/\rho_j\right\}\,.
\end{equation}

The second mode treats the case of purely rotational collapse, i.e. where all velocities are tangential with speeds for circular motion.  For low-mass discs in approximate Keplerian rotation, the accretion timescale at a radius $R$ is given by the Shakura-Sunyaev prescription, $t_{\rm SS} \sim\alpha_{_{\rm SS}}^{-1}(GM_\star R)^{1/2}a^{-2}$, where $\alpha_{_{\rm SS}}$ is the Shakura-Sunyaev viscosity and $a$ is the local sound speed.  A kernel-weighted average of this timescale over all particles in the sink gives 
\begin{equation}\label{EQN:TDISC}
\left<t_{_{\rm DISC}}\right> = \frac{(GM_s)^{1/2}}{\alpha_{_{\rm SS}}{\cal W}}\!\sum\limits_{j}\!\left\{\!\frac{|\Delta{\bf r}_{js}|^{1/2}m_jW(|\Delta{\bf r}_{js}|,H_s)}{\rho_ja_j^2}\!\right\}.
\end{equation}
Since accreting particles will in general fall between these two limits, we use a simple interpolation using a weighted geometric mean to give an overall accretion timescale of 
\begin{equation}\label{EQN:t_ACC}
t_{_{\rm ACC}} = \left<t_{_{\rm RAD}}\right>_s^{(1-f)}\,\left<t_{_{\rm DISC}}\right>_s^f\,,
\end{equation}
where
\begin{equation}
f = \mbox{\sc min}\left\{ 2E_{_{\rm ROT}}/|E_{_{\rm GRAV}}|\;,\,1\right\}\,
\end{equation}
is a simple measure of the centrifugal support using the rotational and gravitational energies of particles inside the sink, where $f = 1$ is expected for circular rotation.

The total mass of gas particles to be accreted in the current time-step is then
\begin{equation}
\delta M_{_{\rm ACC}} = M_{_{\rm INT}}\,\left[ 1 - \exp{\left( - \frac{\delta t_s}{t_{_{\rm ACC}} }\right)} \right]\,.
\end{equation}

\section{Misc} \label{S:MISC}

\subsection{Dust}
The dynamics of dust-gas mixtures have been implemented in \gandalf{} using the `two-fluid' formalism. An additional set of dust particles can be included, which are coupled to the gas motions via drag forces.  The main scheme closely follows \citet{LorenAguilar2015}, who provide expressions for a semi-implicit update for the drag force that avoids the need for small time-steps when the drag forces are very strong.  We refer the reader to \citet{LorenAguilar2015} for details and only briefly outline the scheme. The Equations of Motion for gas and dust particles are 
\begin{align}
\deriv{{\bf v}_g}{t} &= - \frac{\rho_d}{\rho_g}\frac{({\bf v}_g - {\bf v}_d)}{t_s} + {\bf a}_g  - \frac{\nabla P}{\rho_g}\,,  \\
\deriv{{\bf v}_d}{t} &= - \frac{({\bf v}_d - {\bf v}_g)}{t_s} + {\bf a}_d\,,
\end{align}
where $t_s$ is the one-particle stopping time, and the back-reaction of the dust on the gas has been included to conserve the total momentum. 

{\refone To solve these equations over a single time-step $\Delta t$, the hydrodynamic and gravitational forces are first calculated as normal. The semi-implicit update is computed by making the ansatz that these forces, along with the densities and $t_s$, are constant throughout the time-step. The above equations can then be solved  to give the new velocities,}
\begin{align}
 {\bf v}_{\rm d}(t + \Delta t) &= \tilde{\bf v}_{\rm d}(t + \Delta t) - \frac{\rho_g}{\rho_d + \rho_g}{\bf S}_{\rm dg} \label{Eqn:DustEqn} \\ 
 {\bf v}_{\rm g}(t + \Delta t) &= \tilde{\bf v}_{\rm g}(t + \Delta t) + \frac{\rho_d}{\rho_d + \rho_g}{\bf S}_{\rm dg}
\end{align}
{\refone where $ \tilde{\bf v}_{\rm d,g}(t + \Delta t) = {\bf v}_{\rm d,g}(t) + {\bf a}_{\rm d,g}(t) \Delta t$. Writing
$\Delta \tilde{ \bf v} =  \tilde{\bf v}_{\rm d} -  \tilde{\bf v}_{\rm g}$ and $\Delta { \bf a} =  {\bf a}_{\rm d} -  {\bf a}_{\rm g} + \nabla P / \rho_g$, then ${\bf S}_{\rm dg}$ is given by}
\begin{align}
{\bf S}_{\rm dg} =&  \left(1 - e^{\Delta t/t_s}\right) \Delta \tilde{\bf v}(t + \Delta t) \nonumber \\
 & - \left[\left(\Delta t + t_s\right)\left(1 - e^{\Delta t/t_s}\right) - \Delta t\right] \Delta {\bf a}(t). \label{Eqn:DustS}
\end{align}

{\refone To convert this update into SPH form, we project the velocity along the line of sight and sum over the neighbours using a double-hump kernel (which we denote by $\tilde{W}$), in order to ensure angular momentum conservation while computing the drag force accurately \citep{Laibe2012,LorenAguilar2015}. The resulting equations are:}
\begin{align}
{\bf v}^i_{\rm d}(t + \Delta t, {\bf r}_i) &= \tilde{\bf v}^i_{\rm d}(t + \Delta t, {\bf r}_i) \nonumber \\
&- D\sum_a^{\rm Gas} \frac{m_a}{\rho_i + \rho_a}\left({\bf S}_{ia} \cdot \hat{\bf r}_{ia}\right) \hat{\bf r}_{ia} \tilde{W}({\bf r}_{ia}, h_a) \label{Eqn:DustUpdate1}
\end{align}
\begin{align}
{\bf v}^a_{\rm g}(t + \Delta t, {\bf r}_a) &= \tilde{\bf v}^a_{\rm g}(t + \Delta t, {\bf r}_a) \nonumber \\
&+ D\sum_i^{\rm Dust} \frac{m_i}{\rho_i + \rho_a}\left({\bf S}_{ia} \cdot \hat{\bf r}_{ia}\right) \hat{\bf r}_{ia} \tilde{W}({\bf r}_{ia}, h_a).  \label{Eqn:DustUpdate2}
\end{align}

{\refone The drag force dissipates kinetic energy, which may go into heating the gas, dust or be lost from the system depending on the details of the problem.  When using a barotropic equation of state, which is common in astrophysical applications with dust-gas mixtures (e.g. discs, star formation or molecular clouds), we do not explicitly track the kinetic energy dissipated.  However, when using an adiabatic equation of state, we assume that  the dissipated kinetic energy heats the gas directly. }

{\refone To ensure exact conservation, we compute the change in kinetic energy due to drag forces directly from the above equations, }
\begin{equation}
\Delta KE_i = m_i |{\bf v}_i (t + \Delta t) - \tilde{\bf v}_i (t + \Delta t)|^2.
\end{equation}
{\refone The change in kinetic energy of a gas particle is added directly to the change in its internal energy. For dust particles, we spread its change in kinetic energy amongst its neighbouring gas particles, using the same kernel as for the drag force calculation. The total change in a gas particle's internal energy is thus}
\begin{equation}
m_a \Delta u_a = \Delta KE_a + \frac{m_a}{\rho_a} \sum_i^{\rm Dust} \frac{1}{N_i} \Delta KE_i \tilde{W}({\bf r}_{ia}, h_a), \label{Eqn:DustEnergy}
\end{equation}
{\refone where $N_i$ is a normalization factor, }
\begin{equation}
N_i = \sum_a^{\rm Gas} \frac{m_a}{\rho_a} \tilde{W}({\bf r}_{ia}, h_a).
\end{equation}
{\refone Summing \autoref{Eqn:DustEnergy} over all gas particles gives $\sum_a^{\rm Gas} m_a \Delta u_a = \sum_a^{\rm Gas} \Delta KE_a + \sum_i^{\rm Dust} \Delta KE_i$, i.e. manifest energy conservation. Finally, we note that this energy update can be implemented simply. We compute $N_i$ during the drag force calculation for the dust. Once the drag force for the single dust particle has been computed, the change in kinetic energy is then `given back' to its neighbours. In practice we use Equations \ref{Eqn:DustUpdate1}, \ref{Eqn:DustUpdate2} and \ref{Eqn:DustEnergy} to define time-averaged rates of change in the physical quantities which are included in the standard SPH time integration scheme.}

{\refone The dust scheme has been described above in terms of SPH, but can naturally be extended to the MFM integration algorithm. To do this we proceed exactly as in SPH, except that change in velocity is multiplied by the particle mass and added to the change in momentum, $\Delta {\bf p}$. Also, since the MFM method integrates the total rather than the internal energy, only the change in kinetic energy from the dust particles needs to be included. This allows conservation of energy and momentum to machine precision. However, there is one subtlety, in that MFV and MFM use a single hydrodynamical update per time-step, but the gravitational acceleration is treated using the KDK leap frog, i.e. two kicks per time-step. Rather than use two drag kicks per time-step (one with the initial and one with the final gravitational acceleration), we instead take the pragmatic approach of using the time average, $m_1 \bar{\bf a} = (m_0 {\bf a_0} + m_1 {\bf a_1})/2$, where $m_{0,1}$ and ${\bf a_{0,1}}$ are the accelerations and masses computed at the beginning and end of the step. This works well in practice because the drag forces only depend on the difference between the dust and gas accelerations (see \citealt{LorenAguilar2015}), which for gravitational forces is typically close to zero (except perhaps in very poorly resolved regions close to sink particles). Finally, in the meshless the $\nabla P/\rho_g$ term is taken from the change in momentum computed using the Riemann Solver (Equation \ref{Eqn:dQdt}).}

{\refone In addition to full two-fluid scheme above, \gandalf{} also includes a test-particle scheme. The main advantage of this scheme is that, unlike the full two-fluid scheme, it can naturally handle block time-steps, whereas the full two-fluid scheme becomes inaccurate if not used with global time-steps. 
While it would be straight-forward to create a test-particle scheme by setting $\rho_i = 0$ in Equation \ref{Eqn:DustUpdate1} and neglecting Equations \ref{Eqn:DustUpdate2} \& \ref{Eqn:DustEnergy}, in cases where the particle distribution is non-uniform the force accuracy can be improved by using a normalized interpolations scheme, as in \citet{Booth2015}. In this scheme Equation \ref{Eqn:DustUpdate1} is replaced by Equation \ref{Eqn:DustEqn} and ${\bf S}_{\rm dg}$ is computed by interpolating the gas properties to the location of the dust particle and using them directly in Equation \ref{Eqn:DustS}. In formula, any given quantity $A_i$, defined on the gas particles, it is interpolated using}
\begin{equation}
A_d = \sum_i^{\rm Gas} \frac{A_i}{\hat{n}_d} W({\bf r}_{id}, \hat{h}_d)
\end{equation}
{\refone where } 
\begin{equation}
\hat{n}_d = \sum_i^{\rm Gas} W({\bf r}_{id}, \hat{h}_d)
\end{equation}
{\refone and  $\hat{h}_d = \eta_{\rm SPH} (1/\hat{n}_d)^{1/D}$, which is evaluated using the standard Newton-Raphson iteration with the same tolerance as the mass density.}

As with pure hydrodynamics problems with the MFM method, we find that using the quintic kernel can significantly improve the accuracy of the results due to more accurate density estimates and smaller interpolation errors \citep[see, e.g.][]{Price2012, Laibe2012, Price2015}. We thus recommend use of the quintic kernel in problems involving dust, and use it in the tests presented here.


\subsection{Tree} \label{S:KDTREE}

In \gandalf{}, we have implemented a KD-tree to efficiently determine neighbour list for computing all local quantities (e.g. smoothing lengths) and for computing gravitational forces.  Our implementation is loosely based on the one described in \citet{GR2011}; we refer the interested reader to that paper and highlight the differences from our implementation in the following text. The tree is built in a top-down approach; starting from a root cell that contains all the particles, each cell is divided in two subcells along a chosen direction until one is left only with \textit{leaf} cells, i.e. cells containing a number of particles equal or smaller than a set maximum, $N_{_{\rm LEAF}}$. The slice direction is always chosen to be the one along the cell's most elongated axis, in order to avoid having cells with large aspect ratios.  In contrast to \citet{GR2011}, we follow a more traditional KD-tree construction and split cells using the median value of the particle's positions (what they describe as MPS method). This guarantees that the tree is balanced; i.e., if there are $2^{l}$ particles, the tree will contain $l$ levels (for $N_{_{\rm LEAF}} = 1$), which simplifies the memory management.

Once the tree has been constructed, a number of properties can be computed for each cell and propagated upwards to the parent cells, such as the position of the centre-of-mass, the gravitational moments (needed for computing the gravitational acceleration) and the extent of the smallest box containing all the smoothing spheres of the particles. This box will be used during the tree walk to decide if a given cell potentially contains hydrodynamical neighbours of a given particle. 

When including self-gravity, the tree is also used to reduce the expensive $O(N^2)$ calculation to $O(N\log N)$ by grouping the contribution from distant particles together. The tree is walked from the root cell and each cell is tested to see whether the contribution from the cell is sufficiently accurate; if not the cell is opened and its children are tested.  This can be done using the classic geometric opening criterion \citep[e.g.][]{BarnesHut86},
{\refone
\begin{equation}
|{\bf r}\ssi - {\bf r}\ssc|^2 \geq \frac{l\ssc^2}{\theta_{_{\rm MAX}}^2}
\end{equation}
where ${\bf r}\ssc$ is the cell position, $l\ssc$ is the cell `size' (i.e. the centre-to-corner distance of the cell) and $\theta_{_{\rm MAX}}$ is the maximum allowed opening angle of the cell (typically $\sim 0.3$).  The cell approximation can be used if the inequality is satisfied.  Otherwise, we must open the cell and test each of its children cells.}
Optionally a second criterion can be included whereby cells are opened if the contribution to the force from their quadrupole moment is too large.  Either the \citet{Gadget2} criterion,
{\refone
\begin{equation}
|{\bf r}\ssi - {\bf r}\ssc|^2 \geq \left(\frac{G\,M\ssc l\ssc^2}{\alpha\ssc}\right)^{1/2}\,|{\bf a}_{_{\rm GRAV}}|^{-1/2}
\end{equation}
where $M\ssc$ is the cell mass, ${\bf a}_{_{\rm GRAV}}$ is the gravitational acceleration from the previous step and $\alpha\ssc$ is the maximum fractional contribution to the total acceleration from the cell quadrupole term (typically $\alpha\ssc \sim 10^{-4}$). 
}
or the eigenvalue-based criterion of \citep[see][for details]{Hubber2011} can be used in \gandalf{}.

Even with the optimisations provided by using a tree, walking the tree to find neighbours is still an expensive operation that can dominate the total CPU cost of a simulation.  We optimise the walk by retrieving the list of neighbours for each leaf cell rather than for each individual particle \citep{Gasoline}. \gandalf{} caches the list of particles and cells found during the tree walk. When self-gravity is included, the gravitational force contribution from the particles is computed directly for all of the particles in the leaf cell. For the contribution from the distant cells, the gravitational force calculation can be computed in one of two ways: either directly for each particle in the leaf cell or using a Taylor series expansion about the centre of the leaf cell similar to \citet{GR2011}. Both the monopole and quadrupole moments can be included in the force contribution for the cells; when using the Taylor series method we expand the monopole term to second order (as in \citealt{GR2011}), but only include the 1st order term in the expansion of the quadrupole. In practice, because the actual force computation takes only a small fraction of the time spent walking the tree, we find that computing the force directly for each particle and including the quadrupole moments is typically the most efficient (see Section \ref{SS:COLDCOLLAPSE}).   The serial performance and parallel scaling of the tree is found to be sensitive to the choice of value for $N_{_{\rm LEAF}}$.  This is discussed in detail in Section \ref{S:PERFORMANCE}.

Finally, rather than rebuilding the complete tree at every step, we can update the properties of the tree cells bottom-up.  This is particularly relevant for time-steps where only a small fraction of all particles are active, in which case the cost of rebuilding the tree can become comparable to the cost of the hydro step itself. In practice we rebuild the tree after a fixed number of time-steps (specified by the user).  In contrast to \citet{GR2011}, we do not perform an integrity check on the tree since the tree-walking algorithm will always retrieve the correct neighbours even if the particles have moved outside of the initial cell (provided that the extent of the cells is updated accordingly).

\begin{figure*}
\includegraphics[width=\textwidth]{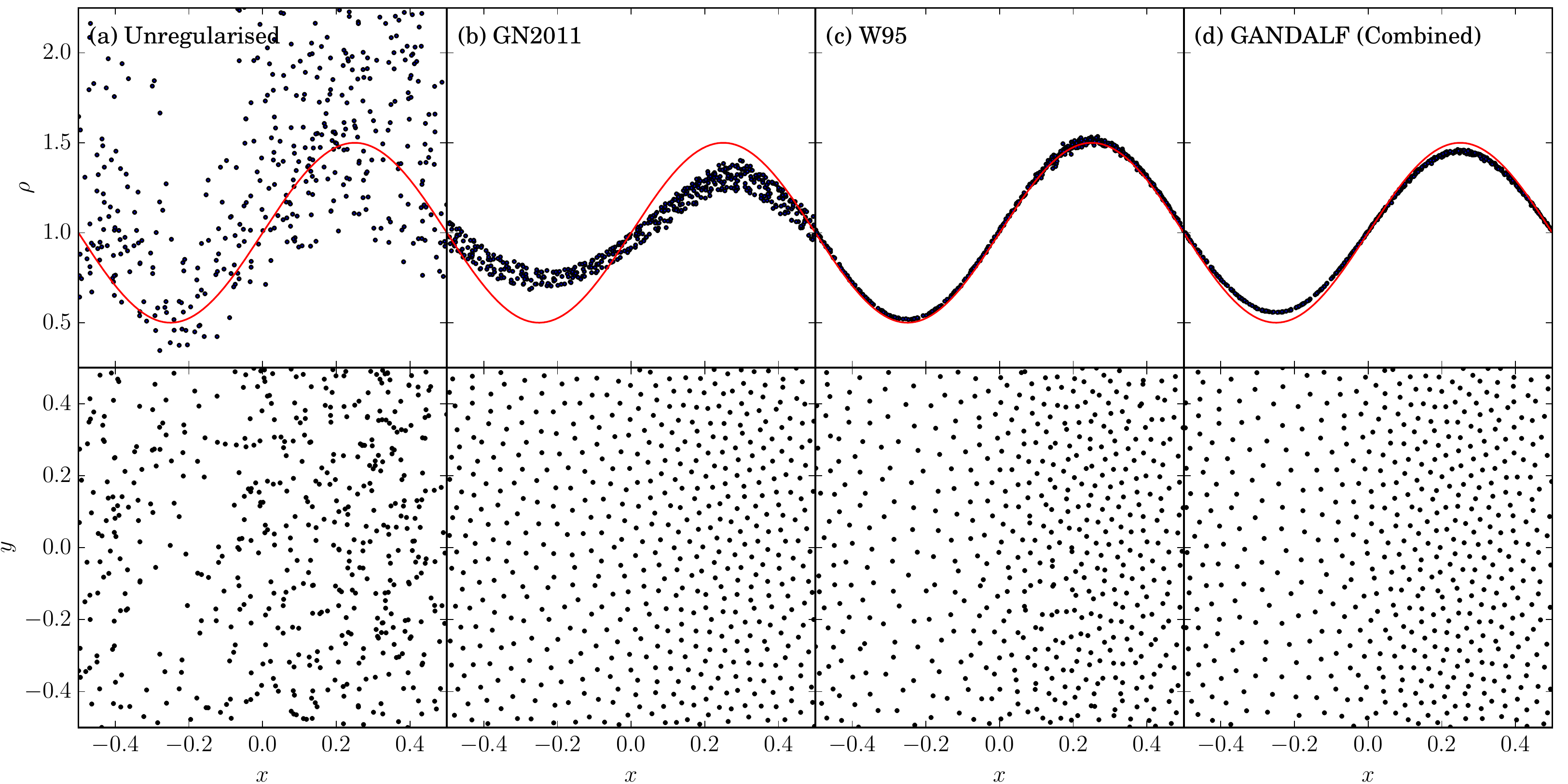}
\caption{The density profile (top row) and particle distribution (bottom row) resulting from generating a simple sine-wave density field using Monte-Carlo rejection sampling (1st column), the \citet{GN2011} regularisation method (2nd column), the \citet{IC1995} density method (3rd column) and the combined approach used in {\small GANDALF} (4th column).}
\label{FIG:REG_IC}
\end{figure*}

\subsection{Boundary conditions}

Both the SPH and MFV schemes can naturally handle isolated systems with no need for explicit boundary conditions. However, boundaries need to be explicitly handled in cases such as the join between computational domains, when modeling systems with reflection symmetry, or in periodic domains. Periodic and reflecting boundaries in \gandalf{} are handled using `ghost particles', which are copies of real particles that fall near the edges of the simulation domain. Depending on the type of boundary, these particles may be direct copies on a different processor (MPI domain boundaries), copies of particles that have been translated to a new position (periodic ghosts) or reflected across a boundary.

The ghost particles are constructed in one of two different ways; they can be computed in advance of time or generated on-the-fly as needed. In \gandalf{} both approaches are used. 
For the density and dust force calculations, both the physical and MPI ghosts are computed ahead of time. This is done because these loops may require the smoothing lengths to be iterated to achieve convergence, resulting in the need to export the particles every time the smoothing lengths are changed. As long as enough ghosts are  constructed initially there is no need to iterate the density. However, in the hydrodynamical and gravitational force calculations, which do not require iteration, ghosts at physical boundaries are constructed on-the-fly. This is done to simplify the gravitational force calculation in periodic simulations. Similar to {\small GADGET-2}, the contribution to the forces from interactions with particles on external processors is handled by exporting the particles to the other processor before computing the forces and sending back the result. 
 
When employing periodic boundaries with self-gravity, we use the Ewald method \citep[e.g.][]{Ewald1991} for computing periodic gravity forces.  This method assumes that the simulation box is infinitely replicated in all Cartesian directions.  A table of periodic gravitational correction terms is generated and used when computing forces between all gravitating particles or tree cells.  \citet{Wunsch2017} has recently adapted the original Ewald method to allow periodic gravitational forces for either 1D or 2D periodicity, which has been implemented in \gandalf{}.  This could be used for example to model an infinitely wide sheet or an infinitely long filament.  Although {\small GANDALF} is a multidimensional code, the periodic gravity can only be employed in 3D, whether using 1D, 2D or 3D periodicity.

\subsection{Generating Initial Conditions} \label{SS:ICS}

Constructing initial conditions for arbitrary density fields is in general more complicated for particle methods than grid methods, which can simply set the density field for each grid cell directly.  The simplest approach is to use Monte-Carlo rejection sampling of the density field, which gives approximately the correct density field but with a considerable amount of noise.  In Figure \ref{FIG:REG_IC}(a) (1st column), we use Monte-Carlo rejection sampling to select particles representing a simple sinusoidal density field, $\rho(x) = 1.0 + \tfrac{1}{2}\sin{\{2\pi x\}}$ in 2D.  As can be seen, the particle distribution is extremely non-regular (bottom row) leading to considerable scatter in the density field (top row), even when smoothed using Equation \ref{EQN:SPHRHO}.

\citet{GN2011} mitigate this problem somewhat by regularising the particle distribution at start-up (i.e. after initial conditions generation) to reduce this noise by making the local particle distribution more glass-like (Figure \ref{FIG:REG_IC}(b)).   Although successful, too many iterations leads to a completely uniform distribution of particles, effectively washing out the original density structure.  After $100$ iterations, while generating a more regular distribution with less noise, the amplitude of the sine-wave has been reduced by approximately a half (Figure \ref{FIG:REG_IC}(b); top row) and will continue to `decay' with successively more iterations.  Alternatively, \citet{IC1995} used a similar method to iterate particle positions towards a given density field (Figure \ref{FIG:REG_IC}(c)).  While giving a good fit to the density field and an improved particle distribution over the original Monte-Carlo sampling, this leads to a imperfect (i.e. not glass-like) distribution of particles with noticeable particle-particle `clumping' at various points in the distribution.

{\small GANDALF} contains a general IC algorithm that effectively combines the two approaches of \citet{IC1995} and \citet{GN2011} by simultaneously iterating towards a given density profile while moving the particles to a more regular distribution.  The full procedure for generating ICs is : 
\begin{itemize}
\item Calculate the total mass contained in the computational domain, $M_{_{\rm TOT}}$, either by analytically or by numerical integration of the density field.  All particles are assigned an equal mass $m = M_{_{\rm TOT}}/N$.
\item Use Monte-Carlo rejection sampling to assign the initial positions of all particles.  Although our algorithm works in principle from any initial distribution, it converges much faster if the particles are already close to their final positions.
\item Iterate the particle positions using 
\begin{equation}
{\bf r}\ssi' = {\bf r}\ssi + h\ssi\,\sum_{j=1}^{N} { \left\{ \alpha_{_{\rm IC}} - \beta_{_{\rm IC}}\,\left( \frac{\rho({\bf r}\ssj) - \rho\ssj}{\rho({\bf r}\ssj)} \right) \right\}  W({\bf r}\ssij, h\ssi) \hat{\bf r}\ssij \, }\,. \label{EQN:ITERATE_IC}
\end{equation}
where $\rho({\bf r}\ssj)$ is the analytical (or tabulated) density at the position of particle $j$, $\rho\ssj$ is the smoothed density of particle $j$, $\alpha_{_{\rm IC}}$ is the weighting of the particle regularisation term and $\beta_{_{\rm IC}}$ is the weighting of the density field term.  In practice, we find values of $\alpha_{_{\rm IC}} = 0.1$ and $\beta_{_{\rm IC}} = 0.9$ give a good balance between giving a regular particle distribution and an accurate density field.  We note that higher values of $\alpha_{_{\rm IC}}$ gives a more regular distribution but can under-resolve density peaks.  
\item Once the positions have converged, assign the remaining particle and hydrodynamical properties (e.g. velocity, specific internal energy).
\end{itemize}

One issue not addressed by this algorithm is creating equilibrium ICs, with the exception of trivial uniform density configurations (such as a uniform glass).  Hydrodynamical forces (due to 2nd order smoothing errors) are not truly represented by any given density gradient, even if the density field is accurate.  Gravitational forces also have a similar (although smaller in magnitude) smoothing error.  Therefore exact hydrostatic equilibrium cannot be obtained with this method.

\section{Implementation details} \label{S:IMPLEMENTATION}

\subsection{General design and structure of the code} \label{S:CLASSES}

\begin{figure}
\includegraphics[width=\columnwidth]{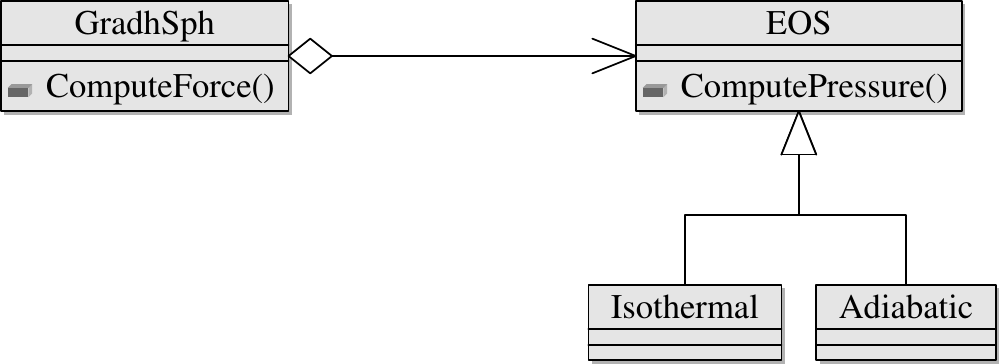}
\caption{An (idealised) example showing how we use the strategy pattern in \gandalf{}. In multiple places the code needs to compute the pressure of a particle; this is accomplished by calling a function defined in an abstract class ``EOS''. At code startup (typically depending on the parameters passed in by the user) it has been decided what the concrete implementation is (e.g., an adiabatic or isothermal equation of state); the code that needs the pressure does not need to be aware of how this is computed.}
\label{fig:strategy_pattern}
\end{figure}

We have followed many object-oriented principles when designing {\small GANDALF}. In this section we show some examples to demonstrate why an object oriented approach is useful for a Astrophysics hydrodynamical code; we refer the interested reader to the userguide and the codebase for more details on the class structure of {\small GANDALF}.  The use of object-oriented design has allowed  \gandalf{} to follow a philosophy of ``compile once for all''; all parameters can be selected at run time from the user, without any need for recompiling the code.

{\small GANDALF} contains multiple implementations of many important algorithmic features, such as hydrodynamics, the SPH smoothing kernel, N-body integration schemes, the spatial decomposition tree and more.  If the code were to inquire about the choice of an algorithm (e.g. how to compute the pressure of a particle) every time it is called, this would require an excessive use of {\it if-else} statements. Moreover, such a code would be inflexible when adding additional algorithms (e.g. a new equation of state); every time a new algorithm is added, every relevant {\it if} statement called in the code-base would need to be modified.  To solve this problem, we use the so-called ``strategy'' pattern proposed in the seminal book of \citet{Gamma}. Different algorithms for performing the same task (e.g. an isothermal or adiabatic equation of state; Figure \ref{fig:strategy_pattern}) are coded as different classes inheriting from a common ``parent'' class (the EOS class). The parent class declares in its interface a virtual pure function (e.g. {\it ComputePressure}), that the different strategies implement. ``Users'' of the algorithm (e.g. the SPH force calculation) only work through a pointer to the parent class, and do not need to behave differently depending on the exact strategy adopted.  Using this approach, we can separate the code where we choose the algorithm (typically done at code start-up) from the location where we invoke it, avoiding a long list of {\it if}s, for the benefit of code clarity and extensibility.

\begin{figure}
\includegraphics[width=\columnwidth]{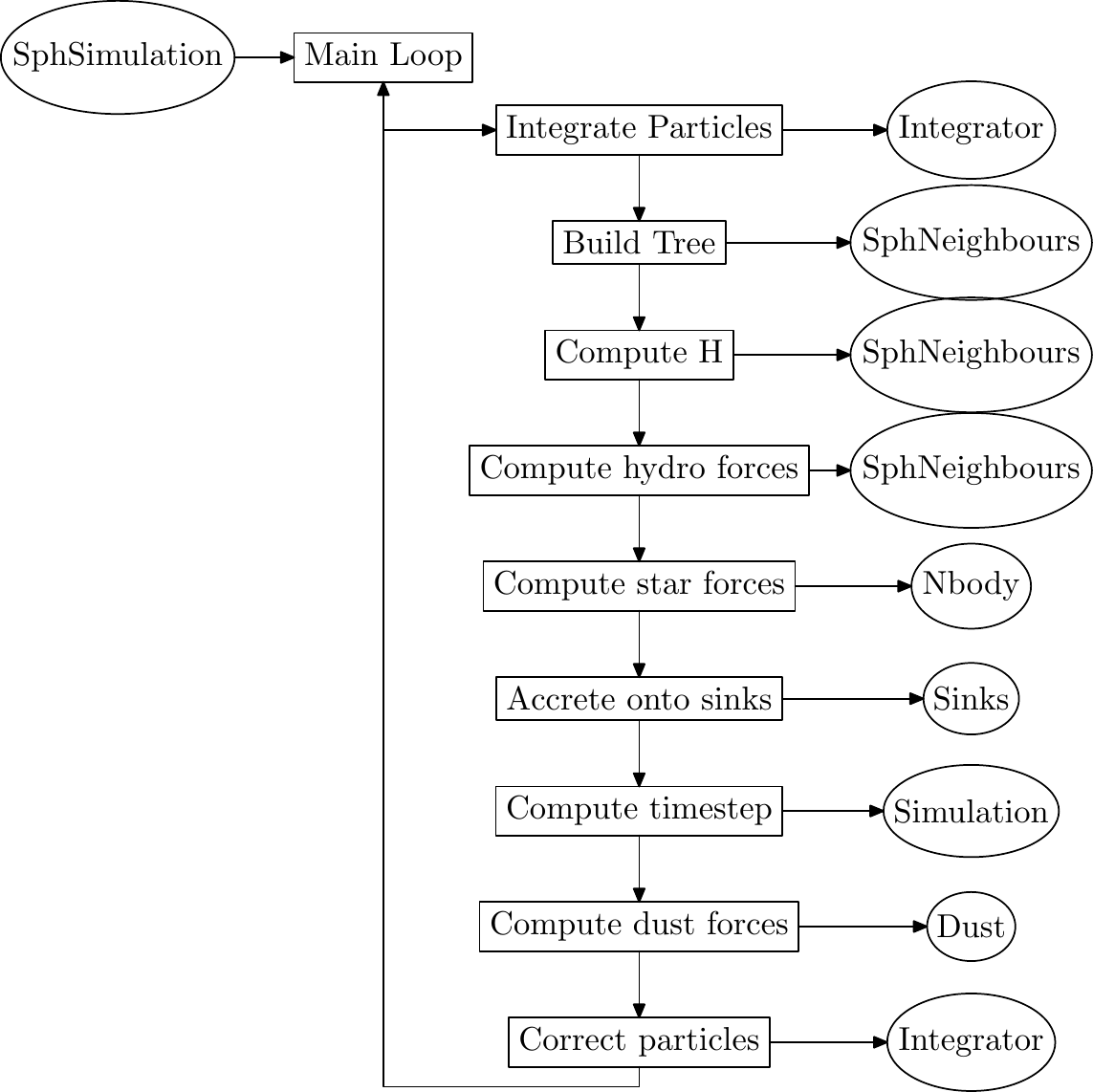}
\caption{A flow chart showing of the flow of the main integration loop of \gandalf{} for the SPH case. The circles indicate the class responsible for each action (shown in the rectangle).}
\label{fig:flow}
\end{figure} 

Another example of object orientedness is the use of a well-known feature of C++ called \textit{templates}. This is a way of expressing polymorphism at compile time rather than at run time, and as such incurs less overheads. Therefore we use this feature in performance critical sections of the code. For example, in a particle based algorithm the smoothing kernel is a critical part of the code. \gandalf{} supports several kernels, and we achieve this by templating the functions that use the kernel with the template class. This has the advantage that the kernel can be inlined (early testing has shown that this can lead to a performance improvement up to 30\%) and we can retain this performance while still being able to select the kernel at run time (i.e., there is no need to recompile the code if one wishes to change the kernel).

Finally, the last example of best object oriented practices is the use of composition over inheritance. The top level class present in the code is the simulation class, which governs for example the flow of the main loop (see the flow chart in figure \ref{fig:flow}). While we do use inheritance to distinguish the meshless algorithms from SPH (e.g., we have a SPHSimulation class and a MeshlessSimulation class), there are many other {\refone individual algorithms available for use} in the code, most of which have different options. This could lead to hundreds of different simulation types. We solve this problem by having multiple classes, each one responsible for one of the main subtasks of the main loop. 

Figure \ref{fig:flow} shows some of these subclasses; the main simulation class stores a pointer to each one of them. The main loop starts with integrating the particles in time (a task handled by a dedicated integrator class). We then build/update the structure used to retrieve neighbours (the tree) and proceed to the core of the algorithm: computing smoothing length and hydro forces. These tasks are also handled by the tree; the actual calculations of smoothing lengths and forces are subsequently delegated to a Sph class once the neighbours of a particle have been retrieved. At this point we compute the acceleration onto the stars and accrete gas onto the sinks. Then we compute the time-step (this is handled by the simulation class itself), compute the dust forces and finally correct the time integration of the particles with the newly computed accelerations (if necessary, depending on the time integration scheme). The meshless loop closely follows the SPH one, with two important differences. The first one is that the force calculation is replaced by two separate loops, one to update the gradient matrices and one to compute the fluxes. The second one is that, while for SPH we compute the gravitational acceleration together with the hydro forces (if both are present), for the meshless we must do it in two independent loops to preserve the second order accuracy in time of the integration.

\subsection{Parallelisation} \label{S:PARALLELISATION}

Our approach to parallelisation in \gandalf{} follows recent trends in high performance computing (HPC). We have parallelised the code using both \href{http://openmp.org/}{OpenMP} and \href{http://www.mpi-forum.org/}{MPI}. This hybrid parallelization allows the code to be used flexibly on different architectures. Modern hardware tends to be composed of few machines (``nodes'') containing each several cores, interconnected by high performance, low latency links (such as InfiniBand). An OpenMP only approach has the disadvantage that it is not possible to use more cores than what is available on a single node. Conversely, a pure MPI approach, while capable of running on any arbitrarily large number of nodes, does not take advantage of the fact that the different threads inside the same node are able to share the same memory, and no communication is needed between them. The use of hybrid parallelization allows us to have the best of both approaches.

\subsubsection{OpenMP parallelisation} \label{SS:OPENMP}
The OpenMP parallelisation strategy in \gandalf{} is straightforward in that the majority of the CPU time is spent in simple loops over the active particles, such as the calculation of the smoothing length (common to both SPH and the MFV schemes) and the calculation of the forces (for SPH) or the calculation of gradient matrices and fluxes (for the MFV schemes).  In these loops the computation for each particle is independent, which makes adding OpenMP parallelisation trivial. Only in very few places we need locks or atomics, which can limit the scaling.  As we mentioned in Section \ref{S:KDTREE}, we walk the tree for the particles in a cell rather than for single particles; a single unity of work for OpenMP is thus an active cell rather than each active particle.

The parallelisation of the KD tree construction is less straightforward.  The tree construction proceeds by bisecting repeatedly the particles on each tree level.  The construction of the first level can be performed only by one thread. On the second tree level, there are two sets of particles, each one of which can be processed independently. This allows us to extract parallelism by assigning a thread to each one.  We apply this strategy recursively to the each level;  we note that, if $N_\mathrm{threads}$ are available, we need $2^l >= N_\mathrm{threads}$, where $l$ is the tree level in order to keep all threads busy and obtain reasonable work-sharing.  Typically $N_\mathrm{particles} \gg N_\mathrm{threads}$, so that eventually all the threads are busy building the tree.  However, the bisection is typically an operation $\mathcal{O} (N)$ (\gandalf{} uses the  algorithm included in the C++ standard library, which is usually introselect), which means that the construction of each level takes roughly the same CPU time. Because the constructions of the first levels is done essentially in serial, it will limit the optimal scaling that can be reached during tree build. Further improvements to our strategy are only possible by parallelizing the select algorithm that performs the bisection.

Finally, for completeness we have also parallelised most of the other operations in the code of order $\mathcal{O}( N_\mathrm{particles})$, {\refone such as time integration, the calculation of the timestep and the calculation of the thermal properties,} although they do not dominate the wall clock time.

\subsubsection{Hybrid OpenMP/MPI parallelisation} \label{SS:MPI}
Typically shared-memory HPC machines contain $16$ cores which limits the problem sizes that can be investigated with \gandalf{}.  In order to extend this to more processors (a few $10$s, if not $\sim 100$ cores), we have implemented a hybrid OpenMP/MPI parallelisation.  The typical usage in \gandalf{} is to use OpenMP inside each shared-memory node and use MPI to communicate between nodes.  

We use domain decomposition via a KD tree to assign the particles to each MPI process.   This imposes the limitation that the number of MPI processes must be a power of 2.  Each MPI node constructs ``pruned'' versions of their trees to send to the other processors.  These are simplified trees, with a smaller number of levels than the full trees. The pruned trees allow each node to have an large-scale approximation of the mass distribution in the other domains, which is useful for many purposes. We note that the pruned trees in our implementation are not locally essential trees; i.e. they are not necessarily deep enough to allow other processor to compute the gravitational force resulting from the domain.

Some steps of the algorithms in \gandalf{} (e.g. the density calculation, the gradient estimation in the meshless, and the dust forces calculation) need information about the neighbours from other domains. This is accomplished by creating ``ghost'' particles on each local domain. Each node uses the pruned tree to establish which of its particles might be ghost particles on other nodes. When using periodic boundaries, we also create MPI ghosts of periodic ghosts. Our algorithm is generic and does not need to treat differently this case.

In other steps, where the ghosts would be modified by the interaction with the local particles (e.g. in the SPH force calculations or the MFV/MFM flux calculations), we have decided to use particle exchange rather than ghosts. This has the advantage that it allows us to treat hydrodynamics and gravity in the same way, and avoids the need to send information about all the ghosts even if only few of them are active. Operationally, when we find that a particle is too close to the boundary or the pruned trees of the other domains are not deep enough for gravity calculations, the particle is sent to the neighbouring domain. The other domains compute the contribution to the force from its local particles and then returns back to the original domain this partial force, which can be added to the total force.

Another significant part of the MPI code deals with transferring particles when they move between domains.  The boundaries of the domains need to be updated regularly to maintain load balancing.  To estimate the new location of the boundary, we assign each particle a fraction of the total CPU work, which depends on its time-step level; the work on each processor is weighted by the CPU wallclock time used by the MPI node to ensure a correct inter-processor normalisation. We use a bisection iteration method to find the best location of the new boundary, using the pruned trees to compute the new work in the domain.  Once the domain boundaries have been updated, particles that are now in different domains are transferred via MPI communication.

\subsection{Automated tests} \label{S:TRAVIS}

\gandalf{} contains many different algorithms and types of physics; it is thus important to make sure that any change to the code does not invalidate pre-existing code. To achieve this goal and ensure that no bugs are introduced in \gandalf{}, we have found invaluable to have a test suite that stresses the different options supported by \gandalf{}. The experience has shown us that such a test suite needs to be automated: it is impossible to inspect manually every time the results of many simulations. We use the python library to inspect the results of the simulations run by the test suite, compare them to analytical (or numerical) solutions and check that the overall error is within a given tolerance. Finally, the last requirement is that the test suite must be invoked automatically, or the execution will be procrastinated. We found that the on-line service {\small TRAVIS-CI}\footnote{https://travis-ci.org/}, which can be automatically linked to a github repository, perfectly matches this requirement by running the test suite every time a commit is pushed. In this way, during development we receive immediate feedback informing us if a newly added feature has broken any of the existing code.

\subsection{Python library} \label{S:PYTHON}

While most of the effort in developing \gandalf{} has been invested in being able to \textit{run} numerical simulations, this is certainly not enough for making science; being able to \textit{visualize} and \textit{analyse} the outputs is equally important. {\small GANDALF} contains a library written in Python dedicated to this task. An excellent software package, called {\small SPLASH} \citep{SPLASH} for the visualization of particle based simulations\footnote{Although {\small SPLASH} is designed for SPH, it can also easily handle outputs from the MFV schemes.} already exists and it is not the purpose of the library to supersede it. We note that {\small GANDALF} snapshot files are fully compatible with SPLASH. While we do provide a very essential subset of the SPLASH functionality in {\small GANDALF} (particle and rendered plots), the design principle of the Python library aims to fill a different gap. The goal of the library is to give the user programmatic access (e.g., save in a variable) to the data in the outputs. The library allows to access the raw data from the simulations (e.g., construct an array containing the smoothing lengths of the particles) and the basic visualisations described before (e.g., construct a 2D array containing a rendered plot). Additional functions permit to compare the simulation with analytical solutions (when known) and to repeatedly apply an analysis function to each snapshot in a simulation, making easy to plot a quantity as a function of time. The goal is to simplify writing analysis scripts. As a bonus, having some plotting capabilities built-in the code allows to inspect the simulation while it is running. We found this feature very convenient while developing the code. In the same way, we hope that future users wanting to add some physics to {\small GANDALF} will find it useful as well. Finally, having interfaced {\small GANDALF} with Python makes it possible to set up the initial conditions directly in Python, in case the user is not familiar with C++.

As already mentioned, following the general trend in scientific computing, the language of choice for this library is Python. This choice is motivated by the extreme flexibility of the language, its easiness to use, and the existence of libraries devoted to numerical analysis and publication-ready plotting (namely matplotlib). As {\small GANDALF} itself is written in C++, we need a ``bridge'' to make the two languages speak. For this purpose we make use of the {\small SWIG} library. With {\small SWIG} {\small GANDALF} can be compiled as a shared library object and therefore loaded into python as any standard python module.

\section{Tests} \label{S:TESTS}
In order to demonstrate the fidelity and limitations of the various components of {\small GANDALF}, we have a performed a wide range of tests of the code.  Many of these test cases deliberately overlap with those performed both with \arepo \citep{AREPO} and \gizmo \citep{GIZMO} in order to more easily compare them to {\small GANDALF}.  Since {\small GANDALF} is aimed more towards Star and Planet Formation problems (as opposed to Galaxy and Cosmological problems), we have substituted some Cosmology-oriented tests for others that are important for Star and Planet Formation scenarios.  In most hydrodynamical test cases, we perform with three different options; (a) Grad-h SPH, (b) MFV and (c) MFM.

\subsection{Soundwave test} \label{SS:SOUNDWAVE}

\begin{figure}
\includegraphics[width=\columnwidth]{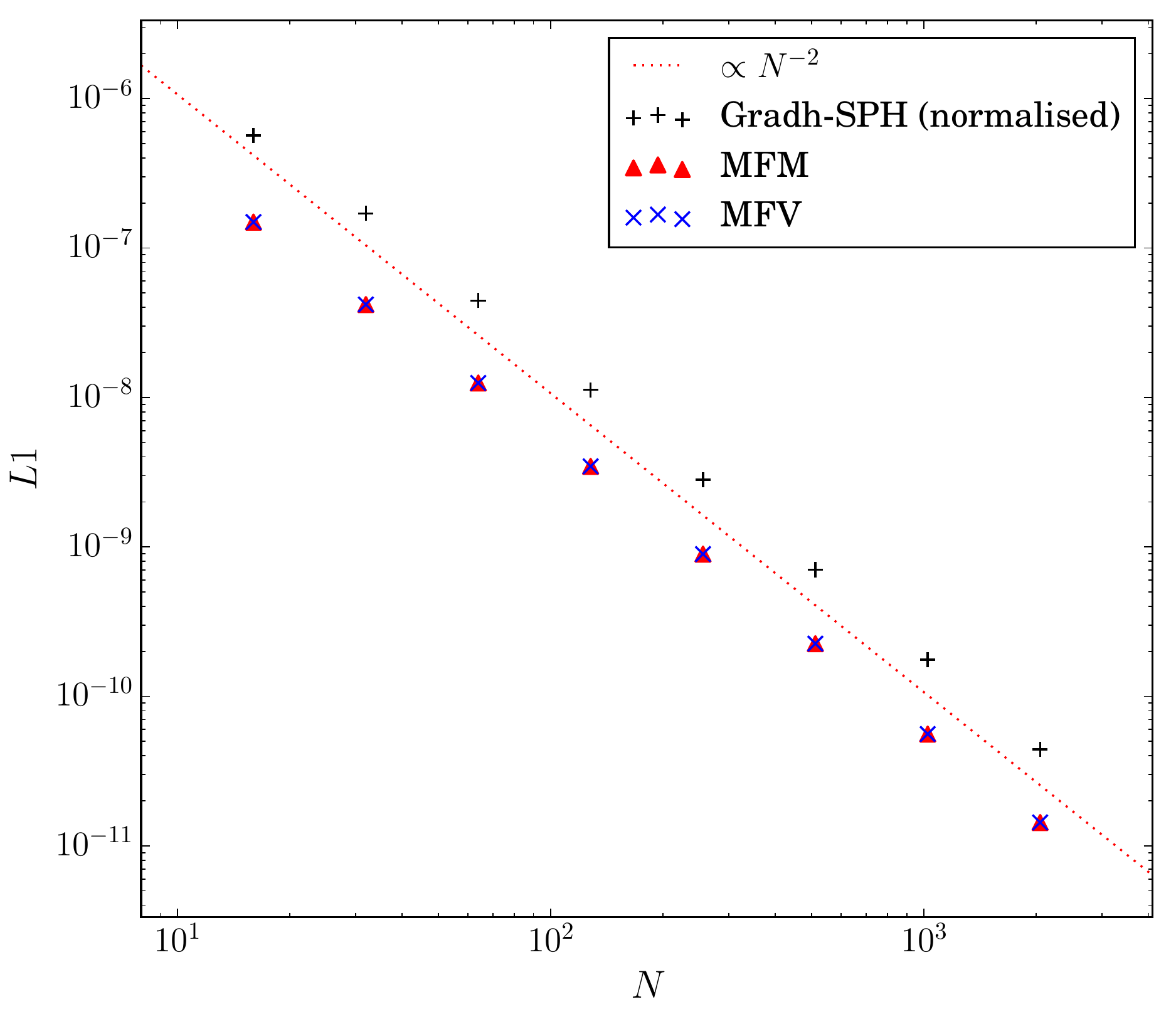}
\caption{The L1 error norm versus the simulation particle number for the soundwave test using the Grad-h SPH (black crosses), MFV (blue crosses) and MFM (red triangles) methods in 1D.  For smooth fluid flows, we would expect the errors to be dominated by the spatial and temporal integration errors of the numerical scheme, which in all cases should be 2nd order.  Therefore the L1 error norm should scale as $\propto N^{-2}$ in 1D (red dotted line).}
\label{FIG:SOUNDWAVE-L1ERROR}
\end{figure}

The goal of this test is to demonstrate that \gandalf{} correctly implements the hydrodynamical and time integration algorithms, preserving 2nd order convergence when dealing with smooth flows. We apply a low-amplitude sinusoidal density and velocity perturbation of the form 
\begin{flalign}
\rho(x) &= \rho_0 \left(1 + A\,\sin{\{kx\}} \right)\,, \label{EQN:SOUNDWAVE-RHO} \\
v(x) &= A\,c_s\,\sin{\{kx\}}\,, \label{EQN:SOUNDWAVE-V} \,
\end{flalign}
where $A$ is the density perturbation amplitude, $c_s$ is the sound-speed of the unperturbed gas and $k = 2\pi/\lambda$ is the wavenumber. To investigate the scaling of the error with resolution, we calculate the L1-error norms of the density field, i.e. 
\begin{equation}
|{\rm L1}| = \frac{1}{N} \sum_{i=1}^{N} {| \rho\ssi - \rho(x) |}\,,
\end{equation}
where $\rho\ssi$ is the particle density and $\rho(x)$ is given by Eqn. \ref{EQN:SOUNDWAVE-RHO}, as a function of particle number, $N$. The L1-error norm is expected to scale as $\propto N^{-2/D}$, where $D$ is the dimensionality.  

The initial conditions are created following \citet{Athena2008}.  A set of $N$ particles are placed in 1D at equidistant intervals along the x-axis between $x=0$ and $x = 1$.  The sinusoidal density perturbation is created by slightly perturbing the positions of the particles along the x-axis to match the correct density profile \citep[see for example][for a description of creating a sinusoidal density field]{Hubber2006}. {\refone We use values $\rho_0 = 1$, $A = 10^{-6}$, $c_s = 1$ and $\lambda = 1$ for our perturbation.}

Figure \ref{FIG:SOUNDWAVE-L1ERROR} shows the L1-error norm as a function of particle number for all simulation modes presented here.  The MFV (blue crosses) and MFM (red triangles) schemes all scale with the expected $L1 \propto N^{-2}$ error norm (red dotted line) for both low and high resolutions, similar to the results found by \citet{GIZMO}. For the SPH simulations, one important caveat is that the SPH density sum (Eqn. \ref{EQN:SPHRHO}) results in a consistent fractional offset/error from the true uniform density of less than one percent {\refone (for the kernels employed in {\small GANDALF})}. Normally this is unimportant in simulations but can affect this test where there is a density perturbation of smaller amplitude.  \citet{GIZMO} attempts to fix this problem by iterating the particle positions; however at high resolutions {\refone this error} eventually dominates, breaking the 2nd order convergence.  Since here we are interested in showing 2nd order convergence in order to test our implementation, we perform our analysis of the SPH simulations by normalising the average density to $\rho_0$ (as measured from the simulation itself); this removes the 0th order error from the L1 norm.  With this normalisation applied, we can see that also the SPH results scale with the expected $L1 \propto N^{-2}$ trend since the spatial error is dominated by the smoothing kernel errors.

\subsection{Shocktube tests} \label{SS:SHOCKTUBES}
Shocktube tests are typically used to test the shock capturing ability of a hydrodynamical code. We use two different equations of state (isothermal and adiabatic) in what follows to test our implementation in both cases (notice that the energy equation is evolved only in the latter case). The initial conditions are set-up in 1D by creating a uniform line of particles in contact to represent the left and right states. The set-up is similar (albeit with slightly higher resolution) to the same test performed by both \citet{AREPO} and \citet{GIZMO} to allow easy comparison with those two papers.  {\refone We use the standard \citet{Monaghan1997} prescription for artificial viscosity without limiters for SPH simulations and the \citet{GIZMO} limiter for MFV/MFM simulations.}  The LHS (i.e. $x < 0$) gas state is $P_L = 1$, $\rho_L = 1$, $v_L = 0$ and the RHS ($x > 0$) is $P_R = 0.1795$, $\rho_R = 0.25$, $v_R = 0$ in a computational domain of size $-20 < x < 20$.  For the adiabatic case, the gas obeys an ideal-gas equation of state, $P = (\gamma - 1)\,\rho\,u$, where $\gamma = 1.4$.  For the isothermal case, the gas obeys an isothermal equation of state where $c_s = 1$ so $P_L = 1$ and $P_R = 0.25$.  We consider two different sets of initial conditions; (i) the LHS contains $240$ particles and the RHS contains $60$ particles (i.e. equal-mass particles); (ii) both the LHS and RHS contains $60$ particles each (i.e. equally-spaced particles).

\subsubsection{Adiabatic shocktube}

\begin{figure*}
\includegraphics[width=\textwidth]{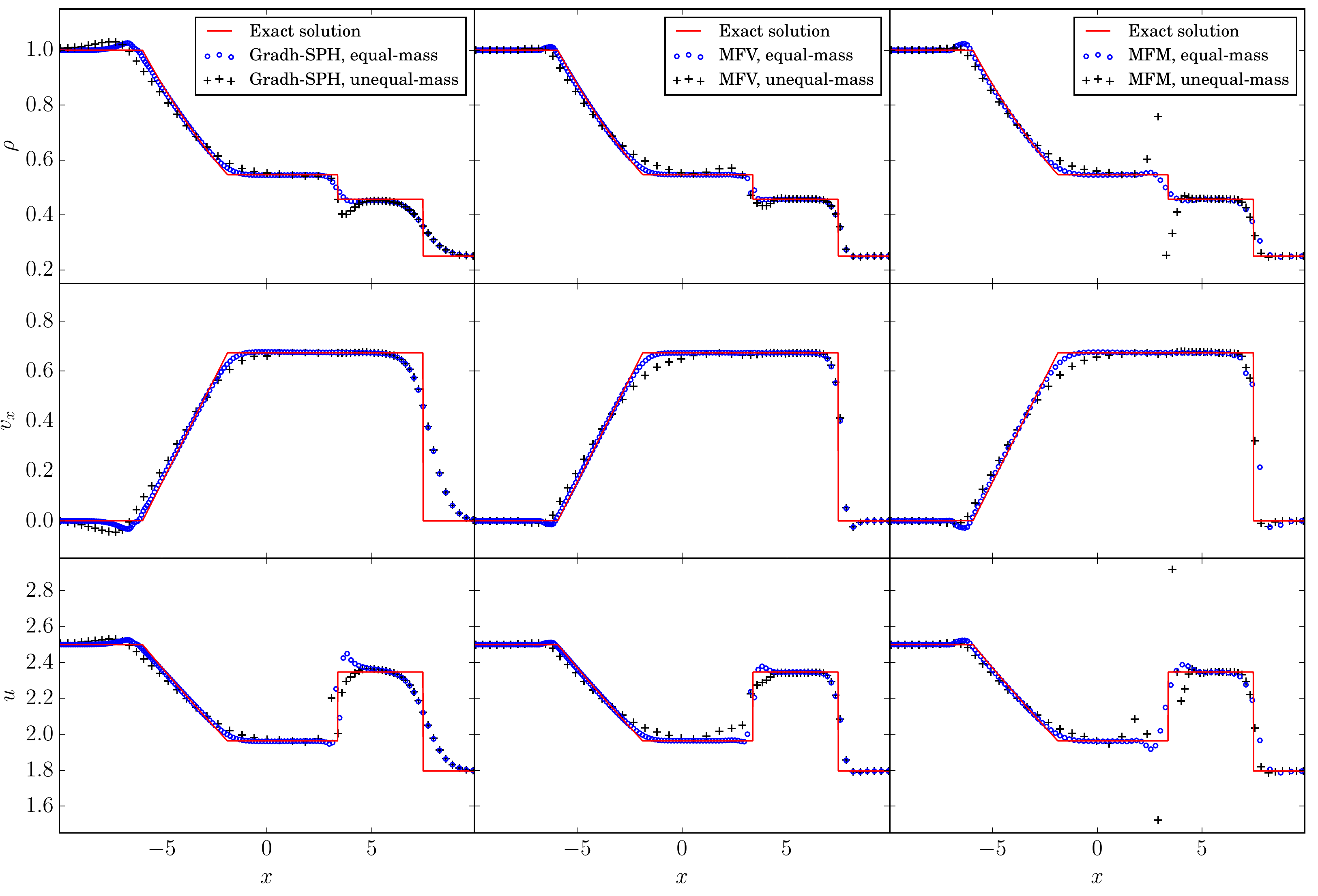}
\caption{Simulations of the adiabatic Sod test using the Grad-h SPH (1st column), MFV (2nd column) and MFM (3rd column) using initial conditions with equal-mass particles (black plus symbols) and equally-spaced particles (blue open circles) at $t = 5.0$.  Plotted are the particle density (1st row), velocity (2nd row) and internal energy (3rd row) for each case including the analytical solution from the Exact Riemann solver (red line).}
\label{FIG:ADSOD}
\end{figure*}

Figure \ref{FIG:ADSOD} shows the results for the adiabatic shocktube for all cases at the final simulation time $t = 5$.  For all simulation types, the general form of the density, velocity and pressure profiles are captured correctly, in line with the results of \citet{GIZMO}, proving the correctness of our implementation of the meshless schemes. We also recover two features noted by \citet{GIZMO}; SPH in general has larger overshoots and undershoots at the discontinuities for equal-mass initial conditions (blue open circles) and a slightly higher diffusivity (the jumps are not as sharp).

For the equally-spaced (non-equal mass) initial conditions (black crosses), we find a more significant dip in the density at the contact discontinuity for SPH and MFV in line with \citet{GIZMO}; however, they did not show results for MFM. We find that this method has a much stronger `blip' in both the density and energy plots at the discontinuity. We interpret this feature as a wall-heating effect; the lack of mass advection in MFM prevents any (artificial) numerical mixing which can smooth out this blip. SPH and MFV are instead more diffusive due to, respectively, artificial viscosity and mass advection. A slightly more diffusive Riemann solver might allow this blip to be diffused away.

\begin{figure}
\includegraphics[width=\columnwidth]{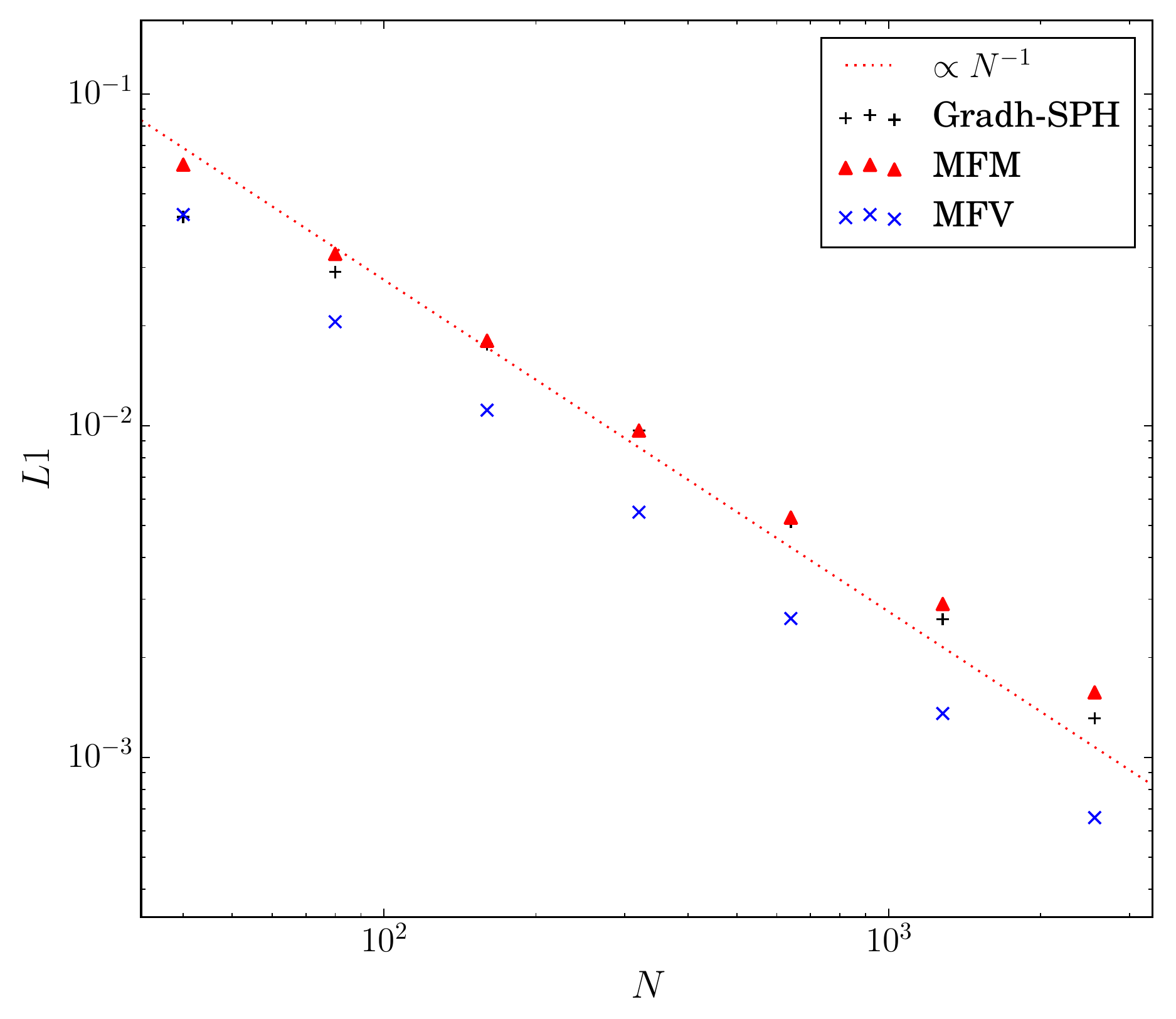}
\caption{Plots of the L1 error norm versus the simulation particle number for the adiabatic Sod test using the Grad-h SPH and Meshless-FV methods in 1D.  For problems involving shocks, the shock error dominates the total error reducing what are nominally 2nd-order schemes to 1st-order. A line scaling as $L1 \propto N^{-1}$ is shown for comparison.}
\label{FIG:ADSOD-L1ERROR}
\end{figure}

We plot the L1 error norms versus the particle number in Figure \ref{FIG:ADSOD-L1ERROR}. In a shocktube problem, errors near the shock-front will dominate the total error in quantities such as the density. In the vicinity of the shock, the numerical schemes should reduce from 2nd (or higher) order to 1st-order since the effect of artificial viscosity, or slope limiters in Godunov codes, is to reduce the scheme to 1st order to satisfy Godunov's theorem \citep[e.g.][]{ToroBook}.  All the methods broadly follow the expected $L1 \propto N^{-1}$ scaling.

\subsubsection{Isothermal Sod shock} \label{SSS:ISOSOD}

\begin{figure*}
\includegraphics[width=\textwidth]{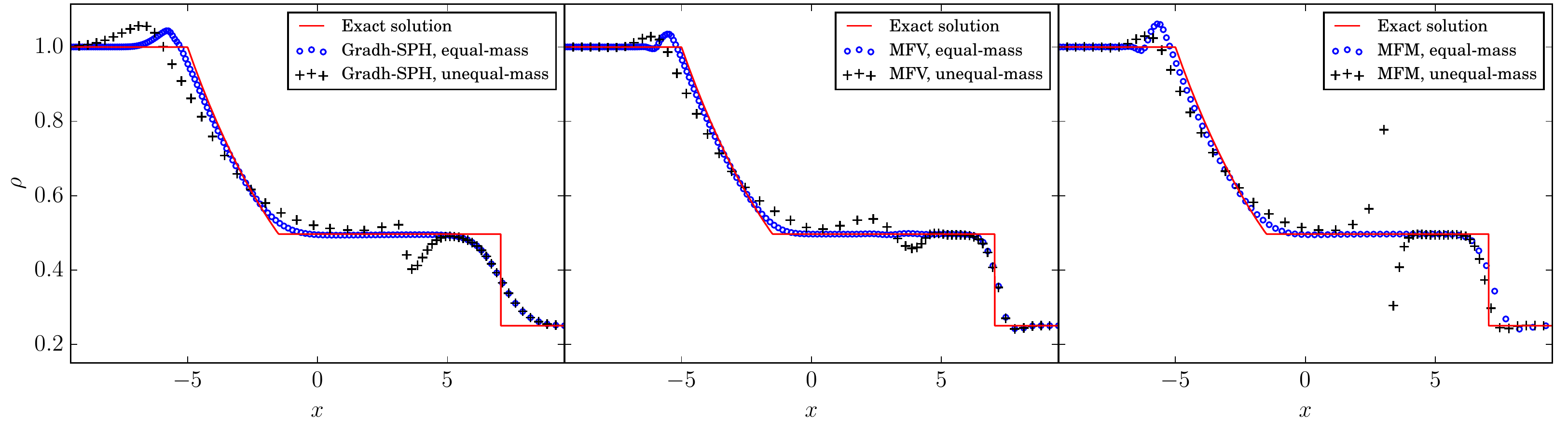}
\caption{Density profile at $t = 5.0$ resulting from simulations of the isothermal Sod test using the Grad-h SPH (1st column), MFV (2nd column) and MFM (3rd column) using initial conditions with equal-mass particles (black plus symbols) and equally-spaced particles (blue open circles).  The exact solution is also plotted (red lines).}
\label{FIG:ISOSOD}
\end{figure*}

We perform the same test using an isothermal equation of state. The purpose of this test is to test our implementation of the isothermal Riemann solver. In Figure \ref{FIG:ISOSOD}, all methods give acceptable results using the equal-mass (blue open circles) initial conditions with similar features to the adiabatic case (but with slightly larger overshoots near the tail of the rarefaction wave). All the methods recover correctly a flat density profile at the original contact discontinuity (although with a small oscillation for the MFV case). For the equally-spaced (non-equal mass) case, the methods show instead more prominent numerical artifacts near the contact discontinuity.

\begin{figure*}
\includegraphics[width=\textwidth]{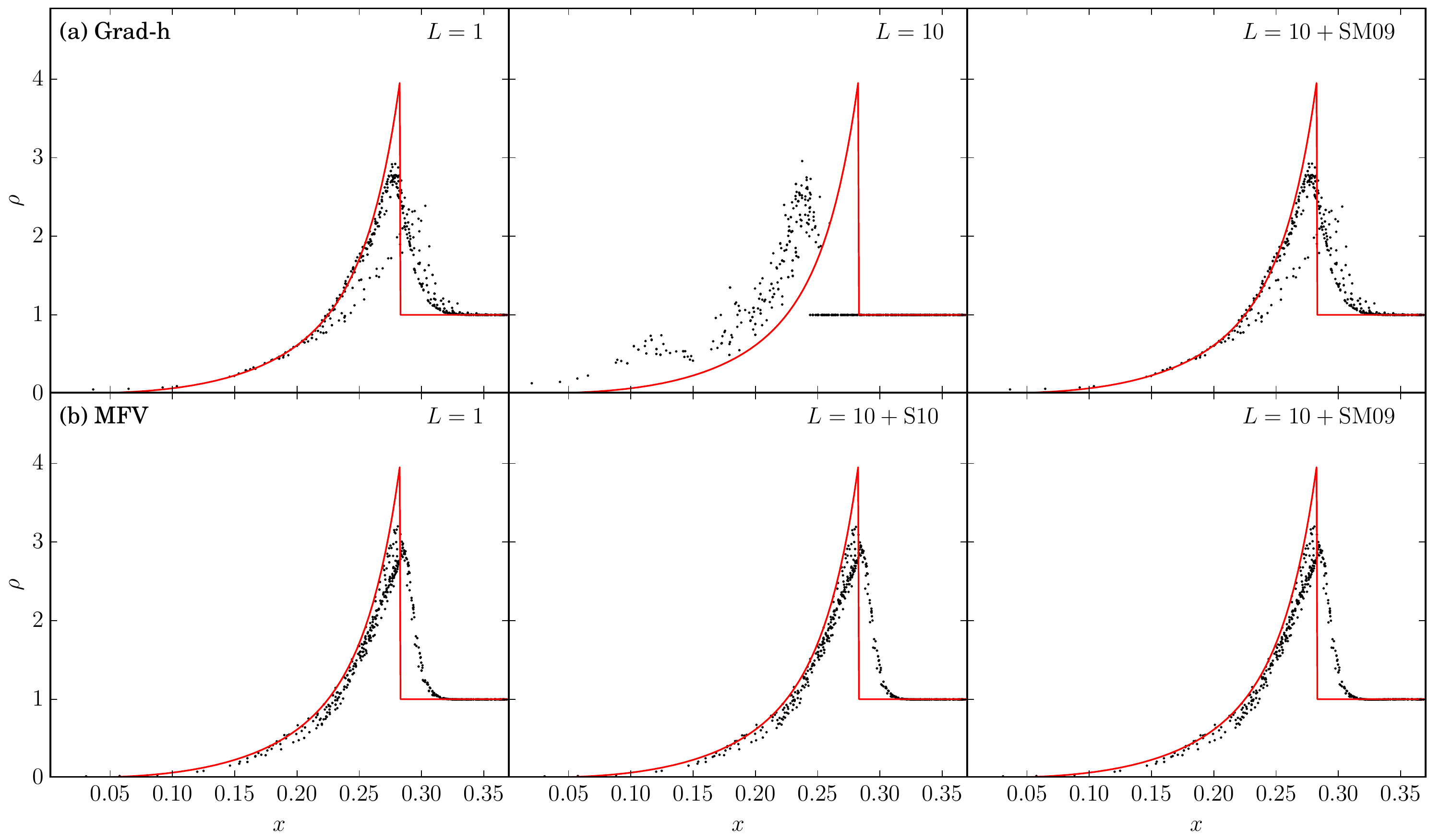}
\caption{Simulations of the 2D Sedov-Taylor blast-wave test at $t = 0.06$ using grad-h SPH (1st row) and the MFV scheme (2nd row). The 1st column shows the results with global time-steps. For SPH, we show in the second column the case with 10 time-step levels and no limiter. We do not show this case for the MFV since it crashes before the end. For MFV, the second column shows instead the results with the \citet{AREPO} limiter. The third column shows the results with the \citet{SM2009} time-step limiter. The semi-analytical solution is plotted for comparison (red line). Both time-step limiters perform very well and the results are indistinguishable from the global time-step run, while using individual time-steps without limiter clearly leads to wrong results.}
\label{FIG:SEDOV2D}
\end{figure*}

\subsection{Sedov blast-wave test}

The Sedov-Taylor blast-wave is a demanding test of the accuracy of energy conservation and of the individual time-stepping algorithm of a particle code; \citet{SM2009} showed that without a time-step limiter one gets catastrophic results. This is important in many astrophysical applications where a sudden energy input may be triggered by supernovae explosions or high-energy feedback from accreting massive stars. In \gandalf{} we provide two different time-step limiters, following \citet{SM2009} and \citet{AREPO}, and we perform this test to benchmark them.

We set-up a 2D Sedov-Taylor blast-wave simulation by creating a cubic lattice containing $64^2$ particles in the region $-1 < x < 1$, $-1 < y < 1$.  The particles are given an equal mass to give a uniform density of $\rho = 1$.  We assign the total energy of the explosion ($E=1$) to the particles within a single smoothing kernel of the origin, where each particle's contribution is weighted by its smoothing kernel value. For both the grad-h SPH and MFV schemes, we perform simulations with (i) global time-steps, (ii) ten time-step levels using the time-step limiter. For SPH the only option is the \citet{SM2009} limiter, while for the meshless we test also the \citet{AREPO} limiter. We also perform additional simulations with multiple time-step levels with no limiter to check that, confirming the results of \citet{SM2009} and \citet{Hubber2011}, in this case we fail to reproduce the analytical result, getting a noisy density field and wrongly predicting the location of the shock. In this case we note that the MFV method is less robust than SPH and it is prone to crash when using multiple time-step levels; we cannot run the test to completion without using a time-step limiter.

Figure \ref{FIG:SEDOV2D} plots the density profile at $t = 0.06$ for all cases along with the semi-analytical solution (red line).  Both the SPH and MFV schemes follow a similar pattern with the various time-step options.  For global time-steps (1st column), they both reproduce the semi-analytical solution reasonably well, including most importantly the shock position.  All the methods under-resolve the peak shock density due to the finite resolution and the use of smoothing kernels.  The MFV scheme resolves the peak slightly better than SPH, with a peak density of just over $3$ (compared to just under $3$ for the SPH), although the difference is smaller than that found by \citet{GIZMO}. We note that the kernel weighting at the base of SPH and the meshless methods will always lead to some smoothing of sharp features. 
Using either of the two implemented time-step limiters, the \citet{SM2009} limiter (2nd column) or the \citet{AREPO} limiter (3rd column), improve the simulation results considerably and are nearly indistinguishable from the single time-step level results, proving the correctness of our implementation. As explained in section \ref{SS:TIMESTEPLIMITER}, the \citet{SM2009} does not enforce energy conservation; for example at the end of the simulation the fractional energy error has gone up to $\sim 10^{-4}$. The \citet{AREPO} time-step limiter instead is conservative and ensures energy conservation at a level of $\sim 10^{-13}$, which is similar to the result we get with global time-steps. This does not come for free though; the test with the conservative time-step limiter is roughly 20\% more expensive in terms of computational time. Even in this case, the time-step limiter still allows a saving of almost a factor of 3 compared with global time-steps ($\sim 11.4$s compared to $\sim 4.2$s).

\begin{figure*}
\includegraphics[width=\textwidth]{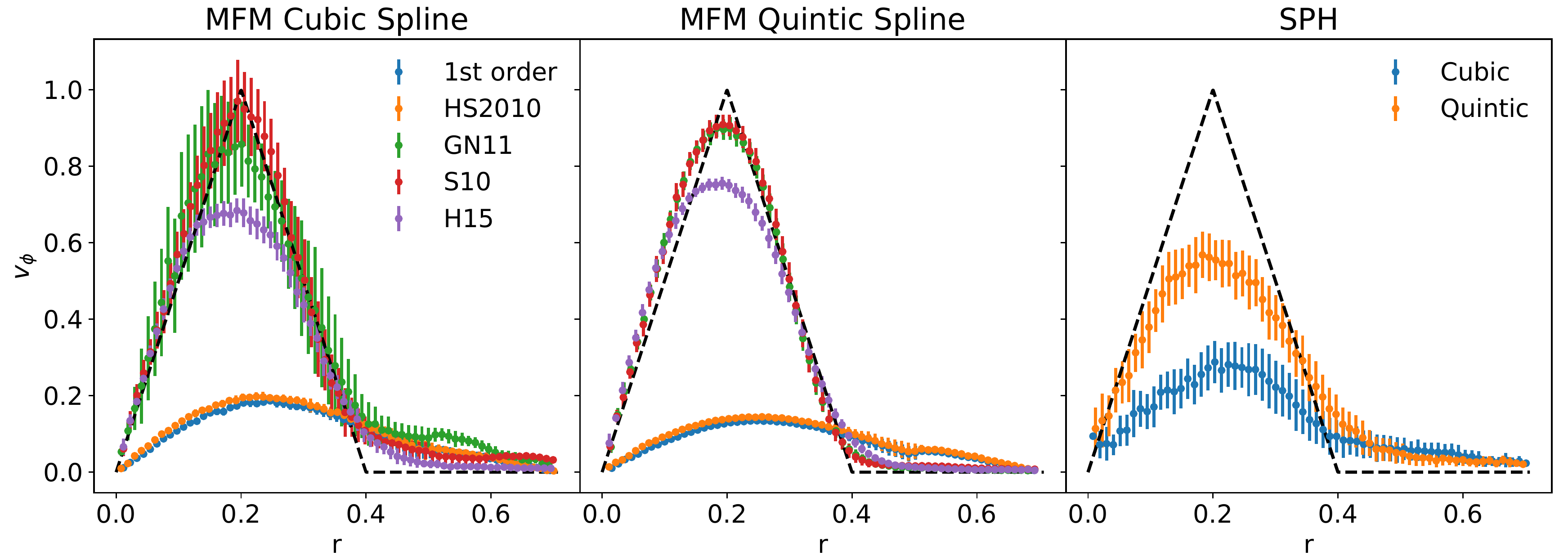}
\caption{The gresho vortex test computed with SPH and the MFM schemes, using both the cubic and quintic spline kernels. For the MFM we also show a range of different slope limiters. For the SPH simulation the \citet{2010MNRAS.408..669C} switch was used with \citet{Price2008} artificial conductivity. Points and error bars show the mean and standard deviation of particles within each radial bin.}
\label{FIG:GRESHO}
\end{figure*}

\subsection{Gresho-Chan vortex}
The \citet{GC1990} vortex test involves a steady rotating vortex profile in which the rotation is supported by pressure. We study this problem in 2D, $64\times 64$ particles on a cubic lattice on a periodic domain with $-0.5 < x,y <0.5$. The initial pressure profile is
\begin{equation}
P(R) = 
\begin{cases}
5 + \frac{25}{2}R^2                   & 0 \leq R < 0.2 \\
9 + \frac{25}{2}R^2 - 20R + 4 \ln{5R} & 0.2 \leq R < 0.4 \\
3 + 4\ln{2}                           & R \geq 0.4\,,
\end{cases}
\end{equation}
and the initial (azimuthal) velocity profile is
\begin{equation}
v_{\phi}(R) = 
\begin{cases}
5R      & 0 \leq R < 0.2 \\
2 - 5R  & 0.2 \leq R < 0.4 \\
0       & R \geq 0.4\,.
\end{cases}
\end{equation}
{\refone The initial density is $\rho = 1$ everywhere and the gas obeys an adiabatic equation of state with $\gamma = 5/3$.}  The initial radial velocity profile is set to zero.  

The azimuthal velocity profile at $t=3$ is shown in Fig.~\ref{FIG:GRESHO} for the both the MFM and SPH methods. We do not show the results for the MFV method, which are essentially the same as those as the MFM method. In the SPH simulations both the \citet{2010MNRAS.408..669C} viscosity limiter and the \citet{Price2008} artificial conductivity were used. For the meshless we show the results for the range of slope limiters included in \gandalf{}. Finally, we explore both the cubic and quintic spline kernels. 

The poor performance of SPH in this test is already well known, with the high artificial dissipation leading to a fast damping of the vortex. The \citet{2010MNRAS.408..669C} switch alleviates this somewhat compared to the behavior of standard SPH (see \citealt{Rosswog2015}), but the dissipation remains large. The performance of the MFM method is very sensitive to choice of the slope limiter (note that this was reported by \citealt{GIZMO}, but they did not show the differences in their figures), with the most diffusive limiters (i.e. the 1st order Godunov scheme, or \citealt{Hess2010}) showing the same poor performance as SPH. The least diffusive limiters (i.e. \citealt{GN2011} and \citealt{AREPO}) show essentially no dissipation, although we do see some broadening of the vortex peak. The \citet{GIZMO} limiter falls between the two extremes, showing a modest level of dissipation.

In addition to running the \citet{GC1990} test with `standard' cubic spline kernel we have also run the test using the quintic spline kernel for both the SPH and MFM schemes. This highlights the importance of accurate volume and gradient estimates in the presence of strong shear, which acts to disrupt the ordered particle positions, as shown by \citet{Rosswog2015}. In the case of SPH the dissipation is reduced considerably, to a level that is only slightly greater than the MFM with the \citet{GIZMO} limiter. 

This test demonstrates that using the quintic spline kernel also significantly improves the performance of the MFM methods. The main effect is a reduced level of noise, which consequently results in the slope limiters being triggered less frequently. In practice this does not much affect the least diffusive methods, where the slope limiters are already triggering very rarely. However, in the case of the \citet{GIZMO} limiter the reduced noise does reduce the level of dissipation. Finally, for the most diffusive cases the reduced noise does not reduce the triggering of the slope limiter, and thus the predominant effect is one of lower effective resolution (due to the large smoothing volume).

\subsection{Gravity tree accuracy} \label{SS:COLDCOLLAPSE}

In this test we set up a {\refone random distribution of particles in a uniform spherical volume of radius $1$.} We compute the gravitational acceleration on each particle using both the tree and direct sum; the comparison between the two informs us on the accuracy of tree and how it varies with the parameters of the tree.  We compute the total net error done in the gravitational acceleration as 
\begin{equation} \label{EQN:GRAVERROR}
|\delta {\bf a}| = \left( \frac{1}{N} \sum_{i=1}^{N} { \left\{ \frac{|{\bf a}\ssi^{\rm TREE} - {\bf a}\ssi^{\rm DIR}|^2}{|{\bf a}\ssi^{\rm DIR}|^2} \right\} } \right)^{1/2}\,,
\end{equation}
where ${\bf a}\ssi^{\rm TREE}$ and ${\bf a}\ssi^{\rm DIR}$ are the accelerations computed via the tree and direct-sum respectively. For this tests we have employed a resolution of 16k particles; the number of particles in each leaf cell has been held fixed to 6.

\begin{figure*}
\includegraphics[width=\textwidth]{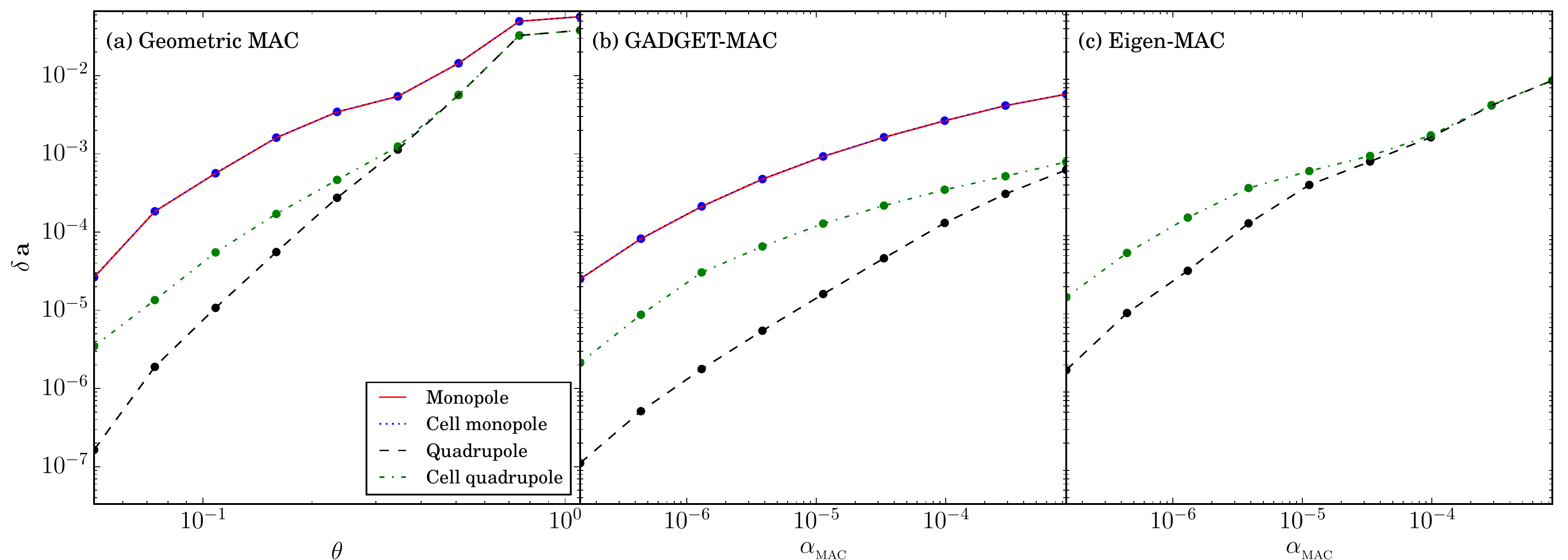}
\caption{The mean gravitational acceleration error (using Equation \ref{EQN:GRAVERROR}) while computing the initial forces for particles in a uniform density sphere using the KD-tree using (a) the geometric opening-angle criterion as a function of the maximum opening angle, $\theta_{_{\rm MAX}}$, and (b) the {\small GADGET-2} \citep{Gadget2} and (c) Eigenvalue-MAC \citep{Hubber2011} as a function of the error tolerance criterion, $\alpha_{_{\rm MAC}}$, while using the monopole (red solid line), cell-monopole (blue dotted line), quadrupole (black long dashed line) and cell-quadrupole (green dot-dashed line) multipole approximations.}
\label{FIG:SPHEREACCELPOT}
\end{figure*}

In Figure \ref{FIG:SPHEREACCELPOT} we show the mean gravitational acceleration error using different tree opening criteria and different multipole approximations. As expected the errors become smaller in all cases when the tree is required to open more cells. In addition, using higher multipole approximations also improves the accuracy of the tree as expected; we see a clear trend when going from the monopole methods to the cell-quadrupole and then to the full quadrupole.

\begin{figure*}
\includegraphics[width=\textwidth]{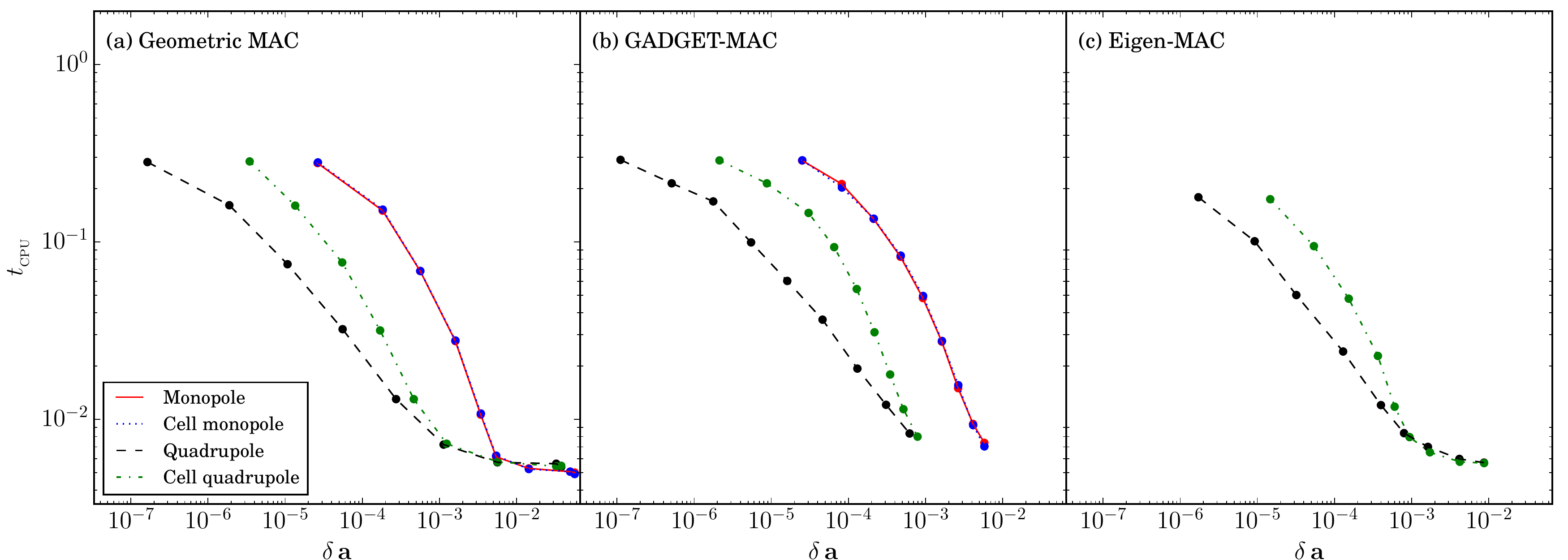}
\caption{The CPU time (relative to the brute-force ${\cal O}(N^2)$ computation) to compute the gravitational forces via the KD-tree compared to the mean gravitational acceleration error (using Equation \ref{EQN:GRAVERROR}) while computing the initial forces for particles in a uniform density sphere using (a) the geometric opening-angle criterion (b) the {\small GADGET-2} \citep{Gadget2} MAC and (c) the Eigenvalue-MAC \citep{Hubber2011}.  Results are plotted for tree-cell expansions using the monopole (red solid line), cell-monopole (blue dotted line), quadrupole (black long dashed line) and cell-quadrupole (green dot-dashed line).}
\label{FIG:SPHEREERRORTIME}
\end{figure*}

Figure \ref{FIG:SPHEREERRORTIME} shows the CPU time to compute the gravitational forces as a function of the accuracy. In our implementation, the quadrupole method calculates the force to a given accuracy with the least amount of CPU time and is therefore the most optimal choice of multipole expansion.  The quadrupole method results in a given accuracy by opening less cells during the tree-walk, but performing more work per cell in computing the extra quadrupole terms.  Whether this is more efficient than opening more cells only using the monopole depends largely on the details of the implementation, and for \gandalf{} the tree is faster doing more iterations over distant cells, rather than opening more cells overall. One reason for this behaviour might be that we make local copies of the quadrupole moments of the distant cells, and hence iterating over them is relatively fast since they are already held in the CPU cache.  In this problem, there is very little difference between the different opening criteria, as they all reach a given accuracy in roughly the same time.  However, this might change with different density fields.

\begin{figure}
\includegraphics[width=\columnwidth]{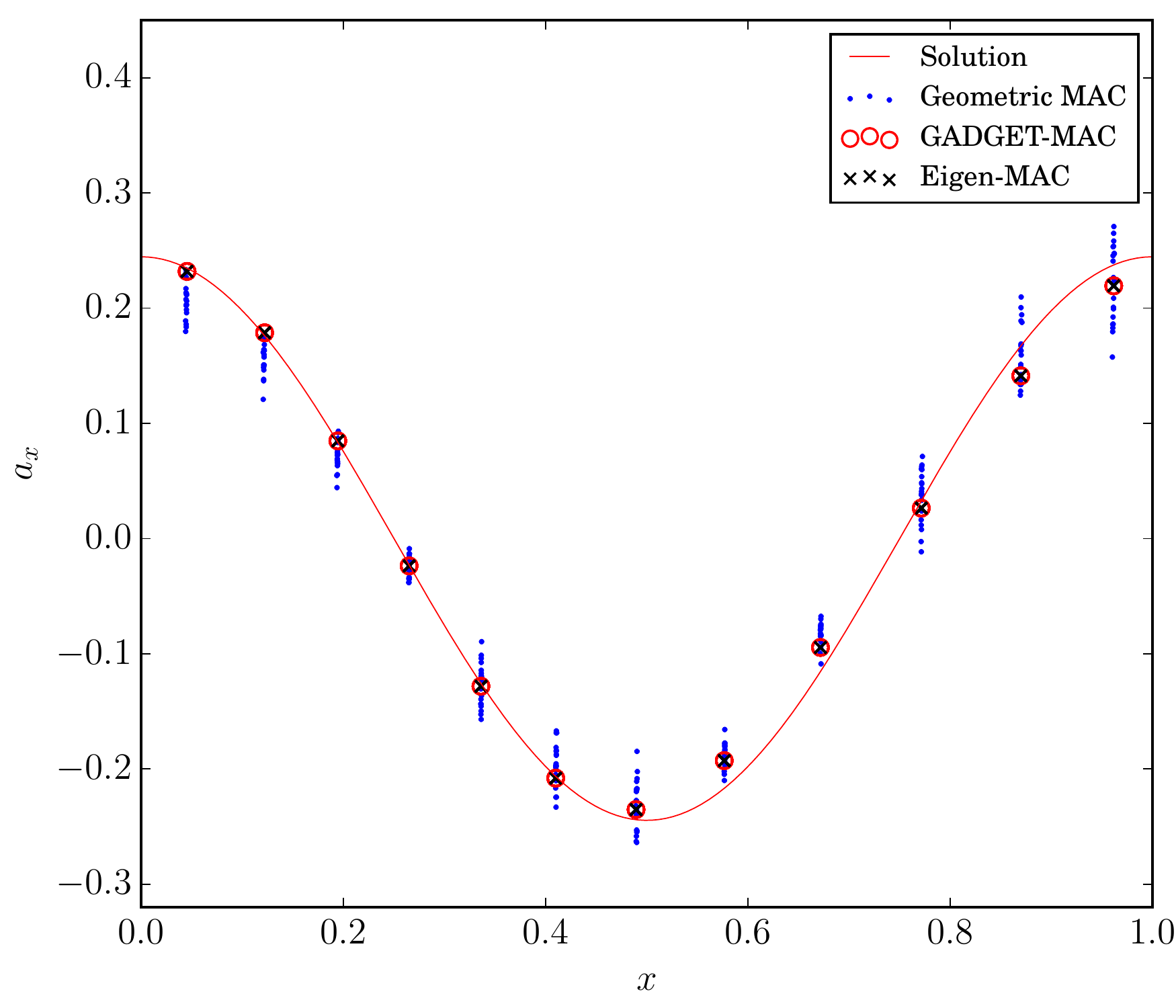}
\includegraphics[width=\columnwidth]{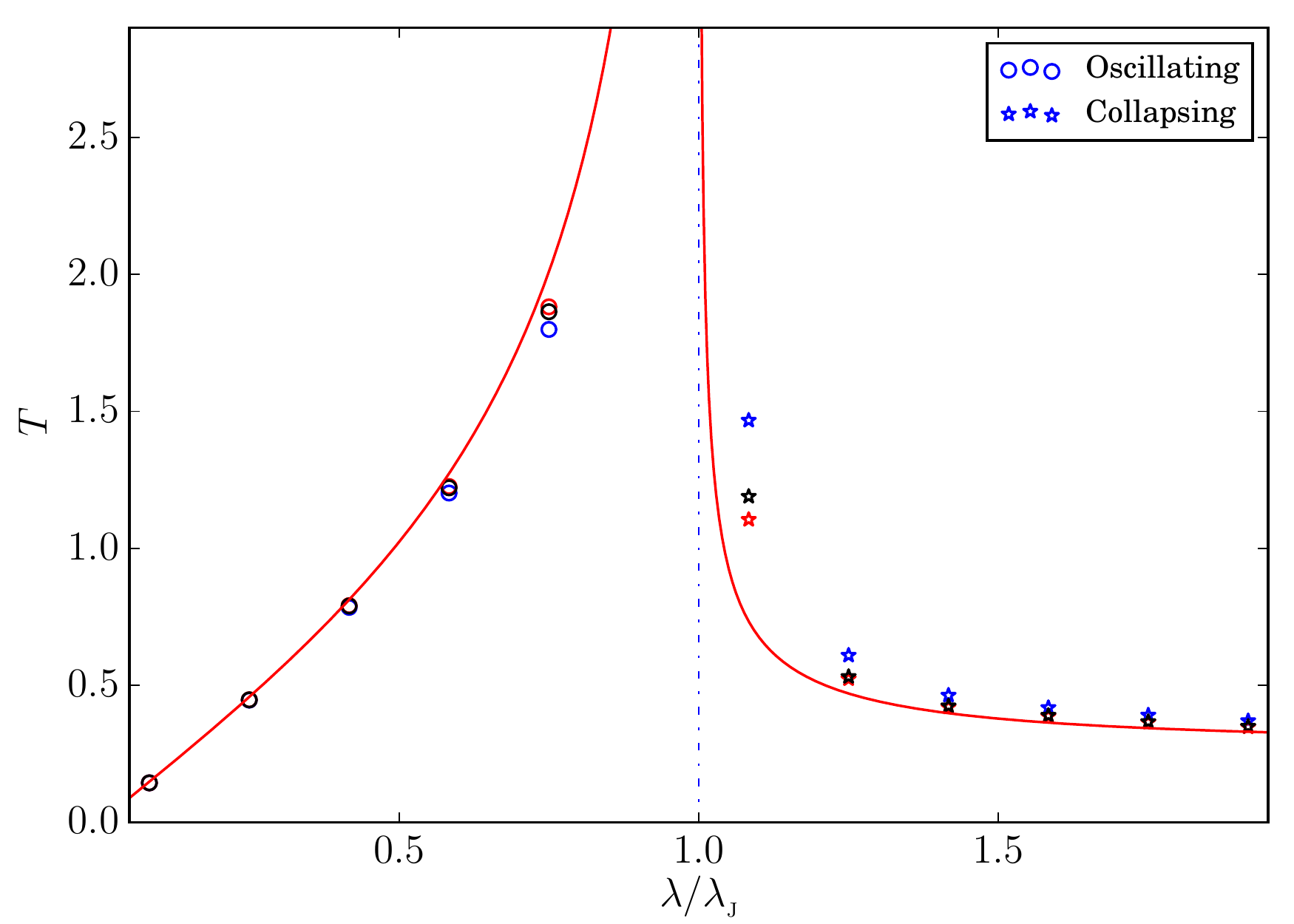}
\caption{(a) The x-component of the gravitational acceleration computed at $t = 0$ for the sinusoidal density perturbation used for the Jeans test using periodic corrections with the kd-tree using (i) the Geometric MAC (blue dots), (ii) the GADGET MAC (red open circles) and (iii) Eigenvalue MAC (black crosses). For reference we plot also the analytical solution.  (b) The characteristic timescales for the evolution of sinusoidal perturbations in the Jeans instability test.  For stable wavelengths (i.e. $\lambda/\lambda_{_{\rm JEANS}} < 1$), the sinusoidal perturbation oscillates with a period given by Equation \ref{EQN:JEANSOSC}.  For unstable wavelengths (i.e. $\lambda/\lambda_{_{\rm JEANS}} > 1$), the perturbations grow with a timescale given by Equation \ref{EQN:JEANSCOL}.  The analytical solutions (Equations \ref{EQN:JEANSOSC} \& \ref{EQN:JEANSCOL}) are plotted in red with the blue-dashed line marking the asymptote where $\lambda = \lambda_{_{\rm JEANS}}$ (and where the oscillation/growth timescales tend to infinity). Oscillating simulations (open circles) and collapsing simulations are plotted using the (i) Geometric MAC (blue), (ii) GADGET MAC (red) and (iii) Eigenvalue MAC (black) criteria for walking the kd-tree.}
\label{FIG:JEANSINSTABILITY}
\end{figure}

\subsection{Jeans instability test}
The Jeans instability test \citep{Hubber2006} is one of the few problems with periodic gravity with known solutions and can be used to validate the Ewald periodic gravity component of the code.  This test sets up a simple sinusoidal density perturbation in an otherwise uniform medium and then monitors the evolution of the density and the velocity field compared to that predicted by the simple Jeans theory \citep[e.g.][]{GalacticDynamics}.

The initial conditions are set-up following \citet{Hubber2006}.  The density field is set-up in a similar fashion to the 1D soundwave test (Equation \ref{EQN:SOUNDWAVE-RHO}), where the particles positions are adjusted to create the required density field (as opposed to altering the particle's masses).  The initial velocity for all particles is zero.  These initial conditions lead to solutions which are standing waves rather than traveling waves as in the classical Jeans solution.  The time-dependent solution is given in \citet{Hubber2006}.  For stable ($\lambda \ll \lambda_{\rm J}$) wavelengths, the perturbations oscillate as sound waves.  The oscillation period is 
\begin{equation} \label{EQN:JEANSOSC}
T_{_{\rm OSC}} = \left( \frac{\pi}{G\,\rho_0} \right)^{1/2} \frac{\lambda}{ \left( \lambda_{_{\rm J}}^2 - \lambda^2 \right)^{1/2}}\,.
\end{equation}
For unstable ($\lambda \gg \lambda_{\rm J}$) wavelengths, the perturbation growth timescale (defined as the time for the perturbation to grow from an amplitude of $A$ to $A\,\cosh{\{1\}} \sim 1.56 A$ is 
\begin{equation} \label{EQN:JEANSCOL}
T_{_{\rm COL}} = \left( \frac{1}{4\,\pi\,G\,\rho_0} \right)^{1/2} \frac{\lambda}{ \left( \lambda^2 - \lambda_{_{\rm J}}^2 \right)^{1/2} }\,.
\end{equation}

Rather than fix the Jeans length and alter the perturbation wavelength, we fix the perturbation wavelength (so the IC setup is always the same) and instead alter the Jeans length via changing the sound speed of the gas.  We perform the simulations only for MFM.

We find that this problem is a stringent test of the tree opening criterion, since the contributions to the gravitational accelerations largely cancel out and sum up to exactly zero for no perturbation. We plot in figure \ref{FIG:JEANSINSTABILITY}(a) the gravitational acceleration computed with different opening criteria.  While the GADGET MAC and the eigenvalue MAC perform quite well in comparison with the analytical solution, the geometric MAC criterion produces a very inaccurate and noisy acceleration. This is not surprising since the criterion does not try to enforce a given error on the acceleration as instead the other two do, which leads to more cells being opened if the acceleration is small.  In Figure \ref{FIG:JEANSINSTABILITY}(b), we plot the oscillation and collapse timescales for various ratios of the perturbation to Jeans wavelength, $\lambda/\lambda_{_{\rm J}}$.  We can see that both evolutionary modes of the perturbation (oscillation and collapse) are correctly realised, i.e. oscillation only for $\lambda < \lambda_{_{\rm J}}$ and collapse only for $\lambda > \lambda_{_{\rm J}}$, similar to the results of \citet{Hubber2006} for so-called `Vanilla' SPH.   As in the previous case we see that the geometric MAC has a worse agreement with the analytic solution. For the other two criteria, the oscillation period and the collapse timescale are extremely well matched by the simulations to the theory although all simulations to some degree underestimate the oscillation timescale and overestimate the collapse timescale due to smoothing and resolution effects.

\subsection{Time Integration accuracy} \label{SS:INTEGRATION}
In this section we investigate how well the different N-body time integration schemes available in \gandalf{}  conserve energy, which we take as a metric of global accuracy. These tests are in an indirect way a test also of the hydrodynamics schemes since they all employ a variant of the leapfrog integrator. We will highlight in particular how in N-body dynamics integrators of order higher than the leapfrog are necessary to guarantee good energy conservation.

\subsubsection{Binary orbits} \label{SSS:BINARYORBIT}

\begin{figure}
\includegraphics[width=\columnwidth]{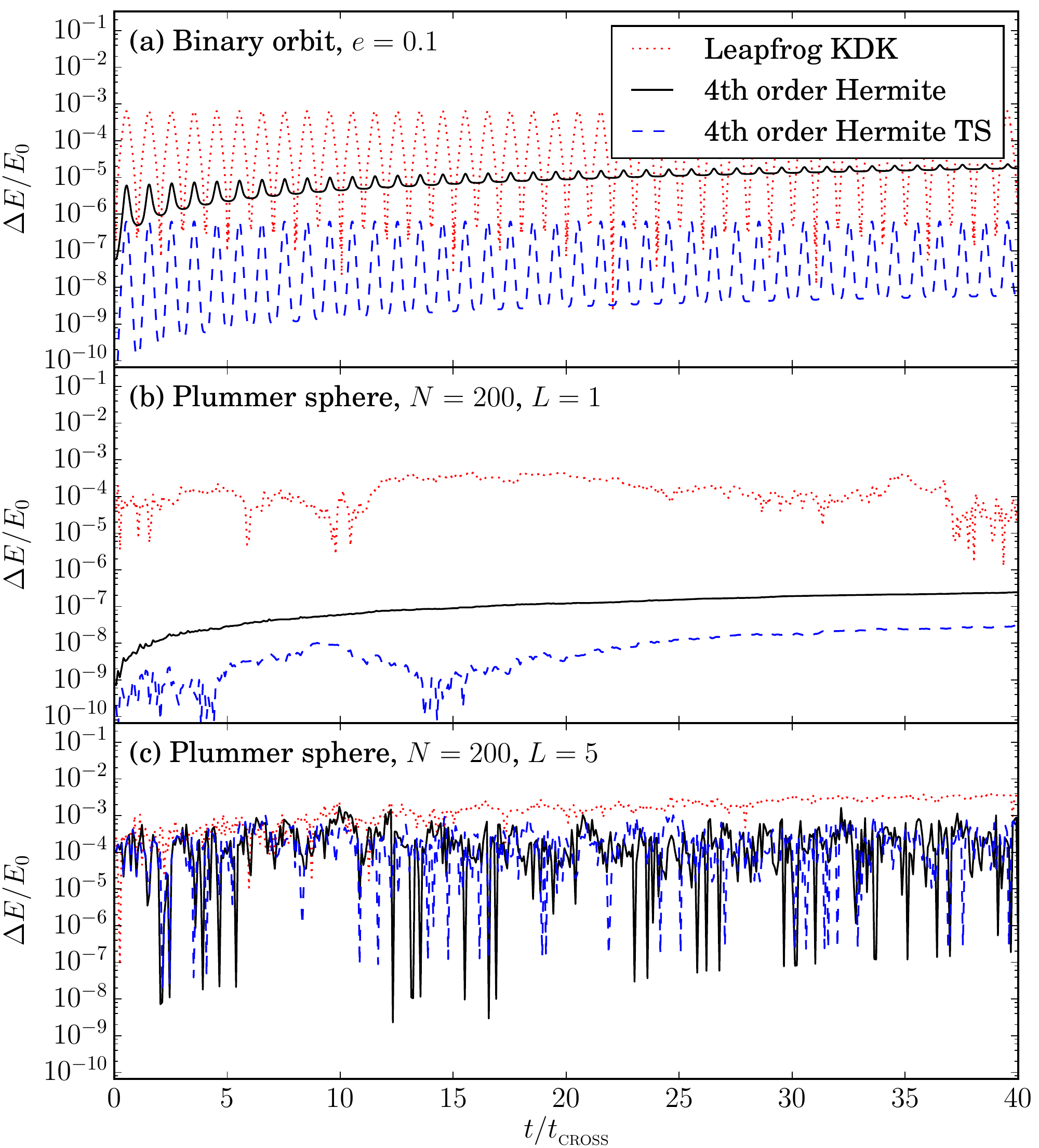}
\caption{The total cumulative fractional energy error for N-body simulations integrating (a) the orbit of an equal mass binary system with a low eccentricity ($e = 0.1$), (b) the evolution of a Plummer sphere containing $N = 200$ stars with global time-steps, and (c) the Plummer sphere using block time-steps with $5$ time-step levels.  For all cases, we perform the integrations using the Leapfrog kick-drift-kick (red dotted line), the standard 4th-order Hermite (solid black line) and time-symmetric 4th-order Hermite (dashed blue line) schemes.}
\label{FIG:INTEGRATIONERROR}
\end{figure}

A binary star with two masses in a bound orbit is the simplest known N-body test problem with an analytical solution and is useful in demonstrating the fidelity of N-body integration schemes.  We have simulated a mildly eccentric ($e = 0.1$) binary orbit for $40$ orbits to highlight the differences in the various schemes.  In Figure \ref{FIG:INTEGRATIONERROR}(a),
we plot the energy error as a function of time for three integration schemes, the Leapfrog-KDK (red dotted line), the standard 4th-order Hermite (solid black line) and the time-symmetric 4th-order Hermite (dashed blue line).  There are two trends to highlight; an oscillation in the energy error (with the same period as the binary orbit) and a long-term error growth.  The two symplectic schemes are characterised by strong oscillations in the energy error which span 3 to 4 orders of magnitude, however they do not show a long-term growth in the error. This is expected since these schemes are time reversible. In contrast, the standard 4th-order Hermite scheme shows a much smaller error oscillation but also a slow long-term increase in the energy error.  Initially the energy error is only slightly higher than the time-symmetric Hermite scheme, but it slowly increases towards the regime occupied by the Leapfrog scheme \citep[cf.][Figure 3.21]{GalacticDynamics}.

\subsubsection{Plummer sphere} \label{SSS:PLUMMERERROR}
A Plummer sphere is a popular and simple stellar cluster profile used often in basic N-body cluster simulations and has been modeled extensively in the literature \citep[e.g.][]{Aarseth1974,Aarseth2003,GalacticDynamics}. The mass density profile for a Plummer sphere is 
\begin{equation} \label{EQN:PLUMMERRHO}
\rho(r) = \frac{3\,M}{4\,\pi\,a^3} \left( 1 + \frac{r^2}{a^2} \right)^{-5/2} \,,
\end{equation}
where $M$ is the total mass and $a$ is the Plummer radius. The 1D velocity dispersion of the stars as a function of radius, $\sigma(r)$, is 
\begin{equation} \label{EQN:PLUMMERV}
\sigma^2(r) = \frac{G\,M}{6\,a} \left( 1 + \frac{r^2}{a^2} \right)^{-1/2} \,.
\end{equation}

A detailed explanation of how to generate initial conditions for a Plummer model with stars is given by \citet{Aarseth1974}.  When including gas, we set-up the Plummer spheres similar to that outlined in \citet{Hybrid2013}.  The positions of the particles are selected with the same Monte-Carlo algorithm, but the gas is given a sound speed equal to the local velocity dispersion.  We perform a simulation of a Plummer sphere containing $200$ equal-mass stars with total (dimensionless) mass $M = 1$ and Plummer radius $R = 1$.  We truncate the Plummer sphere at a radius of $R_{_{\rm MAX}} = 10\,R$.  The Plummer sphere is simulated for $40$ crossing times.

The energy errors (Figure \ref{FIG:INTEGRATIONERROR}(b)) shows markedly different traits to the simple binary orbit.  There is no clear oscillatory error although there are some trends for long term error growth.  The Leapfrog scheme (red dotted line) is the most stable scheme in terms of energy growth, although it also has the largest average energy error: about 2 - 3 orders of magnitude larger than the other schemes.  The Hermite scheme (black line) has a clear long-term growth over the full course of the simulation.  The time-symmetric Hermite also has long term error growth, although about an order of magnitude less than the non-symplectic version. The large energy error with the leapfrog shows why it is important to use higher order, time reversible integrators for the N-body dynamics, in contrast to what is done by most contemporary SPH codes.

\begin{figure}
\includegraphics[width=\columnwidth]{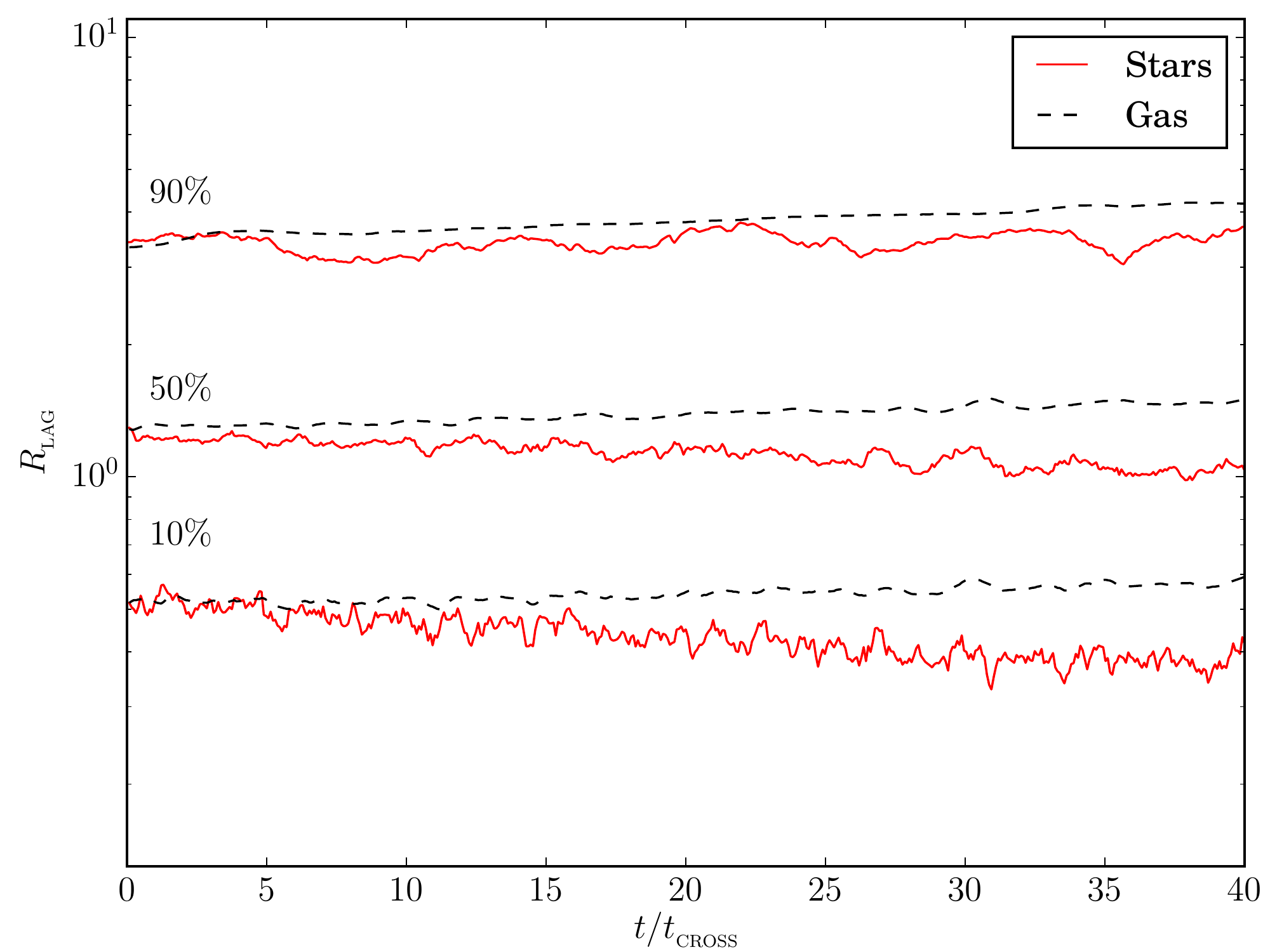}
\caption{Evolution of the $10\%$, $50\%$ and $90\%$ Lagrangian radii in a Plummer sphere containing (a) $N = 500$ equal mass stars and (b) $N = 500$ equal mass stars and $5,000$ SPH gas particles.}
\label{FIG:PLUMMERSPHERE}
\end{figure}

\subsubsection{Plummer sphere with block time-steps} \label{SSS:PLUMMERBLOCKTIMESTEPS}
We simulate the same Plummer sphere as Section \ref{SSS:PLUMMERERROR} using block time-steps (5 time-step levels).  The total global errors for all schemes (Figure \ref{FIG:INTEGRATIONERROR}(c)) are much higher than for the global time-steps simulation. This shows how multiple time-step levels break energy conservation: force calculations are no longer symmetric leading to momentum non-conservation and subsequent energy errors. Overall, the Leapfrog scheme has an energy error starting near $10^{-5}$ growing quickly to $10^{-4}$ and finally almost $10^{-2}$ by the end of the simulation.  The Hermite schemes both tend to have on average a significantly smaller error, of order $10^{-4}$.

\subsection{Hybrid SPH/N-body simulations}
Following \citet{Hybrid2013}, we perform hybrid simulations containing both stars and gas with Plummer profiles.  {\refone The gas is initially set so the local sound speed matches the local velocity dispersion; the initial internal energy is thus $u(r) = \sigma^2(r) / (\gamma - 1)$ and subsequently} evolves according to an adiabatic equation of state.  Differently from \citet{Hybrid2013}, as explained in section \ref{S:HYBRID} in \gandalf{} we take a different symmetrization of particle-particle interactions. In this section we want to show that we still recover the same behaviour in the evolution of a system comprised of gas and stars.

Figure \ref{FIG:PLUMMERSPHERE} shows the evolution of the $10\%$, $50\%$ and $90\%$ Lagrangian radii for both the stellar and gaseous components separately as a function of time.  We find the same qualitative evolution as in \citet{Hybrid2013}: the stellar components decouple from each other and evolve in separate (and opposite) ways.  The stellar Lagrangian radii all contract, most strongly close to the centre.  The gaseous Lagrangian radii on the other hand expands at all radii, leading to a general expansion.  The reason for this difference is whilst there is still energy exchange in interactions, the energy gained by gas from encounters with stars is converted into heat via shocks leading to a one-way expansion of the gas fed by energy from the stars.  After beginning with identical profiles, the two components of several relaxation times eventually decouple.

\subsection{Dust tests}
The two-fluid dust methods included in \gandalf{} are essentially identical to the methods presented in \citet{Booth2015} and \citet{LorenAguilar2015}. For this reason, we refer the reader to those papers and references therein for details on the performance of the method. Here we include a few simple tests to verify the method.

\subsubsection{\textsc{dustybox}}

\begin{figure}
\includegraphics[width=\columnwidth]{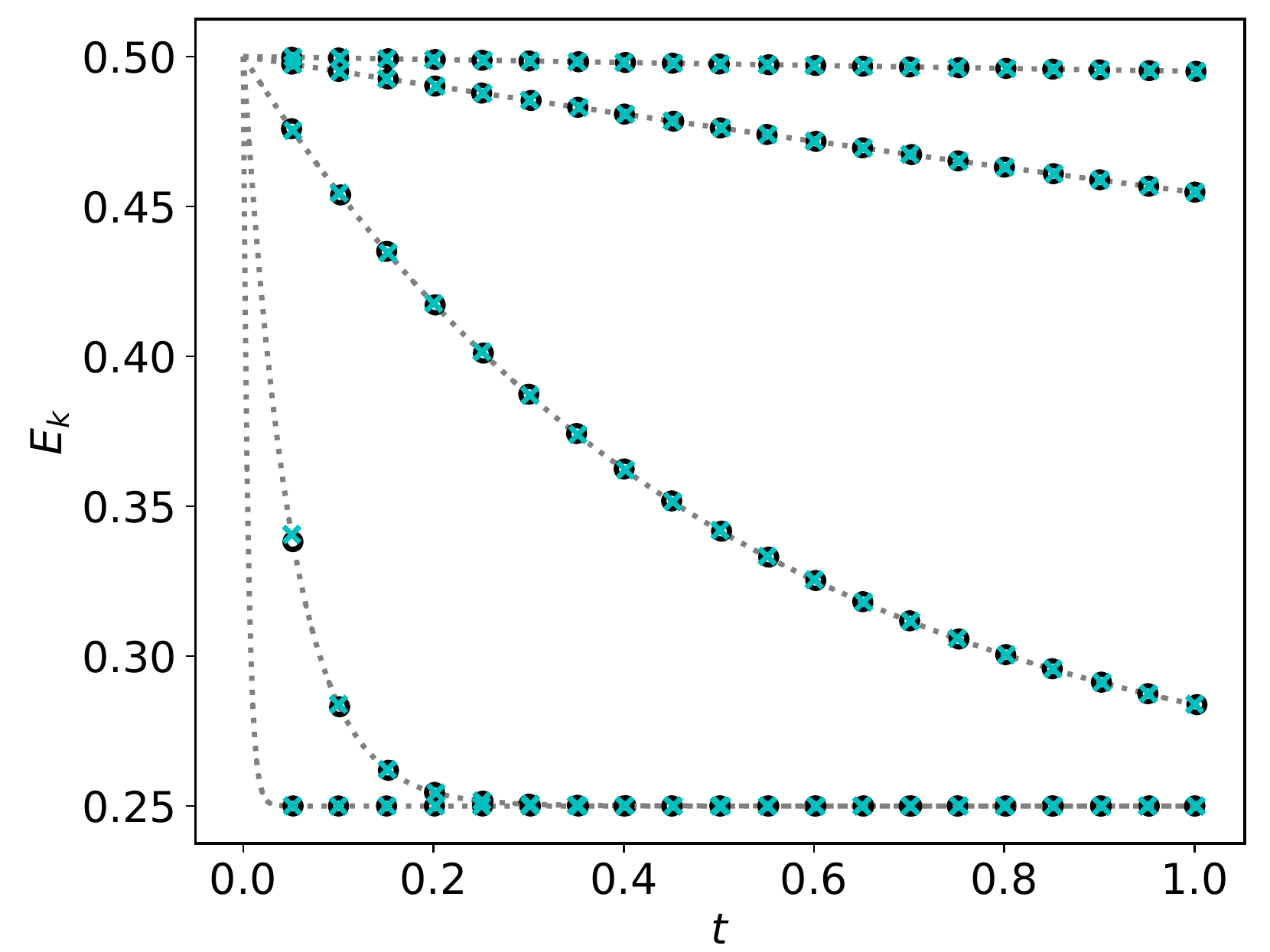}
\caption{Evolution specific kinetic energy in the \textsc{dustybox} test using SPH (circles) and MFM (crosses), with feedback included. {\refone The evolution is shown for stopping times, $t_s$, of 0.01, 0.1, 1, 10 and 100.}}
\label{FIG:DUSTYBOX}
\end{figure}

This test consists of two uniform gas and dust fluids which are set up to initially have a velocity difference. We solve this problem in a 3D periodic box with size $1\times 0.5 \times 0.5$ using $32 \times 16 \times 16$ particles arranged on a cubic lattice. We set the {\refone dust density, gas density, and sound speed to 1}, using a fixed stopping time and taking the initial gas velocity to be at rest while the dust is given a velocity of 1. In Fig.~\ref{FIG:DUSTYBOX} we show the evolution of the kinetic energy for different stopping times computed with the full-scheme including feedback. Both methods produce accurate solutions for all stopping times. We ran this test using an adiabatic equation of state to track the conservation of total energy: in the MFM method the energy is conserved to machine precision, while SPH conserves energy up to time-integration errors ($\sim 10^{-9}$).

\subsubsection{\textsc{dustywave}}

\begin{figure*}
\includegraphics[width=0.8\textwidth]{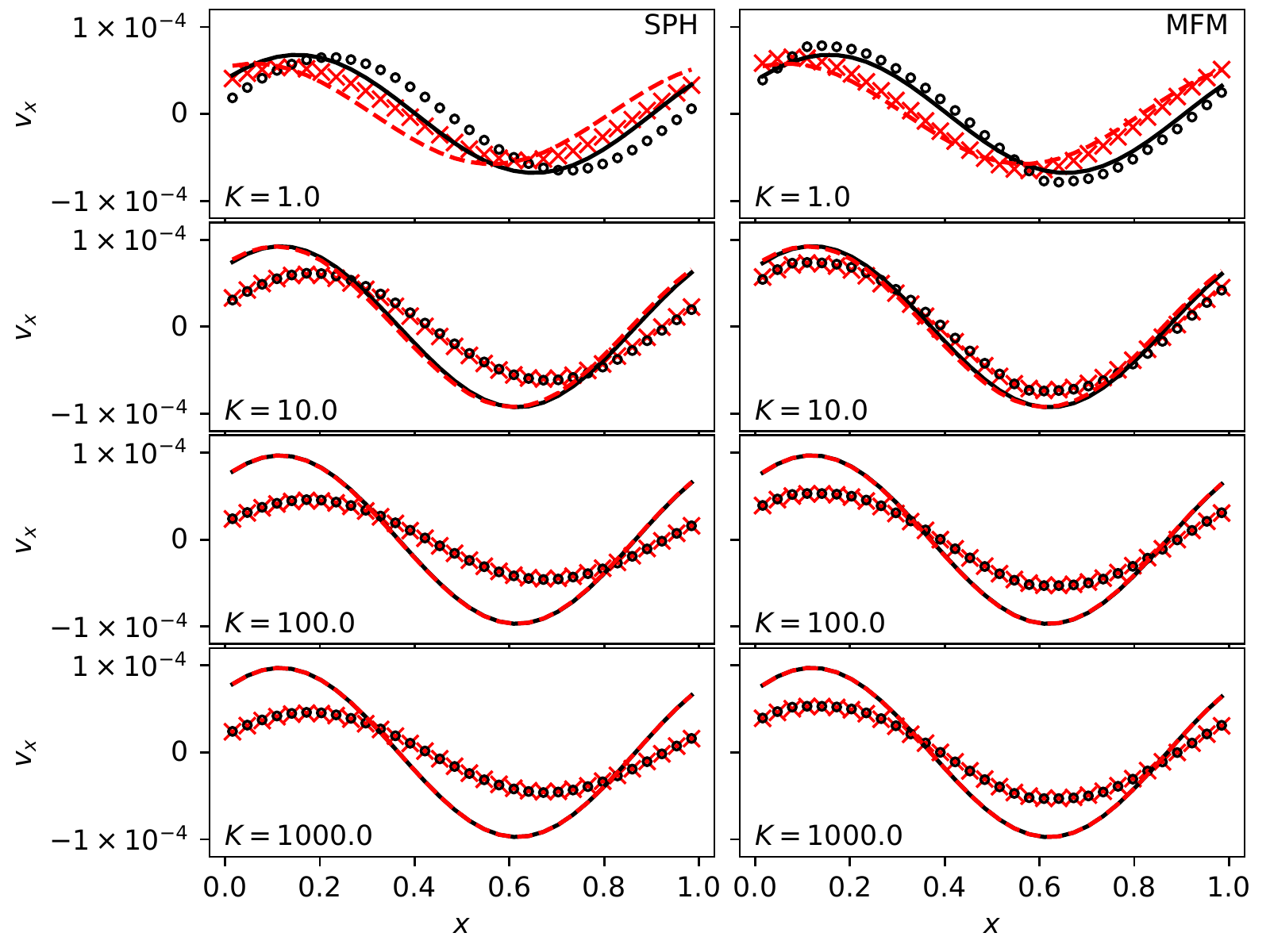}
\caption{Results of the \textsc{dustywave} test at $t=3.0$. Lines show the analytical solution while points show the particle values. Gas particles are shown by circles and the dust particles are shown by crosses.}
\label{FIG:DUSTYWAVE}
\end{figure*}

This is the \textsc{dustywave} test of \citet{Laibe2011}, which involves the evolution of two linear sound waves in a dusty fluid. We solve this problem in 1D using 32 particles per phase and dust-to-gas ratio of 0.1. The gas and dust are both given the same initial velocity, a soundwave with initial velocity of $10^{-4}$. The gas is isothermal with sound speed, $c_s = 1$, and the simulations are evolved for three sound crossing times. The results for models in which the feedback is included are shown in Fig.~\ref{FIG:DUSTYWAVE} for both SPH and MFM, with different values of the drag coefficient, $K$, as defined by \citet{Laibe2011}.

Both methods produce similar results even at this low resolution, but the MFM method reproduces the combined sound-speed more closely, which is partly due to the smaller smoothing length ($\eta_{\rm MFV} = 1$, $\eta_{\rm SPH} = 1.2$). Both methods exhibit the well-known over dissipation of the waves when the stopping time is very small $(c_s t_s \ll h$, \citealt{Laibe2012,LorenAguilar2015}). Here the MFM method shows marginally lower dissipation, which is again mostly due to higher effective resolution. When run with feedback turned off, both the SPH and MFM implementations show essentially no dissipation, which is expected as the gas velocity is not damped \citep[see e.g.][]{Booth2015}

\subsubsection{Shocks in 2D}
\begin{figure*}
\includegraphics[width=0.8\textwidth]{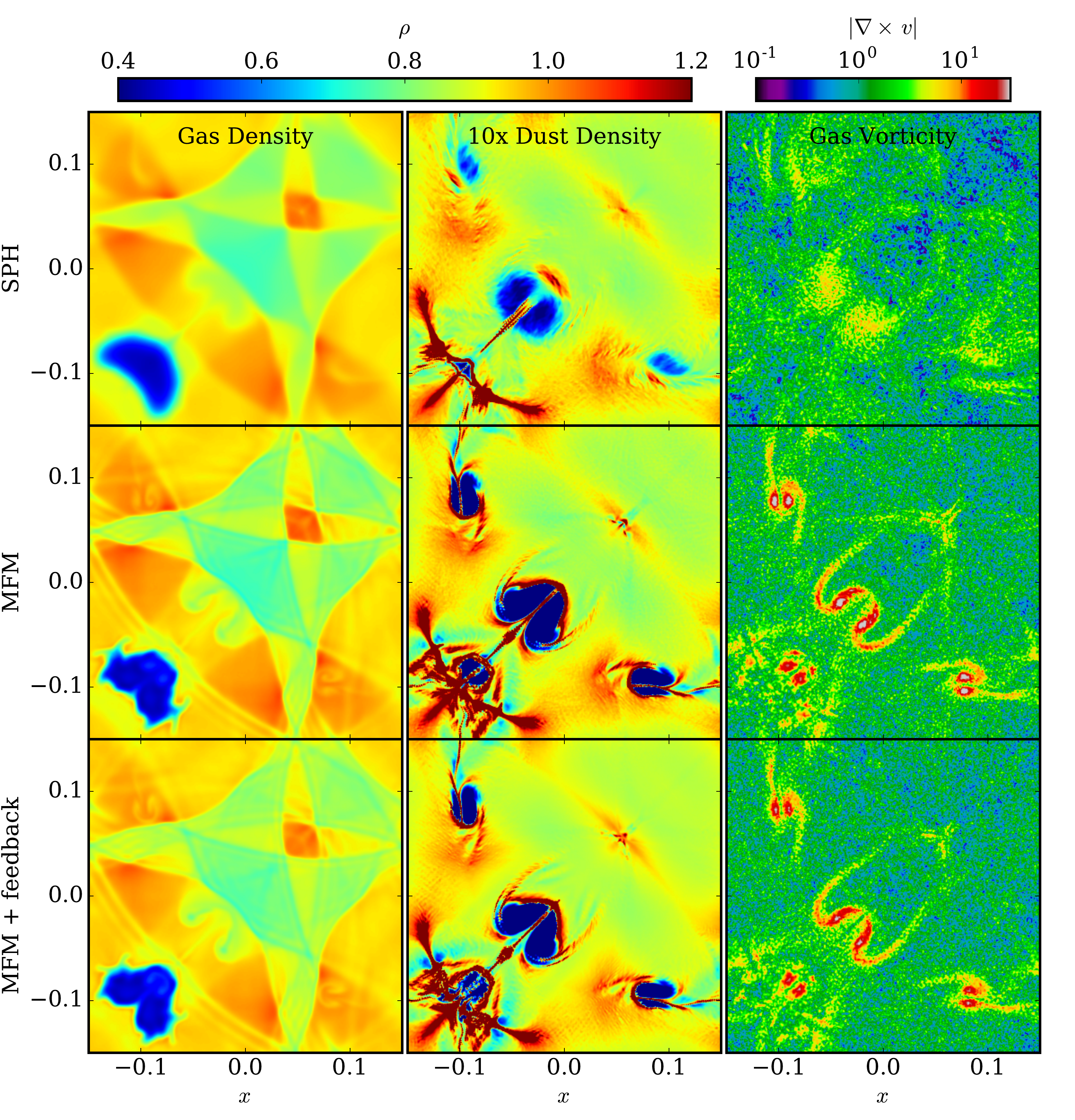}
\caption{The two dimensional shock-tube test including dust for both the SPH and the MFM schemes. The top two rows show results computed with the test particle dust implementation, whereas the bottom row shows the results with feedback included.} 
\label{FIG:SHOCK2D}
\end{figure*}

Here we present the 2D shock problem including dust as set up in \citet{Booth2015}, except a dust-to-gas ratio of 0.1 is used. We show this test in both SPH and the MFM method using the test particle dust implementation, and the full two-fluid scheme with feedback in the MFM scheme.

This test is sensitive to level of noise in the gas velocity distribution, which can hide the underlying gas vorticity field and introduce noise into the density fields of both the gas and dust \citep{Sijacki2012,Booth2015}.  Given the much better performance of the quintic spline kernel in the \citet{GC1990} test, we also employ it here. In SPH the \citet{2010MNRAS.408..669C} switch and \citet{Price2008} artificial conduction are used, while in the MFM the \citet{GIZMO} limiter is employed with the HLLC Riemann solver. 

Fig.~\ref{FIG:SHOCK2D} shows the resulting density and vorticity distributions. The overall features of both SPH and the MFM agree well here, largely due to the improvement of the SPH results that comes from using the quintic kernel. However, the SPH density and vorticity fields are considerably more smoothed than the MFM results. SPH still shows a small level of noise in the dust density. This density noise is nearly absent in the MFM results, which show close agreement with grid based methods \citep[e.g.][]{Sijacki2012,Booth2015}. 

The MFM simulation with feedback included shows very similar results, demonstrating that the dust particles are not introducing noise into the gas dynamics in this problem. The only significant difference between the test particle and full two-fluid results is that with feedback switched on the peak vorticity is reduced, which is likely due to the physical damping by the feedback.


\subsection{Spreading-ring} \label{SS:spread}

The spreading ring test is a standard test \citep{Flebbe94,Artymowicz1994,Murray1996,Kley99} in accretion disc theory to measure the shearing viscosity (either numerical or physical) of a numerical method. The MFM scheme should have a much lower numerical viscosity than SPH and we wish to quantify this effect. We follow \citet{Murray1996} to initialise a ring of particles with a Gaussian density profile $\Sigma \propto \exp (-(r-r_\mathrm{centre})/w))$, where $w$ is the width of the ring and $r_\mathrm{centre}$ its position; the two parameters take the value of 0.033 and 1 respectively. We place the particles in a number of rings (equally spaced by a distance $\Delta r$), with a constant inter-particle separation in the azimuthal coordinate $\Delta \phi$; the number of rings is set such that $r \Delta \phi \simeq \Delta r$. Therefore, to generate the desired density profile we employ particles with different mass. To keep the test as clean as possible, we run it in two dimensions.

Previous works \citep[e.g.,][]{Murray1996} have switched off pressure forces to test only the effect of the artificial viscosity term in SPH. This is not possible to do with the meshless schemes since they do not employ artificial viscosity. Therefore, we run the test with pressure forces. The downside is that pressure forces will contribute to the spreading of the ring. To counteract this problem, we modify the rotation curve of the particles so that the pressure forces are in equilibrium with the gravitational and centrifugal acceleration, preventing spreading due to pressure forces. In addition, we explore different temperatures of the disc (we use a isothermal equation of state), sampling both a cold disc ($c_s=10^{-3}$) where the pressure is too little to cause spreading and a hot one ($c_s=0.05$) where it is potentially a significant contribution. Finally, differently from \citet{Murray1996}, the particles initially have a vanishing radial velocity, since we do not know a priori the magnitude of viscosity in the meshless schemes.

\begin{table}
\begin{tabular}{ccc}
\hline
 & Cold & Hot \\
\hline
SPH & $2 \times 10^{-6}$ & $1.5 \times 10^{-5}$ \\
MFM & $7.7 \times 10^{-9}$ & $8.4 \times 10^{-8}$ \\
\hline
\end{tabular}
\caption{Values of the kinematical viscosity $\nu$ derived from fitting the evolution of the spreading ring after $t=10$. Notice that the values for the meshless should be considered as upper limits rather than measurements.}
\label{table:spread}
\end{table}

\begin{figure}
\includegraphics[width=\columnwidth]{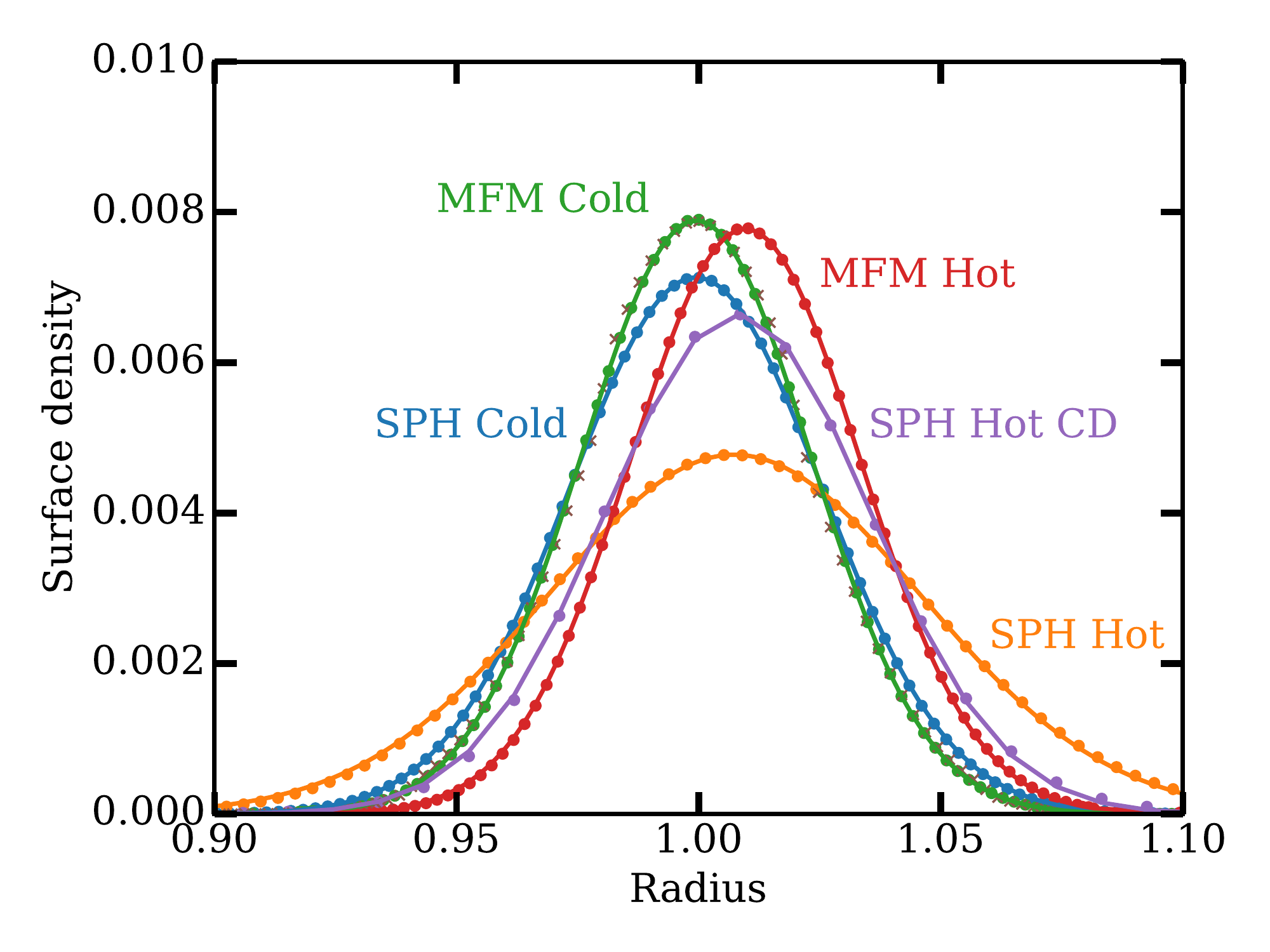}
\caption{Evolution of the density of a spreading ring for the cases explained in the text at time $t=10$. The initial conditions are plotted with the red crosses and they are nearly indistinguishable from the MFM cold case. We show the best fit with the numerical solution with a solid line and the results from the simulation (averaged for each ring of particles) with dots.}
\label{FIG:spread}
\end{figure}

Figure \ref{FIG:spread} shows the evolution of the density. The calculations have employed a resolution of 250000 particles. To measure the value of the kinematic viscosity $\nu$, we {\refone perform a least squares fit. To define the squared residuals, we compare the average density in each ring of particles after a dimensionless time of 10 to the analytical solution (see e.g. eq. 30 in \citealt{Murray1996}). As common in differential equation theory, the analytical solution is a convolution between the kernel of the equation and the initial conditions; to the best of our knowledge the convolution cannot be expressed in closed form and therefore we compute the integral numerically.} Table \ref{table:spread} shows the results of the fit. The difference between SPH and the meshless is already clear by eye. The fact that the meshless has very little viscosity in the cold case is perhaps not surprising; since the pressure forces are weak, the code in this case is effectively a N-body integrator. We can see that instead even in this case the artificial viscosity in SPH ({\refone here used without any switch}), due to the shear, has a significant effect on the evolution of the ring, leading to a relatively high value of $\nu$. In this case, because of the low sound speed the quadratic $\beta$ term dominates the artificial viscosity; setting $\beta=0$ yields a $\nu$ of $6 \times 10^{-7}$, a factor of 3 smaller but still significantly higher than the meshless. {\refone Given that $\beta$ dominates, an artificial viscosity switch would not change the resulting shear viscosity as the switches only operate on $\alpha$}. {\refone The value obtained by our implementation is consistent with the shear viscosity expected from SPH in an accretion disc. According to \citet{Artymowicz1994}, in 2D the shear viscosity expected is $\nu = \frac{1}{8} \alpha_\mathrm{SPH} c_s h$. Substituting the value of $\alpha_\mathrm{SPH}$=1, $c_s=10^{-3}$ and $h=2.7 \times 10^{-3}$, we obtain a value of $3.3 \times 10^{-7}$, which is within a factor of 2 from what we measure.}

Additionally, we have used this test to verify the physical viscosity implementation in the meshless. Including a fixed shear viscosity, $\nu = 2 \times 10^{-6}$, we find that the spreading is consistent to within 5 per cent. This confirms that physical viscosity implementation is working as intended, and that the spreading ring test is good measure of the effective viscosity.

In the hot case, the meshless still performs very well; even in this case the ring remains almost indistinguishable from the initial one\footnote{We have checked in this case that removing the contribution of the pressure forces to the rotation curve leads to a much bigger spread of the ring. Note that in the hot case the centre of the ring moves slightly further out, but we ignore this effect in the analysis since it affects both SPH and the meshless.}. Notice that the value we report for the meshless is effectively an upper limit rather than a measurement; our numerical solution deteriorates for lower value of $\nu$. {\refone For SPH in this case we get a value of $1.5 \times 10^{-5}$. As in the previous case, this compares well to the value expected from the equations in \citet{Artymowicz1994} of $1.7 \times 10^{-5}$}. In this case the dominant term in the SPH artificial viscosity is the linear $\alpha$ term; setting $\beta=0$ leads only to a 10\% reduction of $\nu$. For this reason, it is worth investigating whether a modern viscosity switch can help reducing the numerical viscosity. We have run this test with both the \citet{1997JCoPh.136...41M} switch and the \citet{2010MNRAS.408..669C} one. We find that for this particular test they perform very similarly, {\refone with a small advantage for the latter; they yield a kinematic viscosity of $4 \times 10^{-6}$ and of $3 \times 10^{-6}$, respectively. This is an improvement of a factor of 4-5, clearly visible in the figure (we plot only the \citealt{2010MNRAS.408..669C} case for simplicity)}. We note that this comes though at the cost of increased noise in the particle distribution; when running with either of the two switches, the particles very quickly lose the initial ring structure and rearrange in a more continuous (but noisier) structure. Even when using a viscosity switch in SPH, we conclude that the meshless has a significantly lower numerical viscosity than SPH.

\subsection{Disc-planet interaction}

Having established in the previous section in an idealised test that the meshless scheme has a lower numerical viscosity than SPH, we now wish to assess how the scheme performs in a more realistic simulation. For this goal we have decided to run a simulation of a proto-planetary disc with a planet embedded; the setup is loosely based on \citet{deValBorro}. We have run the simulation both with SPH and the meshless in 3D employing a resolution of 500k particles. Random placement of particles is used to create the initial conditions. The initial surface density scales with radius as $\Sigma \propto r^{-1}$, extending from a radius of 0.4 to a radius of 2.5, while the sound speed scales as $c_s \propto r^{-0.5}$ and the aspect ratio of the disc at the inner boundary is 0.05. We insert a planet with a mass ratio of $10^{-3}$ with respect to the star (i.e., a Jupiter mass for a solar mass star) in a circular orbit with a semi-major axis of 1 and evolve the simulation for 40 orbits. While in SPH we consider only artificial viscosity, in the meshless we add a physical viscosity with $\nu=2 \times 10^{-5}$. Without physical viscosity, a vortex develops outside the orbit of the planet, due to the Rossby Wave Instability arising at the edge of the planetary gap \citep[e.g.,][]{1999ApJ...513..805L,2007A&A...471.1043D}. In SPH instead the much higher numerical viscosity suppresses vortex formation.

\begin{figure*}
\includegraphics[width=\textwidth]{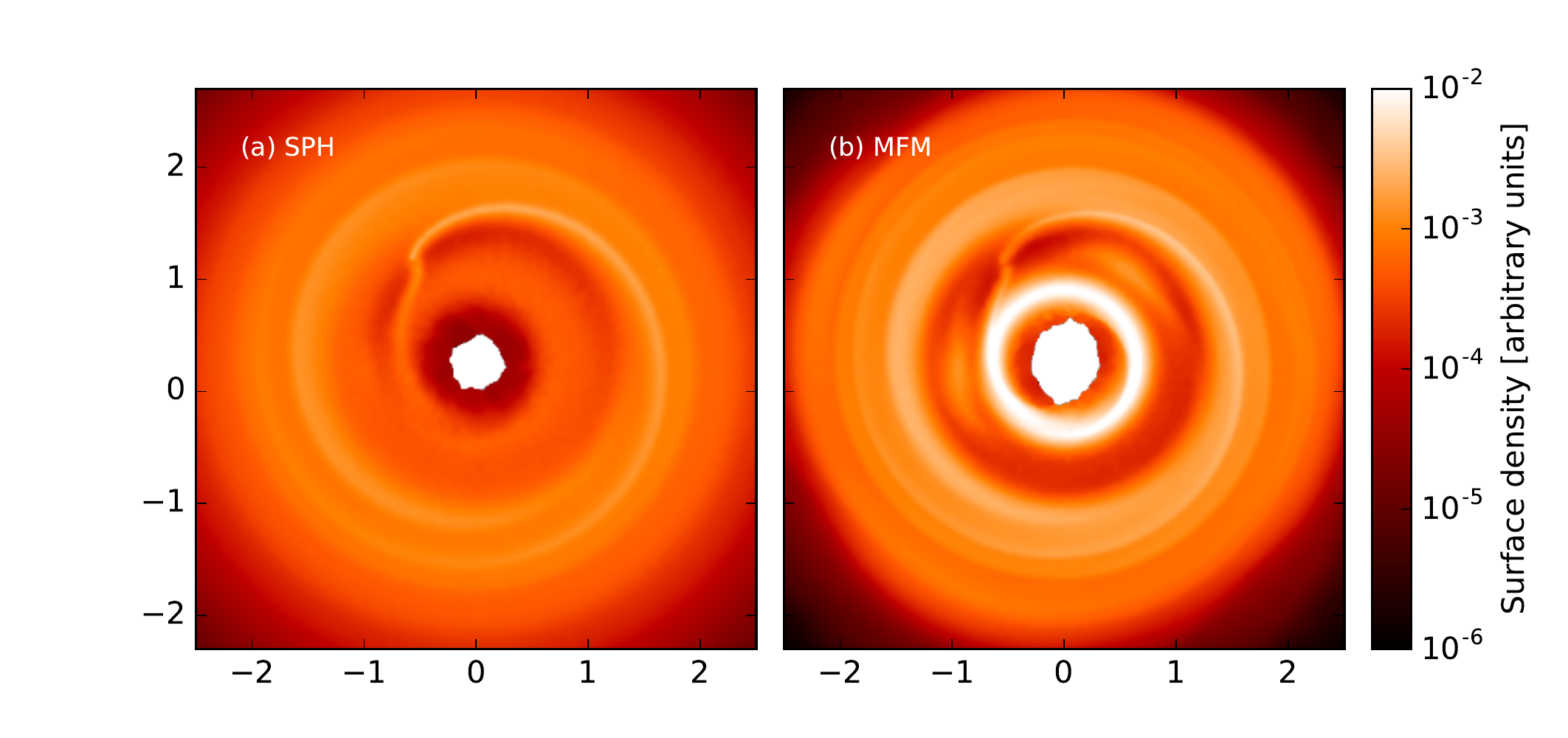}
\caption{A proto-planetary disc with a Jupiter mass planet embedded after 40 orbits. \textbf{Left panel}: SPH. \textbf{Right panel}: MFM method.}
\label{FIG:DISC}
\end{figure*}

Figure \ref{FIG:DISC} shows the surface density of the disc after 40 orbits. It can be seen how in SPH the disc inside the orbit of the planet has a significantly lower mass compared to the meshless case, since the numerical viscosity caused a much higher accretion rate onto the star. Quantitatively, the calculation run with SPH is left with 300k particles at this time, while the one with the  MFV still has 380k particles. The depletion of gas close to the star partially masks the opening of a gap by the planet in the SPH case, which is instead clearly visible in the meshless. In addition, due to the slightly higher effective resolution of the meshless (observed already in the shock tubes, see section \ref{SS:SHOCKTUBES}), the spiral arms created by the planet are much better defined in the meshless case.

\begin{figure*}
\includegraphics[width=\textwidth]{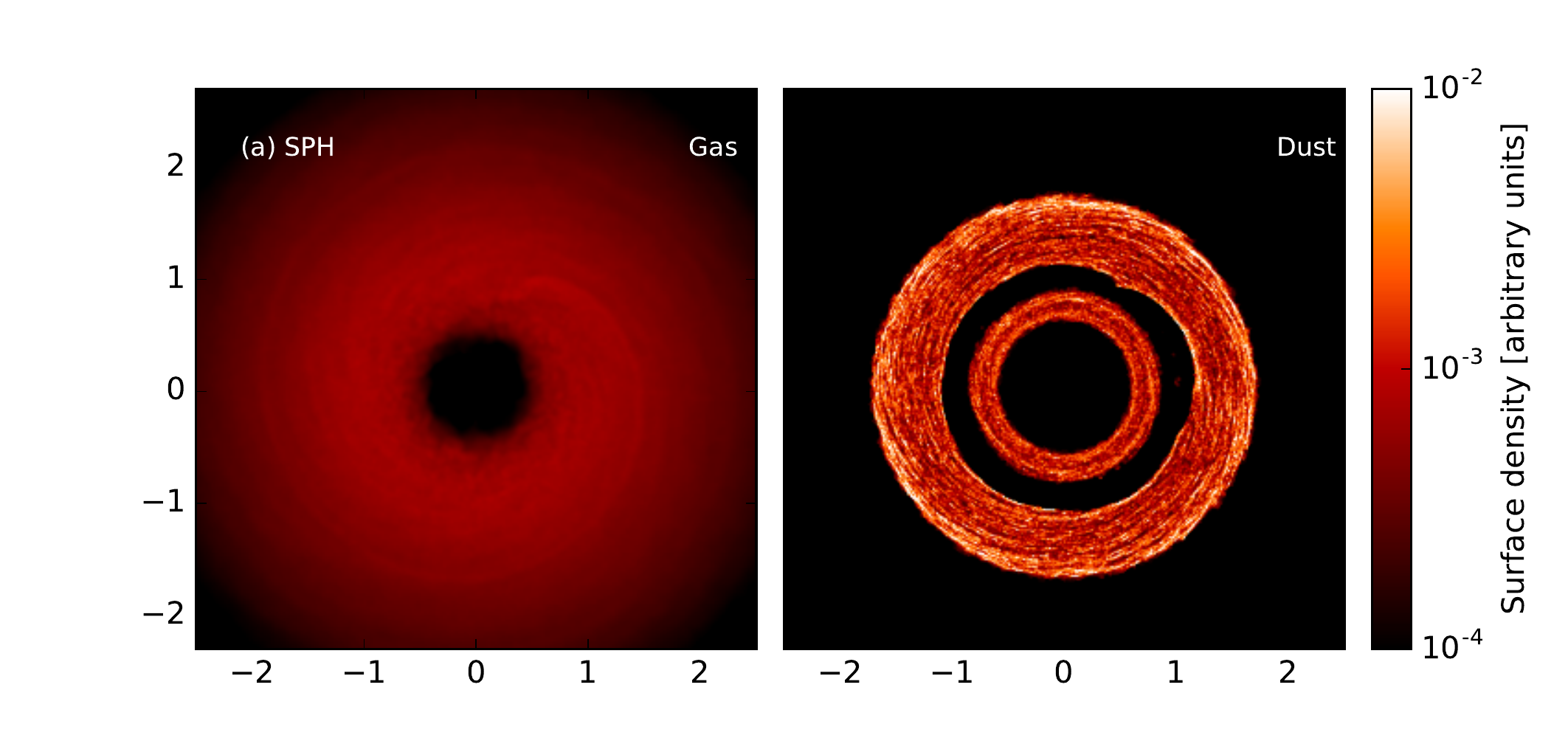}
\caption{A proto-planetary disc with a planet with a mass of $10^{-4}$ with respect to the star. We show results for the gas and the dust distributions. The planet opens up a gap in the dust but not in the gas.}
\label{FIG:DUST}
\end{figure*}

In Figure \ref{FIG:DUST} we show the evolution of a disc containing a planet of a lower mass ($10^{-4}$), a setup similar to \citet{Dipierro}. We now use a shallower surface density $\Sigma \propto r^{-0.1}$ and a sound-speed scaling as $c_s \propto r^{-0.35}$, with an aspect ratio at the inner boundary of 0.075. To reduce the numerical viscosity we set $\alpha_\mathrm{SPH} = 0.1$. We run the simulation both with gas and dust to test our dust implementation. We use 300k particles for the dust, which evolves as test particles. The Stokes number of the dust is 10. We confirm the results of \citet{Dipierro} that such a planet open up a gap in the dust, but not in the gas.

\begin{figure*}
\includegraphics[width=\textwidth]{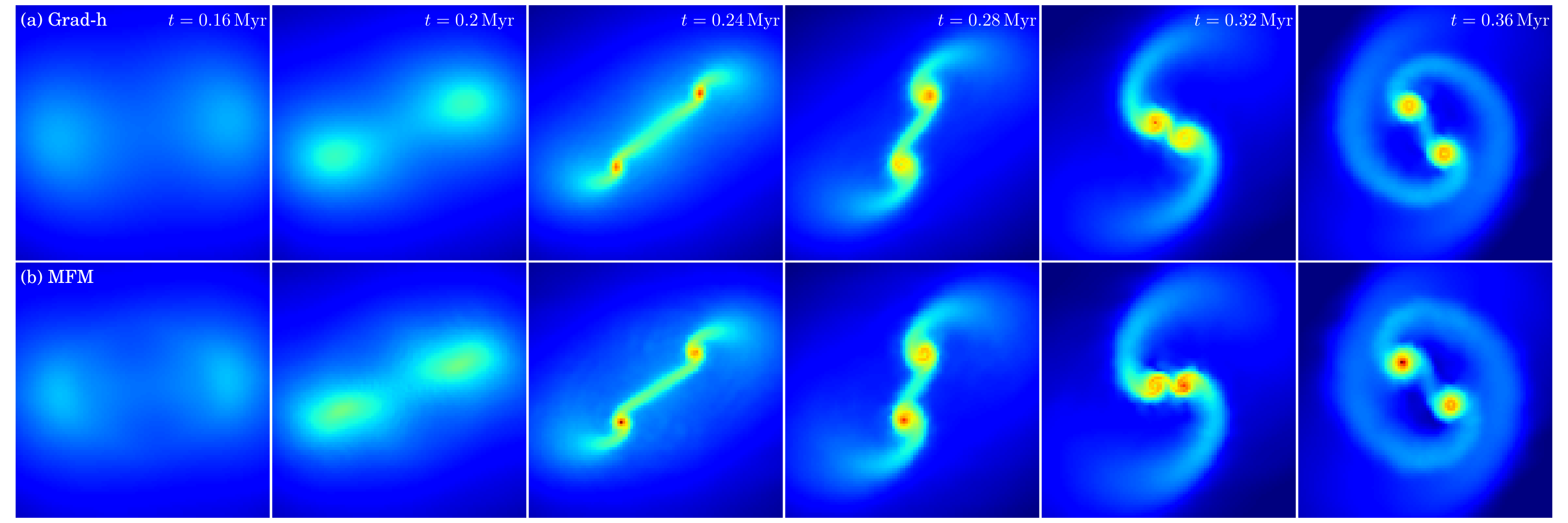}
\caption{Time evolution of the column density profile (low-density : blue; high-density : red) for the Boss-Bodenheimer test using (a) Grad-h SPH (upper row) and (b) MFM (bottom row).  For both methods, the clouds collapse to for a bar-like structure with two denser condensations at either end which collapse to form sink particles.  The two sinks form an accreting binary system with an extended circumbinary disc where mass continually infalls onto the two stars leaving a wake of gas behind each star.}
\label{FIG:BB}
\end{figure*}

\subsection{Boss-Bodenheimer test} \label{SS:BB}

The Boss-Bodenheimer test \citep{BBSIT1979} is a standard test of self-gravitating Astrophysical codes that simulates the collapse and fragmentation of a rotating cloud.  Originally this test was performed with an isothermal EOS.  However, it has also been performed with a barotropic EOS to mimic the optically-thick adiabatic collapse phase during Star Formation.  It provides a simple test-case of combined hydrodynamics with self-gravity in Star Formation and subsequent sink particle formation and evolution.

The initial conditions are set-up similar to that described in \citet{Hubber2011}.  A spherical cloud of total mass $1\,{\rm M}_{\odot}$, radius $0.01\,{\rm pc}$ is created with a density profile
\begin{equation}
\rho = \rho_{_0}\,\left[ 1 + A\,\sin{(m\phi)} \right]\,
\end{equation}
where $\rho_{_0} = 1.44 \times 10^{-17}\,{\rm g}\,{\rm cm}^{-3}$, $A = 0.5$ is the perturbation amplitude, $m = 2$ is the order of the azimuthal perturbation and $\phi$ is the azimuthal angle about the z-axis.  We generate a hexagonal closed-packed array and then cut-out a uniform-density sphere containing the desired number of particles.  The total mass and radius of the sphere is scaled to $1\,{\rm M}_{\odot}$ and $0.01\,{\rm pc}$ respectively.  We finally alter the azimuthal positions of the particles to reproduce the required density field.  {\refone The barotropic equation of state used in this test gives the temperature as a function of density :  
\begin{equation}
T(\rho) = T_0 \left\{ 1 + \left( \frac{\rho}{\rho_{_{\rm AD}}} \right)^{\gamma - 1} \right\}\,,
\end{equation}
where $T_0 = 10\,{\rm K}$, $\rho_{_{\rm AD}} = 10^{-14}\,{\rm g}\,{\rm cm}^{-3}$ and $\gamma = 5/3$.  The gas pressure is given by $P(\rho) = k_{_{\rm B}}\,T(\rho)\rho / (\mu\,m_{_{\rm H}})$ where $k_{_{\rm B}}$ is the Boltzmann constant, $m_{_{\rm H}}$ is the mass of hydrogen and the mean-gas-particle mass, $\mu = 2.35$.
}

We simulate the evolution until a time of $t_{\rm end} = 0.04\,{\rm Myr}$, by which time the cloud should fragments into two stars (or perhaps more) and the binary should have performed several orbits.  The simulations were performed with both SPH and MFM using $32,000$ particles.

\subsubsection{Time evolution} \label{SSS:BB-TIME}
In Figure \ref{FIG:BB}, we show the time evolution of the Boss-Bodenheimer test for both the SPH (top row) and MFM (bottom row) schemes.  The large-scale evolution is the same for both cases as expected with both simulations forming a bar with two density enhancements at either end which gravitationally collapse to form two protostars (i.e. sink particles).  The density enhancements are surrounded by disc-like envelope 

Three noticeable differences between the two simulations are apparent. (i) the evolution of the SPH simulation is slightly slower than the MFM scheme (i.e. it seems to lag slightly behind the MFM scheme) and takes slightly longer for the bar to reach the higher densities where it forms two objects at each end. (ii) Once intermediate densities have been reached and the two ends of the bar have reached some state of centrifugal support, the SPH simulations evolves towards higher densities much more quickly than the MFM simulation.  In fact, the MFM scheme can never reach the sink density if it is too large compared to the adiabatic density.  The main driver of this difference is likely to be the artificial viscosity in the SPH simulations.  The artificial viscosity can efficiently (and artificially) transport angular momentum away from the disc-like object allowing it to collapse to higher densities quicker and hence form sinks rapidly.  As demonstrated in section \ref{SS:spread}, the MFM scheme instead has a much lower effective numerical viscosity, leading to less artificial angular momentum transport.  In this simulation, we lower the sink density enough to allow comparable sink formation times and to allow a meaningful comparison with other features in the simulation.  However this difference highlights that, even though SPH does not artificially cause fragmentation of already unstable regions, other numerical issues can lead to large differences in simulations between SPH and less dissipative methods. This is of particular importance when modeling discs, due to the high shear viscosity of SPH.

Recently \citet{Deng2017} made comparisons between the SPH and MFM by looking at the viscosity-driven angular momentum transport in rotating cores such as the Boss-Bodenheimer test.  We confirm that we obtain similar results to \citet{Deng2017} in that SPH tends to lead to more rapid angular momentum transport than MFM, particularly near the edge of the cloud.

\section{Performance \& Scaling} \label{S:PERFORMANCE}

\subsection{Gravity tree scaling} \label{SS:TREESCALING}

\begin{figure}
\includegraphics[width=\columnwidth]{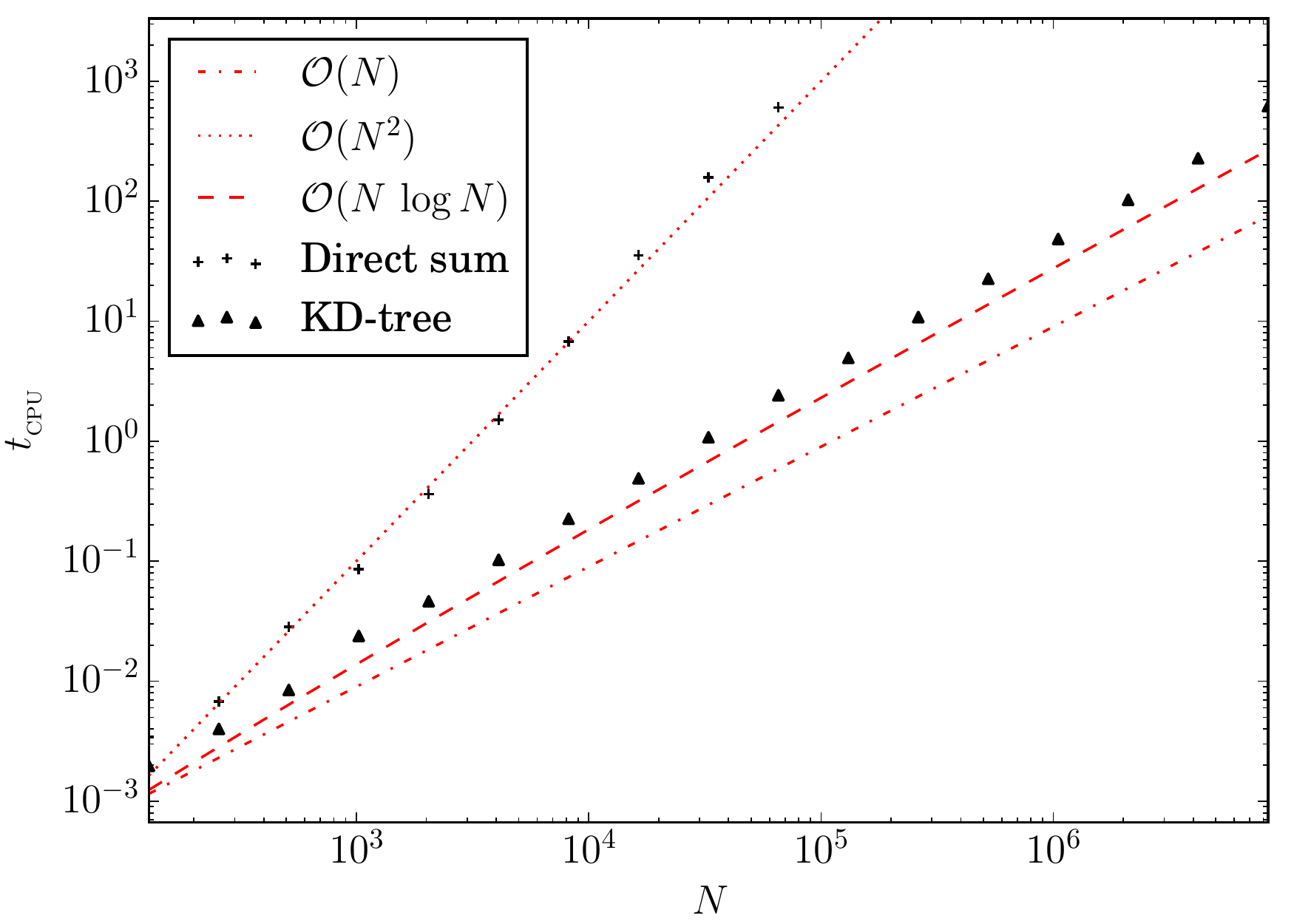}
\caption{Performance and scaling for computing the gravitational acceleration of all particles using the KD-tree in {\small GANDALF} as a function of particle number.  For comparison, we plot lines showing the ${\cal O}(N)$ claimed by \citet{GR2011} (red dot-dashed), ${\cal O}(N^2)$ expected for direct-sum (red dotted) and ${\cal O}(N\log{N})$ expected for tree gravity (red dashed).}
\label{FIG:TREESCALING}
\end{figure}

Gravity trees used in particle codes typically scale as ${\cal O}(N\,\log{N})$, i.e. $N$ particles each requiring an average of ${\cal O}(\log{N})$ computations.  This is mainly because each particle must walk the tree individually, and then compute all contributions to the force from near (i.e. smoothed) neighbours), distant (i.e. non-smoothed) neighbours and distant cells using the COM approximation.  \citet{GR2011} claimed that, if we walk the tree for groups of particles rather than one at a time, we can compute the contributions from the far-cells more efficiently using a multipole expansion around the cell centre and instead approach ${\cal O}(N)$ scaling. As we showed in section \ref{SS:COLDCOLLAPSE}, we do not find a speed benefit in our implementation using the Taylor expansion around the cell centre, implying that our implementation has a different balance of the time spent computing the interaction with near or far particles. Therefore, it is likely that our tree will scale in a different way with the number of particles compared to \citet{GR2011}.

Figure \ref{FIG:TREESCALING} shows the performance of {\small GANDALF} using direct-sum gravity and the tree. We set up a uniform sphere of particles with different numbers of particles and compute the time needed to compute the gravitational acceleration. We plot this CPU time as a function of the particle number. The ${\cal O}(N^2)$ scaling of the direct-sum gravity is evident. Instead, it can be seen that, as hypothesized, our implementation of the tree scales as ${\cal O}(N\log{N})$, and not as ${\cal O}(N)$. We note though that the difference between the two scalings is very small; over the almost 5 orders of magnitude spanned by the plot, the difference in wall clock time is a linear factor of 2--3. It is interesting to note also that \citet{GR2011} comment that their scaling is not perfectly ${\cal O}(N)$, with an extra factor very similar in value to ours. This means that in practical terms the difference in scaling between our implementation and the one presented by \citep{GR2011} is almost negligible.

\subsection{OpenMP parallel scaling} \label{SS:OPENMPSCALING}

\gandalf{} is parallelised using both OpenMP and MPI to allow the code to be used on much larger problem sizes than are achievable on single core machines.  Here we investigate the strong scaling of the OpenMP parallelisation and experiment with the number of particles at the leaf level of the tree to find the most optimal performance. As discussed in detail by \citet{GR2011}, the performance of the KD-tree can be very sensitive to the chosen value of $N_{_{\rm LEAF}}$, the (maximum) number of particles contained in each leaf cell of the tree.  Small values of $N_{_{\rm LEAF}}$ result in more tree-walks being required (since there are fewer leaf cells in the tree) whereas large values of $N_{_{\rm LEAF}}$ can result in much larger neighbour lists being generated for each leaf cell.  \citet{GR2011} empirically determined that the most optimal value of the average number of particles per leaf cell for their tree implementation was $\bar{N}_{_{\rm LEAF}} \sim 12$.  

We use the Boss-Bodenheimer test as a benchmark to test the parallel performance, since it is relatively simple to set-up, has a well-known numerical solution and computes both hydrodynamical and gravitational forces, the two most expensive components of the code.  We run this test with $\sim 10^6$ particles using $N_{_{\rm CORE}} = 1, 2, 4, 8, 16$ and $32$ parallel cores in a shared-memory machine (parallelised with OpenMP) using various values of $N_{_{\rm LEAF}}$ ($1, 4, 8, 16$ and $32$) for 16 steps before terminating the simulation and measuring the time spent in the Main Loop (i.e. ignoring any set-up procedures).  We also run with global time-steps, i.e. one time-step level, in order to demonstrate the best-case scaling for the various parameters.  In Table \ref{TAB::NLEAF}, we show the total CPU wallclock time, $t(N_{_{\rm CORE}})$ for each combination of $N_{_{\rm CORE}}$ and $N_{_{\rm LEAF}}$ and the scaling, $S(N_{_{\rm CORE}}) \equiv t(1)/t(N_{_{\rm CORE}})$.  

We notice some important results from our scaling tests:
\begin{enumerate}
\item For almost all values of $N_{_{\rm CORE}}$, there is a broad minimum in the total CPU wallclock time for the simulation, at $N_{_{\rm LEAF}} = 8$.  This represents our most optimal value and default choice for $N_{_{\rm LEAF}}$ in \gandalf{}. 
\item The scaling of \gandalf{} formally increases with increasing values of $N_{_{\rm LEAF}}$ for all values of $N_{_{\rm CORE}}$ (although we note some fluctuations in the timing routines).  Although this suggests using as high a value of $N_{_{\rm LEAF}}$ as possible, the raw CPU times are a minimum for $N_{_{\rm LEAF}} = 8$ which should be the most important factor.  Although not shown in Table \ref{TAB::NLEAF}, for even larger values of $N_{_{\rm LEAF}}$, achieving good load balancing becomes problematic and the scaling once again drops away.
\end{enumerate}

\begin{table*}
\begin{tabular}{l c  rl rl  rl rl rl}
\hline
\multirow{2}*{$N_{_{\rm LEAF}}$} & \multicolumn{1}{c}{Serial} & \multicolumn{2}{c}{2 cores} & \multicolumn{2}{c}{4 cores} & \multicolumn{2}{c}{8 cores} & \multicolumn{2}{c}{16 cores} & \multicolumn{2}{c}{32 cores} \\
 & Time & Time & Scaling  & Time & Scaling  & Time & Scaling   & Time & Scaling   & Time & Scaling \\
\hline
1	&1388.7	&772.0	&1.80	&405.7	&3.40	&210.6	&6.60	&106	&13.1	&60.2	& 23.1  \\
4	&791.0	&402.8	&1.96	&204.3	&3.88	&104.9	&7.50	&55.2	&14.3	&31.4	& 25.2  \\
8	&732.6	&374.2	&1.96	&192.0	&3.82	&100.0	&7.30	&52.2	&14.0	&28.2	& 26.0  \\
16	&815.0	&416.3	&1.96	&211.4	&3.86	&109.6	&7.50	&56.3	&14.6	&30.7	& 26.6  \\
32	&1066.1	&523.4	&2.04	&271.2	&3.93	&137.6	&7.75	&71.2	&15.0	&37.7	& 28.2 \\
\hline
\end{tabular}
\caption{CPU wallclock times in seconds and parallel scaling for 16 steps of the Boss-Bodenheimer test with $\sim 10^6$ particles using different values of $N_{_{\rm LEAF}}$ ($= 1, 4, 8, 16$ and $32$) using $1, 2, 4, 8, 16$ and $32$ cores in parallel with OpenMP.  For almost all numbers of parallel cores, $N_{_{\rm LEAF}} = 8$ gives the shortest CPU run-times even though the formal parallel scaling for higher values of $N_{_{\rm LEAF}}$ is better.  We note the super-linear scaling of $N_{_{\rm LEAF}} = 32$ with 2 cores is due to fluctuations in CPU performance and in the timing routines.}
\label{TAB::NLEAF}
\end{table*}

\subsection{Hybrid parallel scaling} \label{SS:SCALING}

\begin{figure*}
\includegraphics[width=\columnwidth]{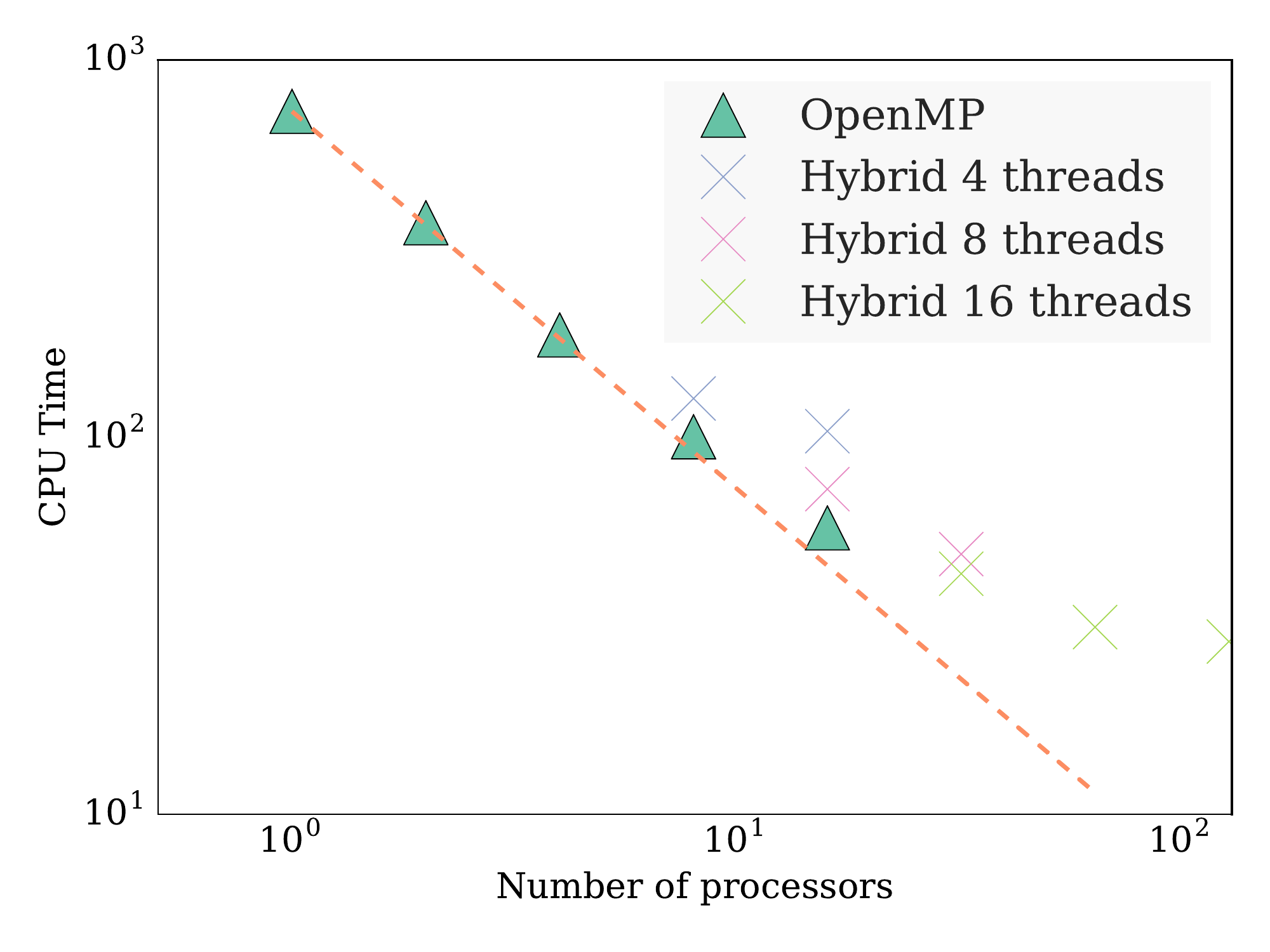}
\includegraphics[width=\columnwidth]{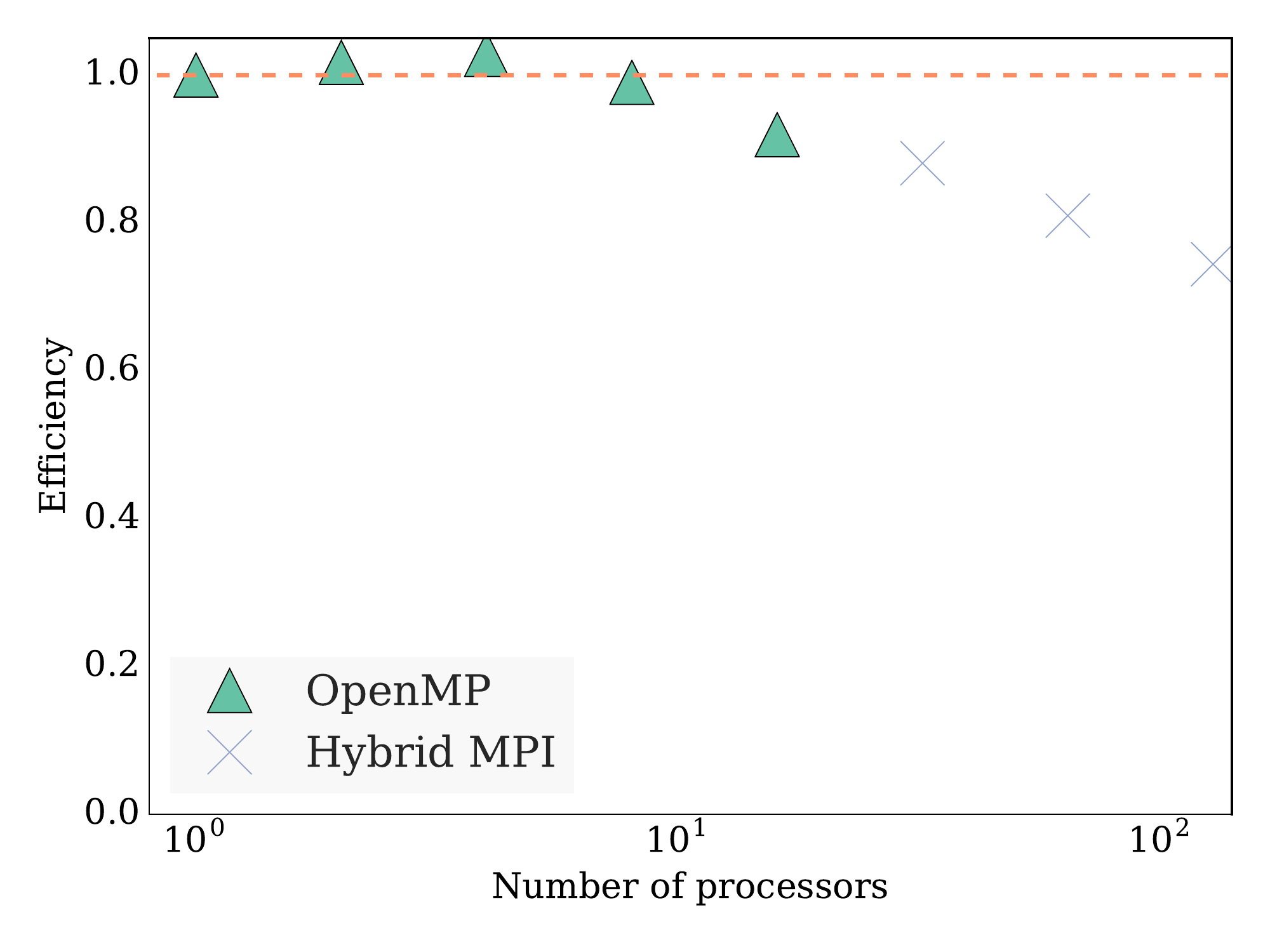}
\caption{\textbf{Left panel}: Strong scaling of \gandalf{} for 16 steps of the Boss-Bodenheimer test with $4 \times 10^6$ particles. The speed-up is relative to the serial version of the code. When using hybrid parallelization, the different colours are for different number of OpenMP threads (as shown in the legend). In all cases we find that the optimal strategy is to use as many OpenMP threads as possible. \textbf{Right panel:} Weak scaling of \gandalf{} for the same test using $2^{18}$ ($\sim$ 250k) particles per processor. The efficiency has been normalised taking into account the $\log N$ scaling of the algorithms.}
\label{fig:scaling}
\end{figure*}

As described in section \ref{S:PARALLELISATION}, \gandalf{} is parallelised both via OpenMP and MPI. The left panel of Figure \ref{fig:scaling} shows the strong scaling of \gandalf{} in pure OpenMP mode and in hybrid MPI-OpenMP mode. We tested the code on the Darwin supercomputer, hosted at the University of Cambridge, using version 12 of the Intel compiler. All the tests have been run for the Boss-Bodenheimer test  as in Section \ref{SS:OPENMPSCALING}. Compared to the previous section we employ here a resolution of $\sim 4 \times 10^6$ particles since we test the code up to 128 processors. Up to 8 threads the speed-up is almost ideal (7.3), and still relatively good with 16 threads (12.7). We note that in both cases most of the time is spent computing the forces (both hydro forces and gravitational forces), with a very good scaling of 14.24 with 16 threads. The bottleneck to the scaling is mostly in the tree building routine and in other serial parts of the code.

With hybrid MPI-OpenMP parallelisation, we experiment using different numbers of OpenMP threads. In general we find that the best performance is achieved by using as many OpenMP threads as possible inside a given node (16 physical cores were available on the supercomputer we used for testing), and using MPI to communicate among the nodes. We interpret this result as a consequence of the fact that the MPI version of the code needs to do more work: pruned trees and ghost particles need to be created and sent to the other processors. This extra work adds to the overhead and limits the parallel scaling. In practice, we find that this particular test problem does not scale well using more than 64 processors, with only minimal improvements on 128 processors.

The real benefit of MPI however is to run simulation at higher resolution than what would be possible otherwise. For this reason we also conduct tests of the weak scaling of \gandalf{} (right panel of Figure \ref{fig:scaling}). The test has been run with a resolution of $2^{18}$ ($\sim$ 250k) particles per processor.  When defining a parallel efficiency, we have taken into account the extra $\log N$ factor demonstrated in section \ref{SS:TREESCALING}. We use OpenMP only up to 16 cores, and switch to hybrid MPI-OpenMP mode using more processors. Based on the previous findings, we employ here 16 OpenMP threads, using MPI only to communicate between the nodes. We can see that the code exhibits very good weak scaling: even with 128 processors, the parallel efficiency is around 70 percent.

\section{Discussion, future development and conclusions} \label{S:FUTUREWORK}

In this paper we have presented the new hydrodynamical code \gandalf{} with details and tests of all implemented algorithms. The code contains the robust and well tested SPH method, as well as the Meshless Finite-Volume numerical schemes presented by \citet{GN2011} and \citet{GIZMO}. In addition \gandalf{} can handle N-body dynamics with higher order collisional integrators than what is commonly employed in SPH simulations and implements an energy conserving scheme for integrating the dynamics of stars and gas.  Both hydrodynamical schemes can also handle dust dynamics, either in the test particle limit or keeping the back reaction of the dust onto gas into account. The object-oriented design of \gandalf{} makes the code flexible, easy to adapt with new physics modules and it is relatively easy to add other particle based schemes.

We have presented an extensive suite of tests to demonstrate the correctness of our implementation, mostly recovering the results of \citet{GIZMO} in terms of the benefits of the MFV schemes compared to SPH. In addition we have conducted a more rigorous test to quantify the numerical viscosity of the method. In the spreading ring test we have shown that the MFM scheme has a much lower numerical viscosity than SPH and is therefore better suited for accretion disc applications, where the numerical viscosity of SPH is typically too high to perform realistic simulations (unless a very high resolution is used). The same conclusion is reached also looking at the evolution of a proto-planetary disc containing a planet, where the inner part of the disc in SPH is rapidly accreted onto the star due to the high numerical viscosity.

The code is publicly available at \href{https://github.com/gandalfcode/gandalf}{this address} under the GPLv2 license. The code is parallelised with OpenMP and MPI for running on modern supercomputers. In addition we provide a python library to facilitate analysis of the results of the simulations and ease code use and development, since the results of a simulation can be inspected live while it is running.

We plan in the future to implement additional algorithms and physics modules in \gandalf{}.  Examples of developments which are underway include algorithms for radiation transport and coupling with existing chemistry codes (e.g. \citealt{KROME}). We encourage users of the code to contact us if there are specific algorithms they are interested in.

We hope that the numerical techniques implemented in \gandalf{}, its ease of use and modularity of design will help future research with this code.

\section*{Acknowledgements}
This research was supported by the DFG cluster of excellence "Origin
and Structure of the Universe", DFG Projects 841797-4, 841798-2 (DAH, GPR),
the DISCSIM project, grant agreement 341137 funded by the European Research Council under
ERC-2013-ADG (GPR, RAB).  Some development of the code and simulations have been carried out on the computing facilities of the Computational centre for Particle and Astrophysics (C2PAP) and on the DiRAC Data Analytic system at the University of Cambridge, operated by the University of Cambridge High Performance Computing Service on behalf of the STFC DiRAC HPC Facility (www.dirac.ac.uk); the equipment was funded by BIS National E-infrastructure capital grant (ST/K001590/1), STFC capital grants ST/H008861/1 and ST/H00887X/1, and STFC DiRAC Operations grant ST/K00333X/1.  We would like to thank the following people for helpful discussions or for contributing code to the public version, including Alexander Arth (discussions on IC generation), Scott Balfour (ionising radiation algorithms), Seamus Clarke (sink particle algorithm refinements and various bug fixes), James Dale (ionising radiation algorithms) Franta Dinnbier (periodic gravity), Stefan Heigl (assisting implementing the MFV schemes), Oliver Lomax (assisting implementing the kd-tree), Judith Ngoumou (assisting implementing the MFV schemes), Margarita Petkova (parallelisation and c2pap support), Paul Rohde (stellar feedback routines), Steffi Walch (supernova feedback routines) and Anthony Whitworth (discussions on trees and radiation algorithms).  {\refone We also thank the anonymous referee for helpful and detailed comments which have improved the clarity and readability of this paper.}

\bibliography{references_gandalf}

\bsp

\label{lastpage}

\end{document}